\documentclass{article}
\usepackage{graphicx} 

\title{The generalized Darboux matrices with the same poles and their applications}
\author{Yu-Yue Li \& Deng-Shan Wang\\
\small Laboratory of Mathematics and Complex Systems (Ministry of Education),\\
\small School of Mathematical Sciences, Beijing Normal University, Beijing 100875, China \\\small wangdsh1980@163.com}
\date{November 2024}

 \usepackage{amssymb}
\usepackage{amsmath}
\usepackage{amsfonts}

\usepackage{bbm}

\usepackage{graphicx}
\usepackage{float}
\usepackage{subfigure}

\usepackage{enumerate}   
\usepackage{CJK}

\usepackage{float}

\usepackage{caption}
\usepackage{xcolor}
\usepackage{mathtools}
\usepackage[all]{xy}

\usepackage{extarrows} 

\newif\ifinexp \inexpfalse

\newcommand{\tr}{\,\hbox{\rm tr}}

\newcommand{\QED}{QED.}
\newenvironment{demo}{\textbf{Proof. }}{\vskip4pt}

\newcommand{\longeq}[3]{\underset{#2}{\overset{#1}{\raise0.3ex\hbox{\hskip0.2em
 \vbox{\hrule width#3mm height0.25pt depth0pt\vskip1.6pt
 \hrule width#3mm height0.25pt depth0pt}\hskip0.4em}}}}

\newcommand{\revddots}{\mathinner{\mkern1mu\raise1pt\vbox{\kern7pt\hbox{.}}\mkern2mu
 \raise4pt\hbox{.}\mkern2mu\raise7pt\hbox{.}\mkern1mu}}

\renewcommand{\arraystretch}{1.1}

\newcommand\ucite[1]{\lower6pt\hbox{\large\cite{#1}}}

\newcommand{\C}{\mathbb{C}}

\newtheorem{lemma}{\bf Lemma}
\newtheorem{theorem}{\bf Theorem}
\newtheorem{remark}{\bf Remark}
\newtheorem{corollary}{\bf Corollary}
\newtheorem{example}{\bf Example}
\newtheorem{proposition}{\bf Proposition}

\begin{document}

\maketitle

\begin{abstract}
Darboux transformation plays a key role in constructing explicit closed-form solutions of completely integrable systems. This paper provides an algebraic construction of generalized Darboux matrices with the same poles for the $2\times2$ Lax pair, in which the coefficient matrices are polynomials of spectral parameter. The first-order monic Darboux matrix is constructed explicitly and its classification theorem is presented. Then by using the solutions of the corresponding adjoint Lax pair, the $n$-order monic Darboux matrix and its inverse, both sharing the same unique pole, are derived explicitly. Further, a theorem is proposed to describe the invariance of Darboux matrix regarding pole distributions in Darboux matrix and its  inverse. Finally, a unified theorem is offered to construct formal Darboux transformation in general form. All Darboux matrices expressible as the product of $n$ first-order monic Darboux matrices can be constructed in this way. The nonlocal focusing NLS equation, the focusing NLS equation and the Kaup-Boussinesq equation are taken as examples to illustrate the application of these Darboux transformations.

\end{abstract}

\section{Introduction}

The Darboux transformation (DT) originates from the so-called ``Darboux theorem" of Jean Gaston Darboux in the nineteenth century, which provides an effective way to obtain soliton solutions of integrable systems \cite{babich1986binary,cieslinski2009algebraic,gu2004darboux,matveev1991darboux,Xuhe2011}. Investigating exact solutions of integrable systems via Lax pairs concerns solving an inverse problem, where the potential functions can be recovered from the spectral data and eigenfunctions \cite{Chen2019,Zhijun2011,xia2016,zhou2008darboux}. A DT often offers a closed-form explicit solution to inverse problems of completely integrable equations. Recent years, much work has been done to explore the exact solutions of various integrable nonlinear equations. Cie{\'s}li{\'n}ski \cite{cieslinski2009algebraic} proposed algebraic construction of Darboux matrix for one-dimensional integrable systems. Shi, Nimmo, Zhang and Zhao \cite{shiying2017-1,shiying2017-2} studied the Darboux and binary DT of discrete integrable systems such as discrete potential KdV and mKdV equations.
Geiger, Horozov and Yakimov \cite{Geiger2017} presented a general theorem to establish the noncommutative bispectral DTs. Mucalica and Pelinovsky \cite{Pelinovsky2022} constructed DT to explore solitons of KdV equation with step-like boundary conditions. More recently,  M{\"u}ller-Hoissen \cite{Muller-Hoissen2023} gave a vectorial binary DT to generate multi-soliton solutions to the first member of the negative part
of AKNS hierarchy. However, there is a lack of unified way to construct Darboux matrices of integrable nonlinear equations. Thus in this work we will propose an algebraic construction of generalized Darboux matrices and take some examples to illustrate their efficiency.
\par
Consider a $2\times 2$ Lax pair of the form
\begin{equation}
\begin{array}{l}
    \Phi_x = U(x,t;\lambda) \Phi,\\
    \Phi_t = V(x,t;\lambda) \Phi,
\end{array}\label{eq:laxpair1}
\end{equation}
where $x,t\in\mathbb{R}$, $\lambda\in\mathbb{C}$ is the spectral parameter, $U(x,t;\lambda)=U_p(x,t)\lambda^p+\cdots+U_1(x,t)\lambda+U_0(x,t)$ is a polynomial of $\lambda$ with degree $p\in\mathbb{N}_+$ and $V(x,t;\lambda)=V_q(x,t)\lambda^q+\cdots+V_1(x,t)\lambda+V_0(x,t)$ is a polynomial of $\lambda$ with degree $q\in\mathbb{N}_+$.
\par
A polynomial matrix $D(x,t;\lambda)$ of $\lambda$ is called Darboux matrix if
\begin{equation}
    \begin{array}{l}
        \widetilde{U}(\lambda) = D_x D^{-1} + DUD^{-1},\\
        \widetilde{V}(\lambda) = D_t D^{-1} + DVD^{-1},
    \end{array}\label{eq:UwilVwilbegin}
\end{equation}
are still polynomial matrices of $\lambda$ of degrees $p$ and $q$, respectively.
\par
Many significant integrable models are compatible conditions of the Lax pair (\ref{eq:laxpair1}). For example, the Ablowitz-Kaup-Newell-Segur (AKNS) hierarchy \cite{ablowitz1974AKNS}, Kaup-Newell (KN) hierarchy \cite{Kaup1978KN}, and their physically important reduction cases such as KdV equation, modified  KdV equation, nonlinear Schr\"{o}dinger equation, Chen-Lee-Liu equation, Gerdjikov-Ivanov equation, etc \cite{gu2004darboux}.
\par
The mathematical structure of the classic DT for Lax pair (\ref{eq:laxpair1}) is a monic diagonalizable matrix polynomial of degree one
\begin{equation}\label{eq:BinaryDTmatrix}
    B(x,t;\lambda) = \lambda I - H(x,t) \left( \begin{matrix} \lambda_1 &0 \\ 0 & \lambda_2   \end{matrix} \right)H^{-1}(x,t),
\end{equation}
which is useful to find the soliton solutions of reduction systems of AKNS and KN hierarchies, see \cite{gu2004darboux,cieslinski2009algebraic} for details. Zakharov and Mikhailov \cite{zakharov1980integrability} found an equivalent form with pole for the classic DT (\ref{eq:BinaryDTmatrix}) of the form
\begin{equation}\label{eq:BinaryDTpole}
    B(x,t;\lambda) = I - \frac{ A_2(x,t,\lambda_1,\lambda_2) }{\lambda - \lambda_2}, \qquad B^{-1}(x,t;\lambda) = I + \frac{ A_1(x,t,\lambda_1,\lambda_2) }{\lambda - \lambda_1},
\end{equation}
where $A_1$ and $A_2$ are constructed according to both the solutions of the Lax pair (\ref{eq:laxpair1}) and the corresponding adjoint Lax pair. Here the equivalence is reflected in the sense that multiply the $B(x,t;\lambda)$ in (\ref{eq:BinaryDTpole}) by $\lambda-\lambda_2$ is just the $B(x,t;\lambda)$ in (\ref{eq:BinaryDTmatrix}), since the multiplication of Darboux matrix by a $(x,t)$-independent scalar function does not change (\ref{eq:UwilVwilbegin}). Notice that the characteristic of the classic DT (\ref{eq:BinaryDTmatrix}) is a diagonal Jordan canonical form, while the equivalent form (\ref{eq:BinaryDTpole}) is meromorphic function with poles $\lambda_1\neq\lambda_2$.
\par
Moreover, there is the third equivalent form of classic DT (\ref{eq:BinaryDTmatrix}). Assume that $B(x,t;\lambda)$ is a monic polynomial of $\lambda$ of degree one, namely $B(x,t;\lambda)=\lambda I - S(x,t)$, then $B(x,t;\lambda)$ is a classic DT if and only if the matrix-valued function $S(x,t)$ solves the linear equations
\begin{equation}\label{eq:BinaryDTLES}
    \left\{\begin{array}{l}
        B(x,t;\lambda_1) h_1(x,t;\lambda_1) = 0,\\
        B(x,t;\lambda_2) h_2(x,t;\lambda_2) = 0,
    \end{array}\right.  \Longleftrightarrow\; S\left(\begin{matrix}h_1 & h_2\end{matrix}\right)=\left(\begin{matrix}\lambda_1h_1 &\lambda_2h_2\end{matrix}\right),
\end{equation}
where $\lambda_1\neq\lambda_2$ with $\lambda_1, \lambda_2\in\mathbb{C}$, and the functions $h_1(x,t;\lambda)$ and $h_2(x,t;\lambda)$ are vector solutions of the Lax pair (\ref{eq:laxpair1}).
\par
However, the left equations of (\ref{eq:BinaryDTLES}) means that $h_1(x,t;\lambda)$ is transformed to a trivial solution of the new Lax pair at $\lambda=\lambda_1$, namely at a zero of ${\rm det}(B(\lambda))$, by the classic DT $B(x,t;\lambda)$. This may cause some difficulties for constructing the DTs for new Lax pair at $\lambda=\lambda_1$ (i.e., the second-fold DT for Lax pair (\ref{eq:laxpair1})) as it does not directly transform a fundamental solution matrix of the original Lax pair to the new Lax pair at $\lambda=\lambda_1$. The generalized DT $B^{\{n\}}(x,t;\lambda)$ is developed by Matveev and Salle \cite{matveev1991darboux} to get the nontrivial DTs for the new Lax pair at the zeros of ${\rm det}(B(\lambda))$. Subsequently, the generalized DT in \cite{guo2012nonlinear} attracts much attention since it constructs higher-order rogue waves solutions of the focusing NLS equation.
\par
The mathematical structure of the generalized DT $B^{\{n\}}(x,t;\lambda)$ is the product of $n$ classic DTs $B_j(x,t;\lambda)$ in (\ref{eq:BinaryDTmatrix}), namely
\begin{equation}
    B^{\{n\}}(x,t;\lambda) = B_n(x,t;\lambda) B_{n-1}(x,t;\lambda) \cdots B_1(x,t;\lambda),
\end{equation}
where each ${\rm det}(B_j(\lambda))$ for $j=1,2,\cdots,n$ has the same zeros $\lambda_1$ and $\lambda_2$. Moreover, $B^{\{n\}}(x,t;\lambda)$ can be rewritten as pole form below
\begin{equation}
    \begin{array}{l}
        B^{\{n\}}(x,t;\lambda) = I + \frac{B^{\{n\}}_1(x,t,\lambda_1,\lambda_2)}{\lambda-\lambda_1} + \cdots + \frac{B^{\{n\}}_n(x,t,\lambda_1,\lambda_2)}{(\lambda-\lambda_1)^n}, \\
        (B^{\{n\}})^{-1}(x,t;\lambda) = I + \frac{((B^{\{n\}})^{-1})_1(x,t,\lambda_1,\lambda_2)}{\lambda-\lambda_2} + \cdots + \frac{((B^{\{n\}})^{-1})_n(x,t,\lambda_1,\lambda_2)}{(\lambda-\lambda_2)^n},
    \end{array}\label{eq:GenaralizedDT}
\end{equation}
where $\lambda_1\neq\lambda_2$. The challenge here is to determine how to build generalized Darboux matrices of the pole type, as described in (\ref{eq:GenaralizedDT}), with the condition that $\lambda_1$ is identical to $\lambda_2$.
\par
Much work has been carried out to consider this topic. The first one is the classification theorem of the first order monic Darboux matrices, which states that if the Jordan canonical form of $D(\lambda)$ is diagonal, $D(\lambda)$ is the classic DT, while if the Jordan canonical form of $D(\lambda)$ is a Jordan block, $D(\lambda)$ is the limit of a series of classic DTs \cite{gu2004darboux}. The second one employs a DT through integration \cite{degasperis2007multicomponent}, yet it presents challenges when it comes to executing calculations and iterating through DTs. The third one is the ``classical" binary DT in \cite{cieslinski2009algebraic}, which states that the coefficient matrix is a nilpotent matrix. The fourth one involves the explicit construction of DTs for specific soliton equations through limit processes, which is exemplified by the construction of dark solitons in the defocusing NLS equation
\cite{han2013determinant}. The fifth one is to set the elements of $\Gamma$ matrix to be zero at the position that the eigenvalues equal to adjoint eigenvalues \cite{ma2021binary}. However, this DT is not well-defined when the union of sets of poles of DT and its inverse contains only one element.
\par
In the present paper, a key modification of the DT is to replace the limit process of the spectral parameter $\lambda$ with the derivative of $\lambda$. For the case that the order of pole is one, the limit version and the derivative version of DT are equivalent. Although this change seems simple, it is fundamentally important and has at least three benefits: (1) it can explain why there is a constant in the $\Gamma$ matrix, see Theorem~\ref{thm:DTonewithC} and Remark~\ref{rmk:DTitwithC}; (2) it allows for obtaining explicit formulas for the DTs with high-order pole, i.e., the Darboux matrix (\ref{eq:Dnpole}) with (\ref{eq:Dninvpole}), which is not reported before; (3) it is commonly considered that our DT and the classic DT are two completely different types of DTs, but in fact, both of them can be unified by a wonderful form, i. e., the formula in (\ref{eq:DTgeneral}).
\par
We are going to investigate the $2\times 2$ Darboux matrix $D(x,t;\lambda)$ in pole form with the properties below:
\par
(i) $D(x,t;\lambda)$ has single pole $\lambda_0\in \mathbb{C}$ of order $n\in \mathbb{N}_+$:
\begin{equation}\label{eq:DTngeneral}
    D(x,t;\lambda) = I + \frac{D_1(x,t,\lambda_0)}{\lambda-\lambda_0} + \cdots + \frac{D_n(x,t,\lambda_0)}{(\lambda-\lambda_0)^n},
\end{equation}
where $I$ is the identity matrix.
\par
(ii) $D(x,t;\lambda)$ is invertible, whose inverse is
\begin{equation}\label{eq:DTInvngeneral}
    D^{-1}(x,t;\lambda) = I + \frac{(D^{-1})_1(x,t,\lambda_0)}{\lambda-\lambda_0} + \cdots + \frac{(D^{-1})_n(x,t,\lambda_0)}{(\lambda-\lambda_0)^n},
\end{equation}
\par
(iii) $\widetilde{\Phi}=D(x,t;\lambda)\Phi$ satisfies the following $2\times 2$ Lax pair
\begin{equation}
\begin{array}{l}
    \widetilde{\Phi}_x = \widetilde{U}(x,t;\lambda) \widetilde{\Phi},\\
    \widetilde{\Phi}_t = \widetilde{V}(x,t;\lambda) \widetilde{\Phi},
\end{array}\label{eq:laxpair2}
\end{equation}
where $\Phi(x,t;\lambda)$ is a fundamental solution matrix of Lax pair (\ref{eq:laxpair1}), $\widetilde{U}(x,t;\lambda)=\widetilde{U}_p(x,t)\lambda^p+\cdots+\widetilde{U}_1(x,t)\lambda+\widetilde{U}_0(x,t)$ is a polynomial of $\lambda$ with degree $p$ and $\widetilde{V}(x,t;\lambda)=\widetilde{V}_q(x,t)\lambda^q+\cdots+\widetilde{V}_1(x,t)\lambda+\widetilde{V}_0(x,t)$ is a polynomial of $\lambda$ with degree $q$. Here $p,q$ are the same as that of the original Lax pair (\ref{eq:laxpair1}).
\par
This work is organized as follow:
\par
(1) The case $n=1$ in (\ref{eq:DTngeneral}) is studied in Section \ref{Se:2}, where the explicit construction of $D(\lambda)$ and a classification theorem of the first-order monic Darboux matrix
are provided. It is shown that this kind of DT is similar to a Jordan block, in which $D(\lambda)$ and $D^{-1}(\lambda)$ have the same pole. By the way, the classification theorem indicates that the compatible equation of the nonlinear ODEs (\ref{eq:S_xdffeq}) and (\ref{eq:S_tdffeq}) is solvable for any initial values $S(x_0,t_0)\in {\rm Mat}(2,\mathbb{C})$.
\par
(2) The case $n\in\mathbb{N}_+$ in (\ref{eq:DTngeneral}) is studied in Section \ref{Se:3}. We investigate
how to fold $n$ first-order DTs to generate a Darboux matrix $D(\lambda)$ of degree $n$, where each first-order DT is  similar to the Jordan block $\left( \begin{matrix}\lambda_0 & 1 \\ 0 &\lambda_0   \end{matrix} \right)$. Then, by using the solutions of the corresponding adjoint Lax pair of (\ref{eq:laxpair1}), the explicit pole form (\ref{eq:DTngeneral}) of Darboux matrix $D(\lambda)$ is derived. Furthermore, this pole form of DT is adopted to construct global multiple-ploe solitons of the $x$-nonlocal focusing NLS equation. These solutions are rational, decay to zero when $x\to \pm\infty$, and  the corresponding spectral parameter is exactly $\lambda=0$. The velocity of the single soliton is zero, while the velocities of the multi-pole solitons are of order $\mathcal{O}\left(\frac{1}{\sqrt{t}}\right)$.
\par
(3) Section \ref{Section4} proposes a theorem to describe the invariance of Darboux matrices regarding pole distributions in $D(\lambda)$ and $D^{-1}(\lambda)$. There are two invariants, one is the order of the DT and the other is the sum of order for poles $\lambda_j$ of $D(\lambda)$ and $D^{-1}(\lambda)$. As an application, the rogue waves and multi-pole solitons in the focusing NLS equation are obtained explicitly.
\par
(4) Section \ref{Section5} offers a unified theorem to construct formal DT in general form, which contains the DTs in (\ref{eq:GenaralizedDT}) and (\ref{eq:DTngeneral}). In generic, the Darboux matrices expressed by the product of $n$ first-order monic Darboux matrices can be constructed by solving the linear equations (\ref{eq:DTgeneral}). The converse is also true when ignoring a constant coefficient. Subsequently, the Kaup-Boussinesq equation serves as an illustration of the application of our theorem, with several new exact solutions presented explicitly. Moreover, some examples that the Darboux matrices can not be expressed by the product of first-order monic Darboux matrices are also given.

\section{The case $n=1$} \label{Se:2}

This section derives the formula of Darboux matrix (\ref{eq:DTngeneral}) for case $n=1$ following the procedure of Zakharov and Mikhailov \cite{zakharov1980integrability}, see also \cite{gu2004darboux,cieslinski2009algebraic}.
\begin{theorem}\label{thm:DT1}
    Let $b(x,t;\lambda)$ be a $2\times 1$ solution of Lax pair (\ref{eq:laxpair1}), and $a(x,t;\lambda)$ be a $1\times 2$ solution of the following adjoint Lax pair
    \begin{equation}
    \begin{array}{l}
    \Psi_x = -\Psi U(x,t;\lambda),\\
    \Psi_t = -\Psi V(x,t;\lambda).
    \end{array}\label{eq:adjointlaxpair}
    \end{equation}
    Denote $a_0 = a(x,t;\lambda_0)$, $b_0 = b(x,t;\lambda_0)$ and $b_0^{(1)}=\left.\frac{\partial}{\partial \lambda}(b(x,t;\lambda))\right|_{\lambda=\lambda_0}$. Suppose $a(x,t;\lambda)$ and $b(x,t;\lambda)$ are analytic at $\lambda=\lambda_0$ with
    \begin{equation}\label{eq:condition1}
        a_0 b_0 = 0,
    \end{equation}
    then the Darboux matrix (\ref{eq:DTngeneral}) and its inverse (\ref{eq:DTInvngeneral}) for case $n=1$ are
    \begin{equation}
    \begin{array}{l}
        D(x,t;\lambda) = I - \frac{b_0(a_0b_0^{(1)})^{-1} a_0}{\lambda-\lambda_0},\\
        D^{-1}(x,t;\lambda) = I + \frac{b_0(a_0b_0^{(1)})^{-1} a_0}{\lambda-\lambda_0},
    \end{array}\label{eq:DD-1order1}
    \end{equation}
    which satisfy

    (i) $D(x,t;\lambda)D^{-1}(x,t;\lambda) = I$,

    (ii) $\widetilde{U}(x,t;\lambda) \triangleq D_x(x,t;\lambda)D^{-1}(x,t;\lambda) + D(x,t;\lambda)U(x,t;\lambda)D^{-1}(x,t;\lambda)$ is a polynomial of $\lambda$ with degree $p$, and $\widetilde{V}(x,t;\lambda) \triangleq D_t(x,t;\lambda)D^{-1}(x,t;\lambda) + D(x,t;\lambda)V(x,t;\lambda)$ $D^{-1}(x,t;\lambda)$ is a polynomial of $\lambda$ with degree $q$. Moreover, $\widetilde{\Phi}=D(x,t;\lambda)\Phi$ solves the new Lax pair (\ref{eq:laxpair2}).
\end{theorem}

\begin{demo}
(i) The definition (\ref{eq:DD-1order1}) shows
\begin{equation}
    D(x,t;\lambda)D^{-1}(x,t;\lambda) = I - \frac{b_0(a_0b_0^{(1)})^{-1} a_0b_0(a_0b_0^{(1)})^{-1} a_0}{(\lambda-\lambda_0)^2} = I,
\end{equation}
in which the second equality is from the condition (\ref{eq:condition1}).
\par
(ii.1) Suppose $\widetilde{U}(\lambda)=\widetilde{U}(x,t;\lambda)$ is an entire function, then $\lambda=0$ is a zero of the entire function $G(\lambda)\triangleq \widetilde{U}(\lambda)-\widetilde{U}(0)$, thus there exists an entire function $G_1(\lambda)$, such that
\begin{equation}
    \widetilde{U}(\lambda)-\widetilde{U}(0) = G(\lambda) = \lambda G_1(\lambda).
\end{equation}
Similarly, construct entire functions $G_k(\lambda)$ of the forms
\begin{equation}
    G_{k-1}(\lambda)-G_{k-1}(0) = \lambda G_k(\lambda),\quad k=2,3,\cdots,p+1,
\end{equation}
then it follows
\begin{equation}
    G_{p+1}(\lambda) = \frac{G_{p}(\lambda)-G_{p}(0)}{\lambda}=\cdots = \frac{\widetilde{U}(\lambda)}{\lambda^{p+1}}-\frac{\widetilde{U}(0)}{\lambda^{p+1}} - \sum_{k=1}^{p}\frac{G_k(0)}{\lambda^{p+1-k}}.
\end{equation}
Since $U(\lambda)$ is a polynomial of $\lambda$ with degree $p$, one has
\begin{equation}
    \lim_{\lambda\rightarrow\infty} G_{p+1}(\lambda) = \lim_{\lambda\rightarrow\infty}  \frac{\widetilde{U}(\lambda)}{\lambda^{p+1}} = 0.
\end{equation}
According to Liouville's theorem, $G_p(\lambda)=0$, then $\widetilde{U}(\lambda)$ is a polynomial of $\lambda$.
\par
Using the conclusion (i) of this theorem and the definition of $\widetilde{U}(\lambda)$, we have
\begin{equation}\label{eq:UDDxDU}
    \widetilde{U}(x,t;\lambda)D(x,t;\lambda) = D_x(x,t;\lambda) + D(x,t;\lambda)U(x,t;\lambda),
\end{equation}
then $\widetilde{U}(\lambda)$ is a polynomial of $\lambda$ with degree $p$ by comparing the order of $\lambda$. Moreover, it is easy to see
\begin{equation}
    \widetilde{\Phi}_x = (D\Phi)_x = D_x \Phi + D \Phi_x = D_x D^{-1} D \Phi + D U D^{-1} D \Phi =\widetilde{U} \widetilde{\Phi}.
\end{equation}

(ii.2) It remains to prove that $\widetilde{U}(x,t;\lambda)$ is an entire function.
\par
Obviously, the function $\widetilde{U}(\lambda)$ is holomorphic everywhere except the pole $\lambda=\lambda_0$. Since $U(\lambda)$ is a polynomial of $\lambda$, it is obvious that the function $U(\lambda)$ is analytic at $\lambda=\lambda_0$ and the Taylor expansion below holds
\begin{equation}
    U(\lambda) = U_0 + U_0^{(1)} (\lambda-\lambda_0) + \frac{1}{2} U_0^{(2)} (\lambda-\lambda_0)^2 + O((\lambda-\lambda_0)^3),
\end{equation}
where $U_0^{(k)}=\left.\frac{\partial^k}{\partial \lambda^k}(U(x,t;\lambda))\right|_{\lambda=\lambda_0}$ with $k\in \mathbb{N}$. Denote $A = b_0(a_0b_0^{(1)})^{-1} a_0$, then one has $D(\lambda) = I - \frac{A}{\lambda-\lambda_0}$, $D^{-1}(\lambda) = I + \frac{A}{\lambda-\lambda_0}$ and
\begin{equation}
    \widetilde{U}(\lambda) = -\frac{A_xA+AU_0A}{(\lambda-\lambda_0)^2} - \frac{A_x-U_0A+AU_0+AU_0^{(1)}A}{\lambda-\lambda_0}+\rm{analytic}\;\rm{part}.
    \label{eq:widetildeU}
\end{equation}
\par
Noticing that $a_0b_0=0$, and $b_0$ and $a_0$ are the solutions of the Lax pair (\ref{eq:laxpair1}) and adjoint Lax pair (\ref{eq:adjointlaxpair}) at $\lambda=\lambda_0$, respectively, it follows
\begin{small}
\begin{equation*}
\begin{array}{ll}
    A_x + AU_0&= U_0b_0(a_0b_0^{(1)})^{-1}a_0 - b_0(a_0b_0^{(1)})^{-1}(a_0b_0^{(1)})_x (a_0b_0^{(1)})^{-1}a_0\\
                 &=  U_0A + b_0(a_0b_0^{(1)})^{-1}a_0U_0b_0^{(1)} (a_0b_0^{(1)})^{-1}a_0 -b_0(a_0b_0^{(1)})^{-1}a_0(b_0^{(1)})_x (a_0b_0^{(1)})^{-1}a_0\\
                 &=  U_0A + b_0(a_0b_0^{(1)})^{-1}a_0U_0b_0^{(1)} (a_0b_0^{(1)})^{-1}a_0 -b_0(a_0b_0^{(1)})^{-1}a_0(Ub)_0^{(1)} (a_0b_0^{(1)})^{-1}a_0\\
                 &= U_0A -b_0(a_0b_0^{(1)})^{-1}a_0U_0^{(1)}b_0 (a_0b_0^{(1)})^{-1}a_0\\
                 &= U_0A - AU_0^{(1)}A.
\end{array}
\end{equation*}
\end{small}
That is, $A_x-U_0A+AU_0+AU_0^{(1)}A=0$. Further, we have $A_xA+AU_0A=U_0A^2-AU_0^{(1)}A^2=0$. Then $\widetilde{U}(x,t;\lambda)$ in (\ref{eq:widetildeU}) is an entire function.
\par
The proof for $\widetilde{V}(x,t;\lambda)$ is similar. \QED
\end{demo}
\par
It seems that the condition $a_0b_0 = 0$ is weird. However, one can convert $D(\lambda)$ into a polynomial of $\lambda$ with degree one, then it is observed that $a(x,t;\lambda)$ is an auxiliary function. For $D(x,t;\lambda)$ in (\ref{eq:DD-1order1}), define
\begin{equation}\label{eq:order1S}
    D_P(x,t;\lambda) = \lambda I - S(x,t) = (\lambda-\lambda_0) D(x,t;\lambda),
\end{equation}
then its inverse is $D_P^{-1}(\lambda) = \frac{1}{\lambda-\lambda_0}D^{-1}(\lambda)$. There are two more equivalent descriptions of $D_P(x,t;\lambda)$. (Since the polynomial form and the pole form of $D(\lambda)$ differ by only a scalar factor, both $D_P(\lambda)$ and $D(\lambda)$ will be represented by $D(\lambda)$ henceforth.)

\begin{theorem}\label{thm:order1threeequi}

    (i) The $2\times2$ matrix $S(x,t)$ in (\ref{eq:order1S}) is the solution of the following linear equations
    \begin{equation}\label{eq:lineareqorder1}
        \left\{
        \begin{array}{l}
            \left.(\lambda I - S)b(\lambda)\right|_{\lambda=\lambda_0} = 0,\\
            \left.\frac{\partial}{\partial \lambda}\Big((\lambda I -S)b(\lambda)\Big)\right|_{\lambda=\lambda_0} = 0,
        \end{array}
        \right.
    \end{equation}
    namely,
    \begin{equation}\label{eq:lineareqofSsmporder1}
        S \left(\begin{matrix} b_0 & b_0^{(1)}   \end{matrix} \right) = \left(\begin{matrix}\lambda_0 b_0 & \lambda_0b_0^{(1)} +b_0  \end{matrix} \right).
    \end{equation}
    (ii) $S=H\Lambda H^{-1}$, where $H=\left(\begin{matrix} b_0 & b_0^{(1)}   \end{matrix} \right)$ and $\Lambda  = \left( \begin{matrix} \lambda_0&1\\
    0& \lambda_0 \end{matrix} \right)$.
\end{theorem}

\begin{demo}
(i) According to the definition of $D(x,t;\lambda)$ in (\ref{eq:DD-1order1}) and $S(x,t)$ in (\ref{eq:order1S}), it follows
\begin{equation}
    S(x,t) = \lambda_0 I + b_0(a_0b_0^{(1)})^{-1}a_0.
\end{equation}
Then one has
\begin{equation*}
\begin{array}{l}
    \left.(\lambda I - S)b(\lambda)\right|_{\lambda=\lambda_0} = (\lambda_0 I-S)b_0 = -b_0(a_0b_0^{(1)})^{-1}a_0b_0 = 0, \\
    \left.\frac{\partial}{\partial \lambda}\Big((\lambda I -S)b(\lambda)\Big)\right|_{\lambda=\lambda_0} = b_0 + (\lambda_0 I-S)b_0^{(1)} = b_0 - b_0 (a_0b_0^{(1)})^{-1}a_0b_0^{(1)}=0.
\end{array}
\end{equation*}
\par
(ii) Direct calculations show that $S=H\Lambda H^{-1}$ is also the solution of equation (\ref{eq:lineareqorder1}). \QED

\end{demo}

\begin{remark}
In principle, three matrices should be invertible. The first one is $a_0b_0^{(1)}$ in (\ref{eq:DD-1order1}), the second one is the coefficient matrix in (\ref{eq:lineareqofSsmporder1}), and the third one is the matrix $H=(b_0~~~~b_0^{(1)})$. However, for the case $n=1$, the second one is the same as the third one. The relationship between the first one and the third one is demonstrated in the following proposition.
\end{remark}

\begin{proposition}\label{propo:threeinvequn1}
     In Theorem~\ref{thm:order1threeequi}, the invertibility of matrix $H=(b_0~~~~b_0^{(1)})$ is equivalent to the invertibility of $a_0b_0^{(1)}$.
\end{proposition}

\begin{demo}
Let $\Psi_0$ be the fundamental solution matrix of the adjoint Lax pair (\ref{eq:adjointlaxpair}) at $\lambda = \lambda_0$, then $\Psi_0 b_0$ is a $2\times 1$ constant vector since $(\Psi_0 b_0)_x = (\Psi_0 b_0)_t = 0$. As $\Psi_0$ is a fundamental solution matrix, $\Psi_0 b_0\neq 0$. Without loos of generality, assume that the first row of $\Psi_0 b_0$ is not zero, and denote the first row of matrix $\Psi_0$ as $a_1$, then it is immediately that $a_1b_0\neq 0$. Now consider the $2\times 2$ matrix
\begin{equation}
    \left(\begin{matrix} a_0 \\ a_1 \end{matrix}\right) H = \left(\begin{matrix} a_0 \\ a_1 \end{matrix}\right) (b_0,b_0^{(1)}) = \left(\begin{matrix} 0 & a_0b_0^{(1)} \\ a_1b_0 & * \end{matrix}\right).
\end{equation}
Notice that $a_0$ and $a_1$ are linearly independent since $a_1b_0\neq 0$ and $a_0 b_0 =0$, then $H$ is invertible if and only if $a_0b_0^{(1)} \neq 0$.
\QED
\end{demo}

\subsection{The constant $C$ in $\Gamma$ matrix}\label{subsec:CinGamman1}

For the people who is familiar with the topic of DT, the term $a_0b_0^{(1)}$ in Proposition~\ref{propo:threeinvequn1} is the $\Gamma$ matrix of DT in pole form for $n=1$. In this subsection, it will be seen that $D(\lambda)$ and $D^{-1}(\lambda)$ in (\ref{eq:DD-1order1}) is still a Darboux matrix if one replaces $a_0b_0^{(1)}$ with $a_0b_0^{(1)}+C$, where $C$ is an arbitrary constant.
The DT with constant $C$ by limit processing can be seen in \cite{han2013determinant}, which provides a convenient way to get abundant solitons. However, the DT with constant $C$ is not essential, since it can be included in the original Theorem~\ref{thm:DT1} by selecting the appropriate constant vector coefficients $\xi(\lambda)$ in the proof of the following theorem.

\begin{theorem}\label{thm:DTonewithC}
    Let $b(x,t;\lambda)$ be a $2\times 1$ solution of Lax pair (\ref{eq:laxpair1}), and $a(x,t;\lambda)$ be a $1\times 2$ solution of adjoint Lax pair (\ref{eq:adjointlaxpair}). Moreover, suppose $a(x,t;\lambda)$ and $b(x,t;\lambda)$ are analytic at $\lambda=\lambda_0$ with
    \begin{equation}
        a_0 b_0 = 0,
    \end{equation}
    then the Darboux matrix (\ref{eq:DTngeneral}) and its inverse (\ref{eq:DTInvngeneral}) for case $n=1$ are
    \begin{equation}
    \begin{array}{l}
        D(x,t;\lambda) = I - \frac{b_0(a_0b_0^{(1)}+C)^{-1} a_0}{\lambda-\lambda_0},\\
        D^{-1}(x,t;\lambda) = I + \frac{b_0(a_0b_0^{(1)}+C)^{-1} a_0}{\lambda-\lambda_0},
    \end{array}\label{eq:DTorder1withC}
    \end{equation}
where $C\in\mathbb{C}$ is a constant.
\end{theorem}

\begin{demo}
Since $b(\lambda)$ is a $2\times 1$ solution of Lax pair (\ref{eq:laxpair1}), then there exists a $2\times 1$ vector $\xi(\lambda)$, which is independent of $x$ and $t$, such that $b(x,t;\lambda)=\Phi(x,t;\lambda)\xi(\lambda)$. Since $a(\lambda)$ is a $1\times 2$ solution of adjoint Lax pair (\ref{eq:adjointlaxpair}), then there exists a $2\times 1$ vector $\eta(\lambda)$, which is independent of $x$ and $t$, such that $a(x,t;\lambda)=\eta(\lambda)\Phi^{-1}(x,t;\lambda)$. (Notice that $\Phi^{-1}(x,t;\lambda)$ is a fundamental solution matrix of adjoint Lax pair (\ref{eq:adjointlaxpair}).)
\par
Now consider another $2\times 1$ vector $\widetilde{\xi}(\lambda)$ as
\begin{equation}
    \widetilde{\xi}(\lambda) = \xi(\lambda_0) + (\lambda-\lambda_0)\xi_v,
\end{equation}
where $\xi_v$ is a $2\times 1$ constant vector which is independent of $x$, $t$ and $\lambda$.
\par
Define a solution of Lax pair (\ref{eq:laxpair1}) as
\begin{equation}
    \widetilde{b}(x,t;\lambda) = \Phi(x,t;\lambda)\widetilde{\xi}(\lambda),
\end{equation}
since $\Phi(x,t;\lambda)$ is a fundamental solution matrix of Lax pair (\ref{eq:laxpair1}). Then one has
\begin{equation}
    a_0 \widetilde{b}_0 = \eta(\lambda_0)\Phi^{-1}(x,t;\lambda_0)\Phi(x,t;\lambda_0)\widetilde{\xi}(\lambda_0) =  \eta(\lambda_0)\xi(\lambda_0) = a_0b_0 = 0.
\end{equation}
Thus according to Theorem~\ref{thm:DT1},
\begin{equation}
    \begin{array}{l}
        D(x,t;\lambda) = I - \frac{\widetilde{b}_0(a_0 \widetilde{b}_0^{(1)})^{-1} a_0}{\lambda-\lambda_0},\\
        D^{-1}(x,t;\lambda) = I + \frac{\widetilde{b}_0(a_0\widetilde{b}_0^{(1)})^{-1} a_0}{\lambda-\lambda_0},
    \end{array}\label{eq:DTwidetildeb}
\end{equation}
provide a Darboux matrix and its inverse.
\par
Notice that
\begin{small}
\begin{equation*}
    \begin{array}{l}
        \widetilde{b}_0 =  \Phi(\lambda_0)\widetilde{\xi}(\lambda_0) = \Phi(\lambda_0)\xi(\lambda_0) = b(\lambda_0) = b_0,\\
        a_0\widetilde{b}_0^{(1)} = \eta(\lambda_0)\Phi^{-1}(\lambda_0)\Big(\Phi_0 \left.\frac{\partial}{\partial \lambda}\left(\widetilde{\xi}(\lambda)\right)\right|_{\lambda=\lambda_0} + \Phi_0^{(1)}\widetilde{\xi}(\lambda_0) \Big) = \eta_0 \Phi_0^{-1}\Big( \Phi_0 \widetilde{\xi}_0^{(1)} + \Phi_0^{(1)}\xi_0  \Big)\\
        =\eta_0 \Phi_0^{-1}\Big( \Phi_0 \xi_v +  b_0^{(1)} - \Phi_0\xi_0^{(1)}  \Big) = (\eta_0\xi_v -\eta_0\xi_0^{(1)} )+ a_0 b_0^{(1)},
    \end{array}
\end{equation*}
\end{small}
where $\eta_0=\eta(\lambda_0).$ Since $\eta_0\neq 0$, there exists a vector $\xi_v$ such that $\eta_0\xi_v -\eta_0\xi_0^{(1)} = C$ for any given $C\in\mathbb{C}$. Select a proper $\xi_v$, then (\ref{eq:DTwidetildeb}) is the same as (\ref{eq:DTorder1withC}). \QED

\end{demo}

\begin{remark}\label{rmk:DTitwithC}
Multiplying the $D(\lambda)$ in (\ref{eq:DTorder1withC}) by a scalar constant $\lambda-\lambda_0$ and denoting it by $D(\lambda)$, it follows
\begin{equation}\label{eq:DT1withCPoly}
    D(\lambda) = (\lambda-\lambda_0)I - b_0(a_0b_0^{(1)}+C)^{-1} a_0, \quad C\neq 0,
\end{equation}
which satisfies
\begin{equation}
        \left\{
        \begin{array}{l}
            \left.D(\lambda)b(\lambda)\right|_{\lambda=\lambda_0} = 0,\\
            \left.\frac{\partial}{\partial \lambda}\Big(D(\lambda)b(\lambda)\Big)\right|_{\lambda=\lambda_0} \neq 0.
        \end{array}
        \right.
\end{equation}
That is to say the equation (\ref{eq:lineareqorder1}) in Theorem~\ref{thm:order1threeequi} is no longer applicable. However, according to the proof of Theorem~\ref{thm:DTonewithC}, the correct representation of the equation (\ref{eq:lineareqorder1}) for DT with $C$ in (\ref{eq:DT1withCPoly}) would be
\begin{equation}
        \left\{
        \begin{array}{l}
            \left.D(\lambda)\widetilde{b}(\lambda)\right|_{\lambda=\lambda_0} = 0,\\
            \left.\frac{\partial}{\partial \lambda}\Big(D(\lambda)\widetilde{b}(\lambda)\Big)\right|_{\lambda=\lambda_0} = 0,
        \end{array}
        \right.
\end{equation}
which means that even if one knows the $D(\lambda)$ of the form (\ref{eq:DT1withCPoly}), one does not know the solutions $\widetilde{b}(\lambda)$ of the Lax pair.
\par
On one hand, the nonzero constant $C$ is useful in constructing non-trivial solutions of integrable systems; on the other hand, the fact that one does not know $\widetilde{b}(\lambda)$ will make some troubles when constructing higher-order Darboux matrices.
\par
In this work, for the sake of simplicity, we use the derivatives $\xi_0$, $\xi_0^{(1)},\cdots$ of $\xi(\lambda)$ as free parameters of $b(x,t;\lambda)=\Phi(x,t;\lambda)\xi(\lambda)$, instead of the constant $C$ in $\Gamma$ matrix. For more information, see the proof of  Theorem~\ref{thm:DTonewithC} above or the Remark~\ref{rmk:DTwithCvarxin} below.

\end{remark}

\subsection{Classification theorem
of the first order monic DTs}
\par
\par
Until now, two kinds of DTs of the form $\lambda I - S(x,t)$ are provided.
\par
Denote the Darboux matrix
\begin{equation}\label{eq:DTsamepole11}
    D[\lambda_0,1,b(\lambda)](x,t;\lambda) \triangleq \lambda I - H(x,t) \Lambda H^{-1}(x,t),
\end{equation}
where $\lambda_0\in\mathbb{C}$, $b(x,t;\lambda)$ is analytic at $\lambda=\lambda_0$, and
\begin{equation}\label{eq:DTsamepole12}
    H(x,t) = \left(\begin{matrix} b_0(x,t;\lambda_0) & b_0^{(1)}(x,t;\lambda_0) \end{matrix}\right),\qquad \Lambda =\left(\begin{matrix} \lambda_0 & 1\\ 0& \lambda_0 \end{matrix}\right).
\end{equation}
\par
Further, denote the classic Darboux matrix
\begin{equation}\label{eq:binaryDT1}
    B[\lambda_1,\lambda_2,1,b_1(\lambda),b_2(\lambda)](x,t;\lambda) \triangleq \lambda I - H(x,t) \Lambda H^{-1}(x,t),
\end{equation}
where $\lambda_1\neq \lambda_2$, $b_1(\lambda)$ and $b_2(\lambda)$ are analytic at $\lambda_1$ and $\lambda_2$ respectively, and
\begin{equation}\label{eq:binaryDT2}
    H(x,t) = \left(\begin{matrix} b_1(x,t;\lambda_1) & b_2(x,t;\lambda_2) \end{matrix}\right),\qquad \Lambda =\left(\begin{matrix} \lambda_1 & 0\\ 0& \lambda_2 \end{matrix}\right).
\end{equation}

Here the letters $B$ and $D$ are used to distinguish two types of DTs, in which $B$ represents classic DT and $D$ represents DT with the same pole. The information on how the DT is constructed is contained within the brackets $[\cdots]$, while the variables $x$, $t$ and $\lambda$, which are the independent variables of the DT, are enclosed in parentheses $(\cdots)$.
\par
Notice that the Jordan canonical form of Darboux matrices in Theorem~\ref{thm:order1threeequi} is a Jordan block of order two (the conclusion (ii) in  Theorem~\ref{thm:order1threeequi}), while the Jordan canonical form of classic Darboux matrices is a diagonal matrix (refer to equations (\ref{eq:binaryDT1}) and (\ref{eq:binaryDT2})). Therefore, one can get the Darboux matrices in Theorem~\ref{thm:order1threeequi} with some limit processes of classic Darboux matrices.
\par
Theorem~\ref{thm:order1threeequi} not only provides explicit expressions, but also completes the classification of the first order monic DTs. Following the procedure of \cite{gu2004darboux}, it is seen that the case with Jordan block can be approximated by the diagonalizable case.
\par
\begin{theorem}[Classification of the first order monic DTs]\label{thm:classificationDT1}
    If $D(\lambda)=\lambda I -S(x,t)$ is a Darboux matrix for the Lax pair (\ref{eq:laxpair1}), then $D(\lambda)$ is one of the following three types:\\
    (i)  Trivial, i.e, $D(\lambda)=(\lambda-c) I$, where $c$ is a scalar constant.\\
    (ii) Classic Darboux matrix of the form (\ref{eq:binaryDT1}) with (\ref{eq:binaryDT2}).\\
    (iii) Darboux matrix with the same pole of form (\ref{eq:DTsamepole11}) with (\ref{eq:DTsamepole12}).
\end{theorem}

\begin{demo}
    The first equation of (\ref{eq:UwilVwilbegin}) is equivalent to
    \begin{equation}\label{eq:UwilDDx+DU}
    \widetilde{U}(\lambda)D(\lambda) = D_x(\lambda) + D(\lambda)U(\lambda).
    \end{equation}
     If $\widetilde{U}(\lambda)$ is a polynomial of $\lambda$, then
    \begin{equation}
        \widetilde{U}(\lambda) = \sum_{j=0}^{p} \widetilde{U}_j \lambda^{j},
    \end{equation}
    which is of degree $p$ since $D(\lambda)$ is monic. By comparing the coefficients of $\lambda$ in (\ref{eq:UwilDDx+DU}), it can be checked that
    \begin{equation}
        \begin{array}{l}
            \lambda^{p+1} :\qquad  \widetilde{U}_p = U_p,\\
            \lambda^{j} : \qquad -\widetilde{U}_j S  + \widetilde{U}_{j-1} = U_{j-1} - S U_{j}, \quad j = 1,\cdots p,\\
            \lambda^0: \qquad -\widetilde{U}_0 S = -S_x - S U_0.
        \end{array}\label{eq:lambdasperaUwilDDx+DU}
    \end{equation}
    Solving $\widetilde{U}_j$ step by step yields
    \begin{equation}
        \begin{array}{l}
            \widetilde{U}_p = U_p,\\
            \widetilde{U}_j = U_j + \left[\sum_{k=j+1}^{p}U_k S^{k-j-1},S  \right],\quad  j = 0,\cdots,p-1,\\
        \end{array}\label{eq:Uwildef}
    \end{equation}
where $S$ solves the ordinary differential equation
    \begin{equation}\label{eq:S_xdffeq}
        S_x = \left[\sum_{k=0}^{p}U_k S^{k},S  \right].
    \end{equation}
    By applying similar processes to the second equation of (\ref{eq:UwilVwilbegin}), one has
    \begin{equation}\label{eq:S_tdffeq}
        S_t = \left[\sum_{k=0}^{q}V_k S^{k},S  \right].
    \end{equation}

    So the equations (\ref{eq:S_xdffeq}) and (\ref{eq:S_tdffeq}) are necessary conditions for that $D(\lambda) = \lambda I -S(x,t)$ is a Darboux matrix. In fact, they are also the sufficient conditions.

    If $S(x,t)$ is a solution of the equations (\ref{eq:S_xdffeq}) and (\ref{eq:S_tdffeq}), $\widetilde{U}_j$ can be defined as (\ref{eq:Uwildef}). Then the equations in (\ref{eq:lambdasperaUwilDDx+DU}) hold, which leads to (\ref{eq:UwilDDx+DU}) with $\widetilde{U}(\lambda)$ being a polynomial of $\lambda$. The proof for the $t$-equation is similar, so $D(\lambda)$ is a Darboux matrix.

    Now, it can be claimed that for any $(x_0,t_0)\in\mathbb{R}^2$ and $C\in{\rm Mat}(2,\mathbb{C})$, there exists a solution $S(x,t)$ of the equations (\ref{eq:S_xdffeq}) and (\ref{eq:S_tdffeq}) with initial condition
    \begin{equation}\label{eq:Sinitial}
        S(x_0,t_0) = C.
    \end{equation}
\par
Considering the Jordan canonical form of $C$ as follows
    \begin{equation}
        J = P^{-1}  C P,
    \end{equation}
there are three possible kinds of $J$ since $C$ is a $2\times 2$ matrix:\\
\par
    (a) If $J=cI$, where $c$ is a constant, then $S(x,t)=cI$ solves the equations (\ref{eq:S_xdffeq}) and (\ref{eq:S_tdffeq}). The corresponding $D(\lambda) = \lambda I -S = (\lambda-c)I$ is trivial.
\par
    (b) If $J=\left( \begin{matrix} \lambda_1 & 0 \\ 0 & \lambda_2   \end{matrix} \right)$ and $P=(p_1~~~p_2)$, where $\lambda_1\neq\lambda_2$, consider the solutions $h_1(x,t)$ and $h_2(x,t)$ of Lax pair (\ref{eq:laxpair1}) at $\lambda=\lambda_1$ and $\lambda=\lambda_2$ respectively with the initial values
    \begin{equation}
        h_1(x_0,t_0) = p_1,\qquad h_2(x_0,t_0) = p_2,
    \end{equation}
    then $B[\lambda_1,\lambda_2,1,h_1(x,t),h_2(x,t)]$ in (\ref{eq:binaryDT1}) with (\ref{eq:binaryDT2})
    is the Darboux matrix, in which $S(x,t)=H(x,t) \Lambda H^{-1}(x,t)$ satisfies the initial condition (\ref{eq:Sinitial}).
\par
    (c) If $J=\left( \begin{matrix} \lambda_0 & 1 \\ 0 & \lambda_0   \end{matrix} \right)$ and $P=(p_1~~~p_2)$, consider the fundamental solution matrix $\Phi(x,t;\lambda)$ of Lax pair (\ref{eq:laxpair1}), and denote
    \begin{equation}
    \begin{array}{l}
        \xi(\lambda) = \Phi(x_0,t_0;\lambda_0)^{-1}p_1 + (\lambda-\lambda_0)\Big( \Phi(x_0,t_0;\lambda_0)^{-1}p_2 \\
        -\Phi(x_0,t_0;\lambda_0)^{-1}\Phi^{(1)}(x_0,t_0;\lambda_0)\Phi(x_0,t_0;\lambda_0)^{-1}p_1   \Big),
    \end{array}
    \end{equation}
then the following equations hold
    \begin{equation}
        \begin{array}{l}
            \Phi(x_0,t_0;\lambda_0) \xi(\lambda_0) = p_1,\\
            \frac{\partial}{\partial\lambda}\left.\Big( \Phi(x,t;\lambda)\xi(\lambda)   \Big)\right|_{\lambda=\lambda_0,x=x_0,t=t_0}= p_2.
        \end{array}
    \end{equation}
Thus $D[\lambda_0,1,\Phi(x,t;\lambda)\xi(\lambda)]$ in (\ref{eq:DTsamepole11}) with (\ref{eq:DTsamepole12}) is the Darboux matrix with initial value (\ref{eq:Sinitial}).
\par
Since the equations (\ref{eq:S_xdffeq}) and (\ref{eq:S_tdffeq}) admit a solution $S(x,t)$ for each initial value (\ref{eq:Sinitial}), the equations (\ref{eq:S_xdffeq}) and (\ref{eq:S_tdffeq}) are compatible first order ODEs, i.e., $S_{xt}=S_{tx}$ holds. Thus for every fixed initial value, the solution of the equations (\ref{eq:S_xdffeq}) and (\ref{eq:S_tdffeq}) is unique. That is to say, all the solutions $S(x,t)$ correspond to trivial DT, classic DT, or DT with the same pole. \QED
\end{demo}

\begin{corollary}
The compatible equation of the nonlinear ODEs (\ref{eq:S_xdffeq}) and (\ref{eq:S_tdffeq}) is solvable for any initial values $S(x_0,t_0)\in {\rm Mat}(2,\mathbb{C})$.
\end{corollary}

\section{The case $n\in \mathbb{N}_+$}\label{Se:3}

In this section, the Darboux matrix (\ref{eq:DTngeneral}) and its inverse (\ref{eq:DTInvngeneral}) for the case $n\in \mathbb{N}_+$ are derived by induction. Our main method is a combination of $n$-fold DT \cite{matveev1991darboux,gu2004darboux} with generalized DT \cite{matveev1991darboux,guo2012nonlinear}. By the way, Theorem~\ref{thm:Generaln} below can be used to obtain the blow up rogue wave for $x$-nonlocal focusing NLS in \cite{yang2019rogue}.

\begin{lemma}\label{thm:induction}
    Let $b(\lambda)$ be a $2\times 1$ solution of the Lax pair (\ref{eq:laxpair1}) and analytic at $\lambda=\lambda_0$. If each $D^{[k]}(x,t;\lambda)$ $(k=1,2,\cdots,n-1)$ with $n>1$ is
    a Darboux matrix of the form in Theorem~\ref{thm:order1threeequi}, namely  $\frac{D^{[k]}(x,t;\lambda)}{\lambda-\lambda_0}$ is a Darboux matrix in Theorem~\ref{thm:DT1} of the form
    \begin{equation*}
        \left( \Phi^{[k]},U^{[k]},V^{[k]}   \right) \xrightarrow{D^{[k]}(x,t;\lambda)} \left( \Phi^{[k+1]}=\frac{D^{[k]}}{\lambda-\lambda_0}\Phi^{[k]},U^{[k+1]},V^{[k+1]}   \right),~~ k=1,\cdots,n-1,
    \end{equation*}
    where $U^{[1]} = U$, $V^{[1]} = V$, $\Phi^{[1]} = \Phi$ and
    \begin{equation}\label{eq:lineareqn-1}
        \frac{\partial^s}{\partial \lambda^s}\left.\left( D^{[n-1]}(\lambda)D^{[n-2]}(\lambda)\cdots D^{[1]}(\lambda)b(\lambda)\right)\right|_{\lambda=\lambda_0} = 0,\quad s=0,1,\cdots,2n-3,
    \end{equation}
    then the following items hold

    (i) $b^{[n]}(\lambda)\triangleq \frac{1}{(\lambda-\lambda_0)^{2n-2}}D^{[n-1]}(\lambda)D^{[n-2]}(\lambda)\cdots D^{[1]}(\lambda)b(\lambda)$ is a solution of Lax pair
    \begin{equation}
    \begin{array}{l}
        \Phi^{[n]}_x = U^{[n]}(x,t;\lambda) \Phi^{[n]},\\
        \Phi^{[n]}_t = V^{[n]}(x,t;\lambda) \Phi^{[n]},
    \end{array}\label{eq:laxpairn}
    \end{equation}
    and analytic at $\lambda=\lambda_0$;

    (ii) $D^{[n]}(\lambda)\triangleq  \lambda I - S^{[n]} = \lambda I - H^{[n]}\Lambda {H^{[n]}}^{-1}$ provides a Darboux matrix, i.e.,
    \begin{equation}
        \left( \Phi^{[n]},U^{[n]},V^{[n]}   \right) \xrightarrow{D^{[n]}(x,t;\lambda)} \left( \Phi^{[n+1]}=\frac{D^{[n]}}{\lambda-\lambda_0}\Phi^{[n]},U^{[n+1]},V^{[n+1]}   \right),
    \end{equation}
    where $H^{[n]}=(b^{[n]}_0~~~(b^{[n]})_0^{(1)})$ and $\Lambda  = \left( \begin{matrix} \lambda_0&1\\
    0& \lambda_0 \end{matrix} \right)$;

    (iii) The formula below holds
    \begin{equation*}
        \frac{\partial^s}{\partial \lambda^s}\left.\left( D^{[n]}(\lambda)D^{[n-1]}(\lambda)D^{[n-2]}(\lambda)\cdots D^{[1]}(\lambda)b(\lambda)\right)\right|_{\lambda=\lambda_0} = 0,\quad s=0,1,\cdots,2n-1.
    \end{equation*}

\end{lemma}

\begin{demo}

(i) Denote $f(\lambda) = D^{[n-1]}(\lambda)D^{[n-2]}(\lambda)\cdots D^{[1]}(\lambda)b(\lambda)$. Condition (\ref{eq:lineareqn-1}) indicates
\begin{equation}\label{eq:fexpan}
    f(\lambda) = \sum_{k=2n-2}^{\infty} \frac{1}{k!}  (\lambda-\lambda_0)^{k} f_0^{(k)},
\end{equation}
then $b^{[n]}(\lambda) = \frac{1}{(\lambda-\lambda_0)^{2n-2}} f(\lambda)$ is analytic at $\lambda=\lambda_0$.
\par
Moreover, $f(\lambda)$ is a solution of Lax pair (\ref{eq:laxpairn}) since $D^{[k]}(\lambda)$ $(k=1,\cdots,n-1)$ provide Darboux matrices and $b(\lambda)$ is a solution of the Lax pair with $U^{[1]}=U$ and $V^{[1]}=V$. As $\frac{1}{(\lambda-\lambda_0)^{2n-2}}$ is independent of $x$, $t$, it is obvious that $b^{[n]}(\lambda)$ is also a solution of Lax pair (\ref{eq:laxpairn}) except $\lambda = \lambda_0$.
\par
At $\lambda=\lambda_0$, one has
\begin{equation*}
    b_{0,x}^{[n]} = \left( \lim_{\lambda\to\lambda_0} \frac{1}{(\lambda-\lambda_0)^{2n-2}}f(\lambda) \right)_x = \lim_{\lambda\to\lambda_0} \frac{1}{(\lambda-\lambda_0)^{2n-2}}f_x(\lambda) = U_0^{[n]} b_0^{[n]}.
\end{equation*}
Similarly, $ b_{0,t}^{[n]}=V_0^{[n]} b_0^{[n]}$ holds.
\par
(ii) The expansion (\ref{eq:fexpan}) and $b^{[n]}(\lambda) = \frac{1}{(\lambda-\lambda_0)^{2n-2}} f(\lambda)$ show
\begin{equation*}
    b_0^{[n]} = \frac{1}{(2n-2)!}f_0^{(2n-2)}, \quad (b^{[n]})_0^{(1)} = \frac{1}{(2n-1)!}f_0^{(2n-1)}.
\end{equation*}
So $b_0^{[n]}$ and $(b^{[n]})_0^{(1)}$ exist and are bounded for every fixed $(x,t)\in\mathbb{R}^2$ due to that $f(\lambda)$ is analytic at $\lambda_0$.
Then according to Theorem~\ref{thm:order1threeequi}, DT can be constructed as
\begin{equation*}
    D^{[n]} = D[\lambda_0,1,b^{[n]}(x,t;\lambda)].
\end{equation*}

(iii) For $s=0,1,\cdots,2n-3$, Leibniz formula shows
\begin{equation*}
    \frac{\partial^s}{\partial \lambda^s}\left.\left( D^{[n]}(\lambda)f(\lambda)\right)\right|_{\lambda=\lambda_0} = \sum_{l=0}^{s} C_s^l (D^{[n]})_0^{(s-l)} f_0^{(l)} = 0.
\end{equation*}
For $s=2n-2$, one has
\begin{equation*}
    \frac{\partial^{2n-2}}{\partial \lambda^{2n-2}}\left.\left( D^{[n]}(\lambda)f(\lambda)\right)\right|_{\lambda=\lambda_0} = (2n-2)!D^{[n]}(\lambda_0)b_0^{[n]} = 0.
\end{equation*}
For $s=2n-1$, one has
\begin{equation*}
    \frac{\partial^{2n-1}}{\partial \lambda^{2n-1}}\left.\left( D^{[n]}(\lambda)f(\lambda)\right)\right|_{\lambda=\lambda_0} = (2n-1)!b_0^{[n]}+(2n-1)!D^{[n]}(\lambda_0)(b_0^{[n]})^{(1)} = 0.
\end{equation*}
\QED
\end{demo}
\par
Theorem~\ref{thm:order1threeequi} provides the first-fold DT $D^{[1]}$ which is similar to Jordan blocks $\left( \begin{matrix}\lambda_0 & 1 \\ 0 &\lambda_0   \end{matrix} \right)$, while Lemma~\ref{thm:induction} ensures that the induction process works. Combining them, the monic DT that is the $n$ folds of DTs similar to Jordan blocks $\left( \begin{matrix}\lambda_0 & 1 \\ 0 &\lambda_0   \end{matrix} \right)$ can be constructed. The specific induction process is proposed as follows.
\par
Let
\begin{equation}
    \begin{array}{l}
        b^{[1]}(\lambda) \triangleq b(\lambda),\\
        H^{[1]} \triangleq (b^{[1]}_0,(b^{[1]})_0^{(1)}),\\
        D^{[1]}(\lambda) \triangleq \lambda I - H^{[1]} \Lambda {H^{[1]}}^{-1},
    \end{array}\label{eq:H1}
\end{equation}
and for $k=2,3,\cdots,n$, define
\begin{equation}
    \begin{array}{l}
        b^{[k]}(\lambda) \triangleq \frac{1}{(\lambda-\lambda_0)^{2k-2}}D^{[k-1]}(\lambda)D^{[k-2]}(\lambda)\cdots D^{[1]}(\lambda)b(\lambda),\\
        H^{[k]} \triangleq (b^{[k]}_0,(b^{[k]})_0^{(1)}),\\
        D^{[k]}(\lambda) \triangleq \lambda I - H^{[k]} \Lambda {H^{[k]}}^{-1},
    \end{array}\label{eq:Hk}
\end{equation}
then it is immediately that
\begin{equation}\label{eq:Dlambdaitera}
    D(\lambda) = D^{[n]}(\lambda)D^{[n-1]}(\lambda)\cdots D^{[1]}(\lambda)
\end{equation}
is a Darboux matrix of order $n$ according to Lemma~\ref{thm:induction}, and
\begin{equation}\label{eq:Dnlineareqs}
        \frac{\partial^s}{\partial \lambda^s}\left.\Big( D(\lambda)b(\lambda)\Big)\right|_{\lambda=\lambda_0} = 0,\quad s=0,1,\cdots,2n-1.
\end{equation}

Now it is ready to provide a Darboux matrix $D(\lambda)$ in (\ref{eq:Dlambdaitera}) with pole form.
\par
\begin{theorem}\label{thm:Generaln}
Let $b(x,t;\lambda)$ be a $2\times 1$ solution of Lax pair (\ref{eq:laxpair1}) and $a(x,t;\lambda)$ be a $1\times 2$ solution of the adjoint Lax pair (\ref{eq:adjointlaxpair}). Further, assume both $a(x,t;\lambda)$ and $b(x,t;\lambda)$ are analytic at $\lambda=\lambda_0$, and the formula below holds
\begin{equation}\label{eq:balancecondiDT}
    \frac{\partial^s}{\partial \lambda^s}\left.\Big( a(\lambda)b(\lambda)\Big)\right|_{\lambda=\lambda_0} = (ab)_0^{(s)}=0,\quad s=0,1,\cdots,n-1.
\end{equation}
Then $D(\lambda)$ in (\ref{eq:Dlambdaitera}) can be written as
\begin{equation}\label{eq:Dnpole}
    D(\lambda) = (\lambda-\lambda_0)^{n}
    \left( I - \left(\begin{matrix} b_0& b_0^{(1)}&\cdots&b_0^{(n-1)} \end{matrix}\right)\Gamma^{-1}  \left(\begin{matrix} \frac{a(\lambda_0)}{\lambda-\lambda_0}\\ \vdots\\ \frac{\partial^{n-1}}{\partial \lambda_0^{n-1}}\left( \frac{a(\lambda_0)}{\lambda-\lambda_0}\right)   \end{matrix}        \right)         \right),
\end{equation}
where the elements of $\Gamma$ matrix are
\begin{equation}\label{eq:Gamma}
    \Gamma_{kj} = \frac{1}{k C_{k+j-1}^{k}} \sum_{s=0}^{k-1} C_{k+j-1}^{s} a_0^{(s)} b_0^{(k+j-1-s)},\quad k,j=1,2,\cdots,n,
\end{equation}
where $C_m^n$ is the combinatorial number. The notation in (\ref{eq:Dnpole}) is
\begin{equation}\label{eq:simplederisym}
    \frac{\partial^{k}}{\partial \lambda_0^{k}}\left( \frac{a(\lambda_0)}{\lambda-\lambda_0}\right) = \left.\frac{\partial^{k}}{\partial \mu^{k}}\left( \frac{a(\mu)}{\lambda-\mu}\right)\right|_{\mu=\lambda_0},\quad k=1,2,\cdots,n-1.
\end{equation}

\end{theorem}

\begin{demo}
The theorem is proved by verifying that $D(\lambda)$ in (\ref{eq:Dnpole}) satisfies the equations (\ref{eq:Dnlineareqs}).

(1) Rewrite $D(\lambda)$ in (\ref{eq:Dnpole}) as the polynomial of $\lambda-\lambda_0$ of the form
\begin{equation}\label{eq:DT-lambda}
    D(\lambda) = (\lambda-\lambda_0)^{n} I + \sum_{p=1}^{n} (\lambda-\lambda_0)^{n-p} D_p,
\end{equation}
then comparing the coefficients of $\lambda$ in (\ref{eq:Dnpole}) and (\ref{eq:DT-lambda}) yields
\begin{equation}\label{eq:samepoleDTforDj}
    D_p = -(p-1)! \left(\begin{matrix} b_0& b_0^{(1)}&\cdots&b_0^{(n-1)} \end{matrix}\right)\Gamma^{-1}  \left(\begin{matrix} 0\\ \vdots\\ 0 \\C_{p-1}^{p-1}a_0^{(p-p)}\\ \vdots\\C_{m-1}^{p-1}a_0^{(m-p)}\\ \vdots \\ C_{n-1}^{p-1} a_0^{(n-p)}   \end{matrix}        \right),
\end{equation}
where $p=1,2,\cdots,n$, the first nonzero row $C_{p-1}^{p-1}a_0^{(p-p)}$  is  the $p$-th row of the rightmost matrix, while $C_{m-1}^{p-1}a_0^{(m-p)}$ is the $m$-th row ($p\le m \le n$). Moreover, it is obvious that
\begin{equation*}
    \frac{\partial^q}{\partial \lambda^q}\left.\Big( D(\lambda)\Big)\right|_{\lambda=\lambda_0} = q!D_{n-q},\quad q=0,1,\cdots,n,
\end{equation*}
where $D_0=I$.
\par
(2) For $s = 0,1, \cdots,n-1$, we have
\begin{equation*}
    \frac{\partial^s}{\partial \lambda^s}\left.\Big( D(\lambda)b(\lambda)\Big)\right|_{\lambda=\lambda_0} = \sum_{q=0}^{s} q!C_s^qD_{n-q}b_0^{(s-q)}=0,
\end{equation*}
where the second equality comes from
\begin{equation*}
    \sum_{q=n-m}^{s} q!(n-q-1)!C_s^qC_{m-1}^{n-q-1}a_0^{(m-n+q)}b_0^{(s-q)} = (m-1)!A_{s}^{n-m}(ab)_0^{(m-n+s)},
\end{equation*}
where $A_{m}^{n}$ is the arrangement number with $n-s\le m\le n$ and $0\le s \le n-1$.
\par
(3) For $s=n,n+1,\cdots,2n-1$, one has
\begin{equation*}
    \frac{\partial^s}{\partial \lambda^s}\left.\Big( D(\lambda)b(\lambda)\Big)\right|_{\lambda=\lambda_0} = n!C_s^nb_0^{(s-n)} +  \sum_{q=0}^{n-1} q!C_s^qD_{n-q}b_0^{(s-q)}.
\end{equation*}
\par
Define
\begin{equation*}
    P_{s,m} \triangleq \sum_{q=n-m}^{n-1}q!(n-q-1)!C_s^qC_{m-1}^{n-q-1}a_0^{(m-n+q)}b_0^{(s-q)},
\end{equation*}
then recalling (\ref{eq:Gamma}) and (\ref{eq:samepoleDTforDj}), it is easy to see that
\begin{equation*}
    P_{s,m} = n!C_s^n \Gamma_{m,s-(n-1)},
\end{equation*}
and
\begin{equation*}
    \frac{\partial^s}{\partial \lambda^s}\left.\Big( D(\lambda)b(\lambda)\Big)\right|_{\lambda=\lambda_0} = n!C_s^nb_0^{(s-n)} -\left(\begin{matrix} b_0& b_0^{(1)}&\cdots&b_0^{(n-1)} \end{matrix}\right)\Gamma^{-1}  \left(\begin{matrix} P_{s,1}\\ \vdots\\ P_{s,n} \end{matrix}        \right)=0.
\end{equation*}
\QED
\end{demo}
\par
\begin{proposition}
    The inverse of $D(\lambda)$ in (\ref{eq:Dnpole}) is
\begin{equation}\label{eq:Dninvpole}
    D^{-1}(\lambda) = (\lambda-\lambda_0)^{-n}
    \left( I + \left(\begin{matrix} \frac{b(\lambda_0)}{\lambda-\lambda_0}& \cdots&\frac{\partial^{n-1}}{\partial \lambda_0^{n-1}}\left( \frac{b(\lambda_0)}{\lambda-\lambda_0}\right) \end{matrix}\right)\Gamma^{-1}  \left(\begin{matrix}a_0\\a_0^{(1)}\\ \vdots\\ a_0^{(n-1)}   \end{matrix}        \right)         \right).
\end{equation}
\end{proposition}

\begin{demo}
By using the notations from Theorem~\ref{thm:Generaln}, $D(\lambda)$ can be written as
\begin{equation*}
    D(\lambda) = (\lambda-\lambda_0)^n\left( I+\sum_{p=1}^n \frac{1}{(\lambda-\lambda_0)^p} D_p  \right),
\end{equation*}
where $D_p$ is given by (\ref{eq:samepoleDTforDj}). Rewrite $D^{-1}(\lambda)$ in (\ref{eq:Dninvpole}) as
\begin{equation*}
    D^{-1}(\lambda) = (\lambda-\lambda_0)^{-n}\left( I+\sum_{q=1}^n \frac{1}{(\lambda-\lambda_0)^q} (D^{-1})_q \right),
\end{equation*}
where
\begin{equation}\label{eq:DefinitionDinvelement}
    (D^{-1})_q = (q-1)! \begin{array}{l}\quad\\ \left(\begin{matrix} 0&\cdots&0& C_{q-1}^{q-1}b_0^{(q-q)}& \cdots &  C_{n-1}^{q-1} b_0^{(n-q)} \end{matrix}\right)\\ \qquad\qquad\quad\;\;\qquad q\qquad\qquad\qquad\qquad\; n\end{array}\Gamma^{-1}  \left(\begin{matrix} a_0\\a_0^{(1)} \\ \vdots\\ a_0^{(n-1)}  \end{matrix}        \right).
\end{equation}
\par
Then the expression (\ref{eq:Dninvpole}) holds if $D(\lambda)D^{-1}(\lambda)=I$. If fact, $D(\lambda)D^{-1}(\lambda)=I$ is equivalent to
\begin{equation}\label{eq:DDinverseD}
\begin{array}{l}
    \frac{1}{\lambda-\lambda_0} \left( D_1+(D^{-1})_1 \right) + \sum\limits_{p=2}^n  \frac{1}{(\lambda-\lambda_0)^p}\left( D_p + (D^{-1})_p + \sum\limits_{s=1}^{p-1} D_s(D^{-1})_{p-s} \right) \\
    +\sum\limits_{p=n+1}^{2n}  \frac{1}{(\lambda-\lambda_0)^p} \left(  \sum\limits_{s=p-n}^{n} D_s(D^{-1})_{p-s} \right)  =0.
\end{array}
\end{equation}
\par
It can be verified directly that the coefficients of $\frac{1}{(\lambda-\lambda_0)^p}$ for $p=1,n+1,n+2,\cdots,2n$ in (\ref{eq:DDinverseD}) are zero. To prove the vanishing of the coefficients of $\frac{1}{(\lambda-\lambda_0)^p}$ for $p=2,3,\cdots,n$ in (\ref{eq:DDinverseD}), setting $\Lambda_{n-p+1} = {\rm diag} \Big( C_{p-1}^{p-1},\cdots, C_{m-1}^{p-1},\cdots,C_{n-1}^{p-1} \Big)$ and denoting
\begin{small}
\begin{equation*}
    W_p \triangleq \sum_{s=1}^{p-1}(s-1)!(p-s-1)! \left(\begin{matrix} 0\\ \vdots\\ 0 \\C_{s-1}^{s-1}a_0^{(s-s)}\\ \vdots\\C_{m-1}^{s-1}a_0^{(m-s)}\\ \vdots \\ C_{n-1}^{s-1} a_0^{(n-s)}   \end{matrix}        \right)\begin{array}{l}\quad\\ \left(\begin{matrix} 0&\cdots&0& b_0& \cdots &  C_{n-1}^{p-s-1} b_0^{(n-(p-s))} \end{matrix}\right)\\ \qquad\qquad\quad\;\; p-s\qquad\qquad\qquad\; n \end{array},
\end{equation*}
\end{small}
it follows that
\begin{equation*}
    W_p = (p-1)! \Gamma\left(\begin{matrix} 0 & \Lambda_{n-p+1} \\ 0 & 0  \end{matrix} \right)_{n\times n}- (p-1)!\left(\begin{matrix}  0 & 0 \\ \Lambda_{n-p+1} &  0 \end{matrix} \right)_{n\times n}\Gamma.
\end{equation*}
\par
So with these expressions in mind and according to the definitions of $D_p$ in (\ref{eq:samepoleDTforDj}) and $(D^{-1})_p$ (\ref{eq:DefinitionDinvelement}), it is concluded that
\begin{equation*}
D_p + (D^{-1})_p + \sum\limits_{s=1}^{p-1} D_s(D^{-1})_{p-s}=0~~~{\rm for}~p=2,3,\cdots,n.
\end{equation*}
\QED
\end{demo}
\par
The DT in Theorem \ref{thm:Generaln} is a higher-order pole version of Theorem~\ref{thm:DT1}, which extends the order of pole from one to $n$. Similar to the form of equation (\ref{eq:DTsamepole11}), denote the DT in equation (\ref{eq:Dnpole}) as
\begin{equation}\label{eq:DTsamepolenn}
    D[\lambda_0,n,b(\lambda)](x,t;\lambda),
\end{equation}
where the $n\in\mathbb{N}_+$ inside the brackets $[\cdots]$ indicates the order of the DT.
\par
\begin{remark}\label{rmk:DTwithCvarxin}
Just like Section~\ref{subsec:CinGamman1}, a new version of DT with some constant $C$ can be constructed. However, it is more advisable to consider $\xi_0^{(j)}$ as free parameters directly. The process is given below:
\par
Firstly, take $\xi_0,\; \xi_0^{(1)},\cdots,$ of $\xi(\lambda)$ as free parameters and ensure that $\xi(\lambda)$ is analytic at $\lambda_0$. Since DT always takes finite derivatives of $\xi(\lambda)$, the analytic properties of $\xi(\lambda)$ can be achieved by defining $\xi(\lambda)$ as a polynomial of $\lambda$. For example, if one needs that $\xi(\lambda)$ is analytic at $\lambda_0$ with derivatives $\xi_0^{(j)} = v_j$ ($j=0,1,\cdots,m$), where $v_j$ are arbitrary constant column vectors independent of $x,t,\lambda$, one can define $\xi(\lambda)$ as a polynomial of $\lambda$ below

\begin{equation*}
        \xi(\lambda) = \sum_{j=0}^{m} \frac{1}{j!} (\lambda-\lambda_0)^j v_j.
\end{equation*}
\par
Secondly, define $b(x,t;\lambda) = \Phi(x,t;\lambda)\xi(\lambda)$, where $\Phi(x,t;\lambda)$ is a fundamental solution matrix of Lax pair (\ref{eq:laxpair1}) and analytic at $\lambda = \lambda_0$;
\par
Finally, construct DT with free parameters $\xi_0^{(j)}$ according to Theorem~\ref{thm:Generaln}.
\par
Moreover, according to the equations in (\ref{eq:Dnlineareqs}), $D(x,t;\lambda)$ in Theorem~\ref{thm:Generaln} is uniquely determined by $b(x,t;\lambda)$. Thus the freedom of $a(x,t;\lambda)=\eta(\lambda)\Phi^{-1}(x,t;\lambda)$ dose not change the DT if $\eta(\lambda)$ satisfies
\begin{equation}\label{eq:constrainxieta}
        0 = (ab)_0^{(s)} =  (\eta\xi)_0^{(s)}, \quad s=0,1,\cdots,n-1,
\end{equation}
which are equations for the derivatives of $\xi(\lambda)$ and $\eta(\lambda)$ at $\lambda_0$ with order not exceeding $n-1$.

\end{remark}

\begin{remark}
In principle, three kinds of matrices above should be invertible. The first kind is $H^{[k]}$ in (\ref{eq:H1}) and (\ref{eq:Hk}) for $(k=1,2,\cdots,n)$, the second kind is the coefficient matrix of the linear equations (\ref{eq:Dnlineareqs}) and the third kind is the $\Gamma$ matrix in (\ref{eq:Gamma}). Is
the invertibility of any kind equivalent to the other two? Although we can not answer this question directly for general DTs in Theorem~\ref{thm:Generaln}, some related results for other DTs with poles are illustrated:
\par
(i) For the two-dimensional affine Toda equation \cite{zhou2008darboux}, the iteration of classic DTs on different spectral parameters has been proved that the invertibility of the second kind is equivalent to that of the third kind.
\par
(ii) The third kind matrix is invertible but the first kind is not, which will be explained in detail in Remark~\ref{rmk:singularfirstDTNLS} of Section~\ref{Section4}.

\end{remark}
\par
\par
{\bf Example: The $x$-nonlocal focusing NLS equation}
\par
In the previous discussion, it is only required that the DTs preserve the fact that the matrices $\widetilde{U}(\lambda)$ and $\widetilde{V}(\lambda)$ are polynomials of $\lambda$. However, if the compatible conditions of Lax pair (\ref{eq:laxpair1}) are physically interesting soliton equations such as KdV equation and NLS equation, there are usually some constrains between the elements of $U(\lambda)$ and $V(\lambda)$ in Lax pair (\ref{eq:laxpair1}). Thus DTs should keep such constrains for matrices $\widetilde{U}(\lambda)$ and $\widetilde{V}(\lambda)$ correspondingly.
\par
In 2013, Ablowitz and Musslimani \cite{ablowitz2013integrable} introduced the ($PT$-symmetric) $x$-nonlocal NLS equation. Recently, much work has been done for finding exact solutions of this equation \cite{yang2019rogue,xu2022binary}. In what follows, the $x$-nonlocal focusing NLS equation is taken as an example to show how to make the DTs in Theorem~\ref{thm:Generaln} preserving the constrains between the elements of $U(\lambda)$ and $V(\lambda)$. As a result, some new global multiple-pole solitons are found.
\par
The Lax pair of the $x$-nonlocal focusing NLS equation \cite{ablowitz2013integrable} is
\begin{equation}
\begin{array}{l}
\Phi_x = U(\lambda)\Phi=\left( \begin{matrix} \lambda &  u \\ -\hat{u}& -\lambda   \end{matrix}\right) \Phi,\\
\Phi_t = V(\lambda)\Phi=\left( \begin{matrix} -2{\rm i} \lambda^2-{\rm i} u\hat{u} & -2{\rm i}\lambda u-{\rm i} u_x \\ 2{\rm i}\lambda \hat{u}-{\rm i} \hat{u}_x& 2{\rm i} \lambda^2+{\rm i} u\hat{u}   \end{matrix}\right) \Phi,
\end{array}\label{eq:laxpairnonlocalNLS}
\end{equation}
where $\hat{u}(x,t) \triangleq u^*(-x,t).$ The compatible condition $\Phi_{xt}=\Phi_{tx}$ yields the $x$-nonlocal focusing NLS equation
\begin{equation}\label{eq:nonlocalNLS}
    {\rm i} u_t = u_{xx} + 2 u^2 \hat{u}.
\end{equation}
\par
Obviously, the matrices $U(\lambda)$ and $V(\lambda)$ in (\ref{eq:laxpairnonlocalNLS}) have symmetries
\begin{equation}\label{eq:symmetries}
    U(-x,t;\bar\lambda)^* = \sigma_3 U(x,t;\lambda) \sigma_3,\qquad -V(-x,t;\bar\lambda)^* = \sigma_3 V(x,t;\lambda) \sigma_3,
\end{equation}
where the symbol $*$ means conjugate transport and $\sigma_3$ is the third Pauli matrix. Then the following lemma holds.

\begin{lemma}\label{lemma:nonlocalNLSadsolu}
    If $\Phi(x,t;\lambda)$ is a solution of the Lax pair (\ref{eq:laxpairnonlocalNLS}), then $\Phi^*(-x,t;\bar\lambda)\sigma_3$ is a solution of the adjoint Lax pair of (\ref{eq:laxpairnonlocalNLS}). $($Recall that the adjoint Lax pair of Lax pair (\ref{eq:laxpair1}) is (\ref{eq:adjointlaxpair}).$)$
\end{lemma}
\par
Lemma~\ref{lemma:nonlocalNLSadsolu} indicates that
\begin{equation}\label{eq:indext}
    \partial_x \left(\Phi^*(-x,t;\bar\lambda) \sigma_3 \Phi(x,t;\lambda) \right) = \partial_t \left(\Phi^*(-x,t;\bar\lambda) \sigma_3 \Phi(x,t;\lambda) \right) = 0.
\end{equation}
For general fundamental solution matrix $\Phi(x,t;\lambda)$, the matrix
$\Phi^*(-x,t;\bar\lambda) \sigma_3 \Phi(x,t;\lambda)$ may be a function of $\lambda$.
However, for special initial value $\Phi(0,0;\lambda)=I$, it is easy to check that
\begin{equation*}
    \Phi^*(-x,t;\bar\lambda) \sigma_3 \Phi(x,t;\lambda)  =  \Phi^*(0,0;\bar\lambda) \sigma_3 \Phi(0,0;\lambda) = \sigma_3,
\end{equation*}
which leads to
\begin{equation}\label{eq:PhiInvPhihat}
     \Phi^{-1}(x,t;\lambda)=\sigma_3 \Phi^*(-x,t;\bar\lambda) \sigma_3.
\end{equation}
\par
In order to construct DT in Theorem~\ref{thm:Generaln}, define
\begin{equation}\label{eq:nonlocalabdef}
    b(x,t;\lambda)\triangleq \Phi(x,t;\lambda)\xi(\lambda),\qquad a(x,t;\lambda)\triangleq \eta(\lambda)\Phi^{-1}(x,t;\lambda),
\end{equation}
where $\xi(\lambda)$ and $\eta(\lambda)$ are undetermined constant vectors and $a(\lambda)$ is well-defined since
\begin{equation}
    \Big(\Phi^{-1}\Big)_x = - \Phi^{-1} \Phi_x \Phi^{-1} = - \Phi^{-1} U \Phi \Phi^{-1} = - \Phi^{-1} U,
\end{equation}
that is, $\Phi^{-1}(x,t;\lambda)$ is a fundamental solution matrix of adjoint Lax pair. Then the condition (\ref{eq:balancecondiDT}) is reduced to the constrains of $\eta(\lambda)$ and $\xi(\lambda)$, i.e.,
\begin{equation*}
    \frac{\partial^s}{\partial \lambda^s}\left.\Big( \eta(\lambda)\Phi^{-1}(x,t;\lambda)\Phi(x,t;\lambda)\xi(\lambda) \Big)\right|_{\lambda=\lambda_0} = (\eta\xi)_0^{(s)}=0,\quad s=0,1,2,\cdots,n-1.
\end{equation*}

\begin{lemma}\label{eq:Lemma-lambda}
    If $\lambda_0\in\mathbb{R}$, $\eta(\lambda) = \xi^*(\bar\lambda) \sigma_3$, and
\begin{equation}\label{eq:PTNLSsymmCondiXi}
    \frac{\partial^s}{\partial \lambda^s}\left.\Big( \eta(\lambda)\xi(\lambda) \Big)\right|_{\lambda=\lambda_0} = (\eta\xi)_0^{(s)}=0,\quad s=0,1,2,\cdots,2n-1,
\end{equation}
hold, then the DT in Theorem~\ref{thm:Generaln} with (\ref{eq:nonlocalabdef}) makes $\widetilde{U}(\lambda)$ sharing the same symmetry of $U(\lambda)$ in (\ref{eq:symmetries}).
\end{lemma}

\begin{demo}
    Let $\widetilde{U}(x,t;\lambda)= \lambda \widetilde{J}(x,t) + \widetilde{P}(x,t)$ since the order of $\widetilde{U}(x,t;\lambda)$ and $U(x,t;\lambda)=\lambda J(x,t) + P(x,t)$  is the same. Comparing the coefficients of $\lambda$ in
    \begin{equation}
        \widetilde{U}(\lambda) D(\lambda) = D_x(\lambda) + D(\lambda)U(\lambda),
    \end{equation}
where $D(\lambda)=\lambda^n I+\sum\limits_{j=1}^{n}D_j\lambda^{n-j}$, it leads to
    \begin{equation}
        \begin{array}{l}
            \lambda^{n+1}:\qquad  \widetilde{J}(x,t) = J = \sigma_3      ,\\
            \lambda^{n}:\qquad  \widetilde{P}(x,t) = P(x,t) + [D_1(x,t),\sigma_3],
        \end{array}
    \end{equation}
so the diagonal elements of $\widetilde{P}(x,t)$ is $0$. Let $\widetilde{P}(x,t) =\left(\begin{matrix} 0 & \widetilde{u}(x,t) \\ \widetilde{v}(x,t) & 0 \end{matrix}\right) $, then from the expression of $D_1(x,t)$
in equation (\ref{eq:samepoleDTforDj}), the new solutions are
\begin{equation}\label{eq:new-solutions}
    \begin{array}{l}
        \widetilde{u}(x,t) = 2l_1(x,t) \Gamma^{-1}(x,t) r_2(x,t), \\
        \widetilde{v}(x,t) = -2 l_2(x,t) \Gamma^{-1}(x,t) r_1(x,t),
    \end{array}
\end{equation}
where $l_1(x,t)$ and $l_2(x,t)$ are the first and second row of $\left( \begin{matrix}  b_0 & \cdots & b_0^{(n-1)}  \end{matrix} \right)$, respectively, and $r_1(x,t)$ and $r_2(x,t)$ are the first and second column of $\left( \begin{matrix}  a_0^{T} & \cdots & a_0^{(n-1)T}  \end{matrix} \right)^{T}$, respectively.
\par
By $\eta(\lambda) = \xi^*(\bar\lambda) \sigma_3$ in Lemma \ref{eq:Lemma-lambda}, the equations (\ref{eq:PhiInvPhihat}) and (\ref{eq:nonlocalabdef}) show
\begin{equation}\label{eq:astarbsigma3}
    a^*(-x,t;\bar\lambda) = \sigma_3 b(x,t;\lambda).
\end{equation}
\par
Expanding $a(x,t;\lambda)$ and $b(x,t;\lambda)$ at $\lambda=\lambda_0$ yields
\begin{equation}
    \begin{array}{l}
        a(x,t;\lambda) = \sum_{j=0}^\infty \frac{1}{j!}  a_0^{(j)}(x,t) (\lambda-\lambda_0)^j, \\
        b(x,t;\lambda) = \sum_{j=0}^\infty  \frac{1}{j!}  b_0^{(j)}(x,t) (\lambda-\lambda_0)^j.
    \end{array}\label{eq:nonNLSabexpand}
\end{equation}
Further letting $x\rightarrow -x$, $\lambda\rightarrow  \bar\lambda$ and taking conjugate transport of the first equation in (\ref{eq:nonNLSabexpand}), it follows
\begin{equation}\label{eq:nonNLSabexpand-aconj}
    a^*(-x,t;\bar\lambda) = \sum_{j=0}^\infty  \frac{1}{j!} \left( a_0^{(j)}(-x,t)\right)^* (\lambda-\bar\lambda_0)^j.
\end{equation}
If $\lambda_0\in\mathbb{R}$, the equations in (\ref{eq:astarbsigma3})-(\ref{eq:nonNLSabexpand-aconj}) show that
\begin{equation}\label{eq:sigma3b0a0}
    \sigma_3 b_0^{(j)} = \widehat{a_0^{(j)}},\quad j=0,1,\cdots,
\end{equation}
where $\widehat{f(x,t)}=f^{*}(-x,t)$ with $*$ representing conjugate transport. In fact, the equations in (\ref{eq:sigma3b0a0}) are just the following constrains
\begin{equation}
\begin{array}{l}
    l_1(x,t) = r_1^* (-x,t),    \\
    l_2(x,t) = - r_2^*(-x,t).
\end{array}
\end{equation}
\par
Recalling the elements of $\Gamma$ matrix in (\ref{eq:Gamma}), i.e.,
\begin{equation}
    \Gamma_{pq} = \frac{1}{p C_{p+q-1}^{p}} \sum_{s=0}^{p-1} C_{p+q-1}^{s} a_0^{(s)} b_0^{(p+q-1-s)},\quad p,q=1,2,\cdots,n,
\end{equation}
according to (\ref{eq:sigma3b0a0}), one has
\begin{equation}\label{eq:hatGamma}
    \hat\Gamma_{pq} =\widehat{\left( \Gamma_{qp} \right)}=\frac{1}{q C_{p+q-1}^{q}} \sum_{s=0}^{q-1} C_{p+q-1}^{s} a_0^{(p+q-1-s)} b_0^{(s)},\quad p,q=1,2,\cdots,n.
\end{equation}
Since $ \frac{1}{p C_{p+q-1}^{p}} = \frac{1}{q C_{p+q-1}^{q}}$, taking the sum index $s\rightarrow p+q-1-s$ in (\ref{eq:hatGamma}) yields
\begin{equation}
    \Gamma_{pq} +  \hat\Gamma_{pq} = \frac{1}{p C_{p+q-1}^{p}} \sum_{s=0}^{p+q-1} C_{p+q-1}^{s} a_0^{(s)} b_0^{(p+q-1-s)},\quad p,q=1,2,\cdots,n,
\end{equation}
that is
\begin{equation}
    \Gamma_{pq} +  \hat\Gamma_{pq} = \frac{1}{p C_{p+q-1}^{p}} (ab)_0^{(p+q-1)} = \frac{1}{p C_{p+q-1}^{p}}(\eta\xi)_0^{(p+q-1)} = 0,\quad p,q=1,2,\cdots,n.
\end{equation}
Then $\Gamma(x,t) + \Gamma^*(-x,t) = 0 $, which leads to $\widetilde{u}(x,t) +  \widetilde{v}^*(-x,t) =0$. \QED

\end{demo}

\begin{lemma}
    If $\widetilde{U}(\lambda)$ keeps the symmetry, then $\widetilde{V}(\lambda)$ also keeps the  symmetry.
\end{lemma}
\par
Now it is ready to derive the exact soliton solutions of the $x$-nonlocal focusing NLS equation (\ref{eq:nonlocalNLS}), which is just an inverse problem of the Lax pair (\ref{eq:laxpairnonlocalNLS}).
\par
Considering the zero seed solution and zero spectral parameter
\begin{equation}
    u_0(x,t) = 0 , \qquad \lambda_0 = 0,
\end{equation}
the fundamental solution matrix of Lax pair (\ref{eq:laxpairnonlocalNLS}) is
\begin{equation}\label{eq:phixtnonNLS}
    \Phi(x,t;\lambda) = \left(\begin{matrix} e^{\lambda x - 2{\rm i} \lambda^2t} & 0 \\ 0 & e^{-\lambda x + 2{\rm i} \lambda^2t}  \end{matrix}\right),
\end{equation}
which satisfies $\Phi(0,0;\lambda)=I$.
\par
The Schur polynomials for the first two variables $S_k(x_1,x_2)$ is defined by
\begin{equation}
    e^{\lambda x_1 + \lambda^2 x_2} = \sum_{k=0}^{\infty}  S_k(x_1,x_2) \lambda^k,
\end{equation}
where
\begin{equation}
    S_k(x_1,x_2) = \sum_{m=0}^{\lfloor \frac{k}{2} \rfloor} \frac{x_1^{k-2m} x_2^m}{(k-2m)!m!}.
\end{equation}
Then we have
\begin{equation}
\begin{array}{l}
    \Phi_0^{(k)}(x,t) = \left(\begin{matrix} k!S_k(x,-2{\rm i} t) & 0 \\ 0 &k!S_k(-x,2{\rm i} t)   \end{matrix}\right), \\
    \left(\Phi^{-1}\right)_0^{(k)}(x,t) = \left(\begin{matrix} k!S_k(-x,2{\rm i} t) & 0 \\ 0 &k!S_k(x,-2{\rm i} t)   \end{matrix}\right).
\end{array}\label{eq:Phi0PhiInv0nonlocalNLS}
\end{equation}

Let $\xi_0 = (1,-{\rm i})^T$, $\xi_0^{(1)}=(0,-1)^T$, $\xi_0^{(2)}=(1,0)^T$, $\xi_0^{(3)}=(0,0)^T$, $\xi_0^{(4)}=(-3,0)^T$, $\xi_0^{(5)}=(0,0)^T$, $\xi_0^{(6)}=(45,0)^T$, $\xi_0^{(7)}=(0,0)^T$, $\xi_0^{(8)}=(-1575,0)^T$, $\xi_0^{(9)}=(0,0)^T$, which can be derived by the following polynomial
\begin{equation}\label{eq:xi-lambda}
    \xi(\lambda) = \sum_{k=0}^9 \frac{1}{k!} \xi_0^{(k)} \lambda^k,
\end{equation}
and satisfy the symmetry conditions (\ref{eq:PTNLSsymmCondiXi}) in Lemma \ref{eq:Lemma-lambda} for $n\le 5$ with $\eta(\lambda) = \xi^*(\bar\lambda) \sigma_3$.
\par
Substituting (\ref{eq:Phi0PhiInv0nonlocalNLS}) and (\ref{eq:xi-lambda}) into (\ref{eq:nonlocalabdef}) and recalling the formula (\ref{eq:new-solutions}), the global multiple-pole soliton solutions of the $x$-nonlocal focusing NLS equation (\ref{eq:nonlocalNLS}) can be obtained.
\par
\begin{figure}[ht]
\centering
\includegraphics[height=5cm]{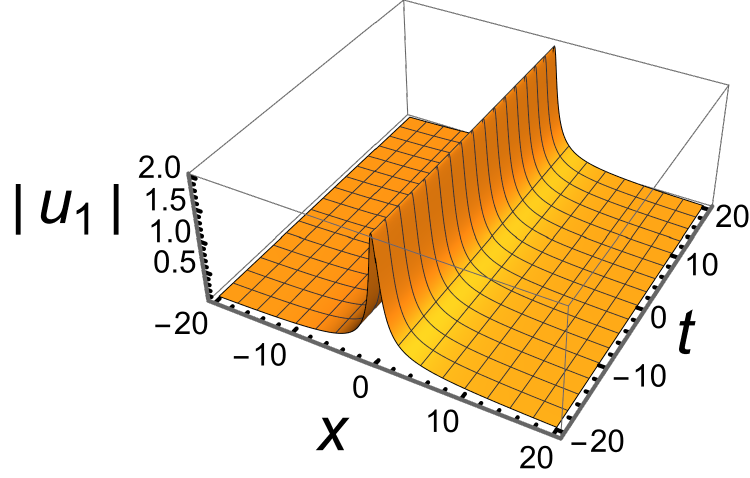}
\caption{The single-pole soliton (\ref{eq:single-pole-soliton-solution}) in the $x$-nonlocal focusing NLS equation (\ref{eq:nonlocalNLS}), whose velocity is zero and the corresponding spectral parameter is $\lambda=0$.}
\label{eq:single-pole-fNLS}
\end{figure}

For $n=1$, the single-pole soliton solution is
\begin{equation}\label{eq:single-pole-soliton-solution}
    u_1(x,t)=-\frac{2 {\rm i}}{{\rm i} + 2 x},
\end{equation}
whose density distribution $|u_1|$ is shown in Fig. \ref{eq:single-pole-fNLS}.

\par
\begin{figure}[H]
\centering
\includegraphics[height=5cm]{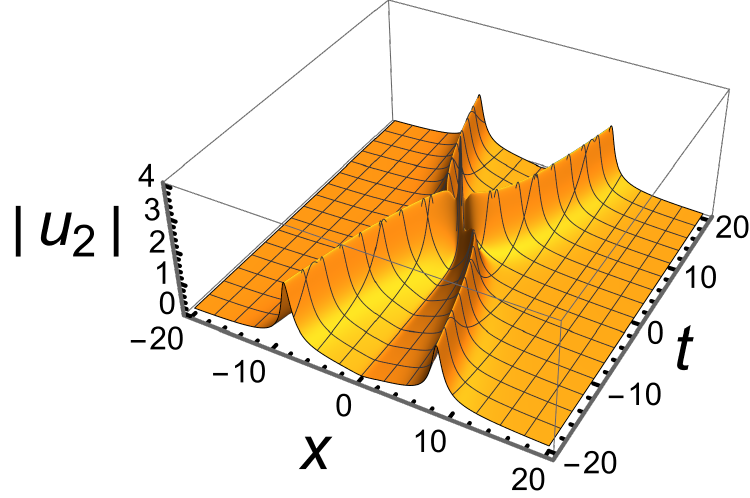}
\caption{The global double-pole soliton (\ref{eq:double-pole-soliton-solution}) in the $x$-nonlocal focusing NLS equation (\ref{eq:nonlocalNLS}), in which the velocities of the two branches are of order $\mathcal{O}\left(\frac{1}{\sqrt{t}}\right)$ instead of zero and the corresponding spectral parameter is also $\lambda=0$.}
\label{eq:double-pole-fNLS}
\end{figure}

For $n=2$, the double-pole soliton solution is
\begin{equation}\label{eq:double-pole-soliton-solution}
    u_2(x,t)=\frac{4 ( 6 {\rm i} x + 12 x^2 - 8 {\rm i} x^3 + 24 t ({\rm i} + 2 x)-3)}{3 + 192 t^2 +
 24 x^2 - 32 {\rm i} x^3 - 16 x^4},
\end{equation}
whose density distribution $|u_2|$ is shown in Fig. \ref{eq:double-pole-fNLS}.
\par
\begin{figure}[H]
\centering
\includegraphics[height=5cm]{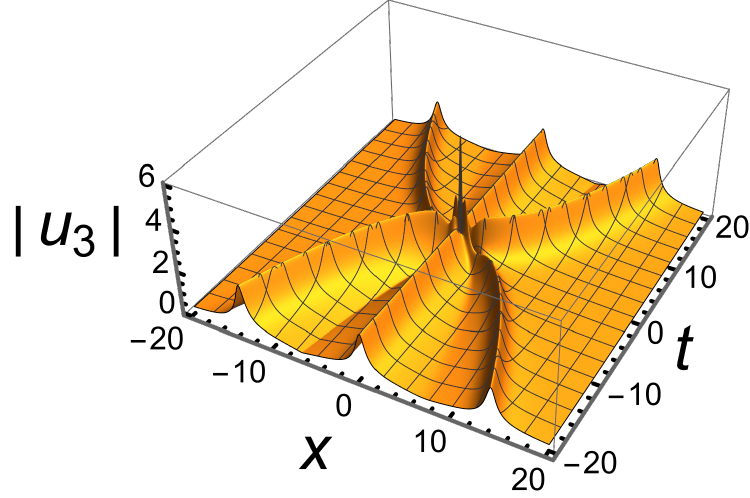}
\caption{The global triple-pole soliton (\ref{eq:triple-pole-soliton-solution}) in the $x$-nonlocal focusing NLS equation (\ref{eq:nonlocalNLS}), in which the velocity of the middle branch is zero, while the velocities of the other two branches are of order $\mathcal{O}\left(\frac{1}{\sqrt{t}}\right)$ instead of zero.}
\label{eq:triple-pole-fNLS}
\end{figure}

For $n=3$, the triple-pole soliton solution is
\begin{equation}\label{eq:triple-pole-soliton-solution}
    u_3(x,t)=6i\frac{P_1}{ P_2},
\end{equation}
where $P_1$ and $P_2$ are
\begin{footnotesize}
\begin{equation*}
    \begin{array}{l}
    P_1 = 184320 t^4 + 720 x^2 - 960 {\rm i} x^3 + 480 x^4 -
     1536 {\rm i} x^5 - 1792 x^6 + 1024 {\rm i} x^7 + 256 x^8-135\\
     \quad \quad +1920 t^2 (3 + 24 x^2 - 32 {\rm i} x^3 - 16 x^4) +
     64 t (45 {\rm i} + 180 {\rm i} x^2 + 240 x^3 - 240 {\rm i} x^4 - 192 x^5 +
        64 {\rm i} x^6),\\
        P_2 = 135 {\rm i} + 270 x  + 1440 x^3 +
   1440 {\rm i} x^4 + 576 x^5 + 3840 {\rm i} x^6 + 4608 x^7 - 2304 {\rm i} x^8 -
   512 x^9\\
    \quad \quad + 2160 {\rm i} x^2+ 552960 t^4 ({\rm i} + 2 x) +
   1152 t^2 (45 {\rm i} + 90 x - 120 {\rm i} x^2 - 80 x^3 + 80 {\rm i} x^4 + 32 x^5),
    \end{array}
\end{equation*}
\end{footnotesize}
whose density distribution $|u_3|$ is shown in Fig. \ref{eq:triple-pole-fNLS}.
\par
\begin{figure}[H]
	\centering  
	\vspace{-0.35cm} 
	\subfigtopskip=2pt 
	\subfigbottomskip=2pt 
	\subfigcapskip=-5pt 
		\includegraphics[width=0.32\linewidth]{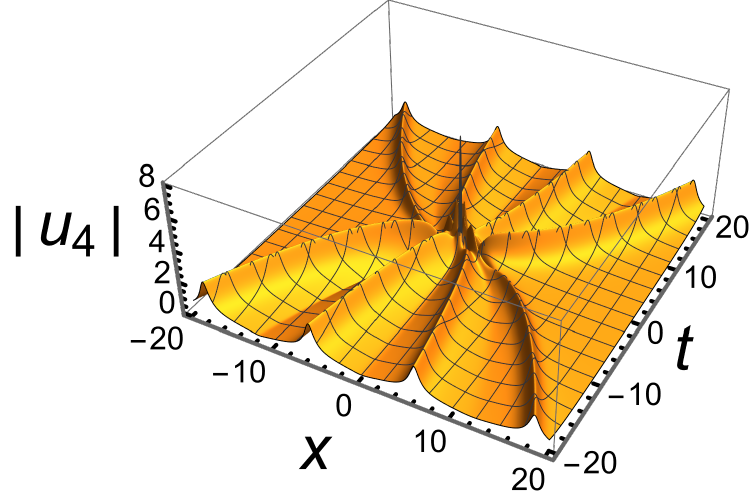}
	\quad 
		\includegraphics[width=0.32\linewidth]{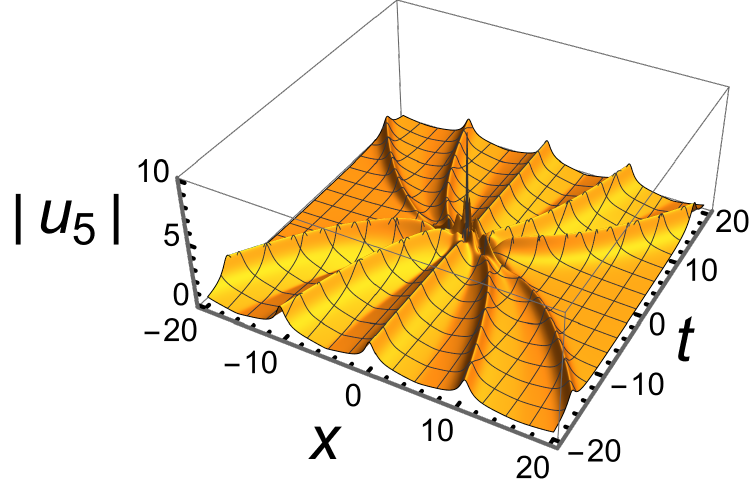}
\caption{The global quadruple-pole and quintuple-pole solitons in the $x$-nonlocal focusing NLS equation (\ref{eq:nonlocalNLS}), which are of order $\mathcal{O}\left(\frac{1}{x}\right)$ for $x\to\pm\infty$.}
\label{level}
\end{figure}

For $n=4$ and $n=5$, the quadruple-pole and quintuple-pole soliton solutions can also be derived, whose expressions are lengthy and we omit them for simplicity. The density distributions $|u_4|$ and $|u_5|$ are shown in Fig. \ref{level}.
\par
The spectral parameter of all the above solitons is exactly $\lambda=0$. The velocity of the single-pole soliton (\ref{eq:single-pole-soliton-solution}) is zero, while the velocities of the multi-pole solitons are not zero but of order $\mathcal{O}\left(\frac{1}{\sqrt{t}}\right)$. For any fixed $t$, these multi-pole solutions decay to zero for $x\to \pm \infty$ with order $\mathcal{O}\left(\frac{1}{x}\right)$.
\par
\begin{remark} It has been checked by Mathematica that the denominators of the solutions $u_j~(j=1,2,\cdots,5)$ don't have zeros in $(x,t)\in\mathbb{R}^2$ and are bounded for $(x,t)\in\mathbb{R}^2$. So they are global soliton solutions to the $x$-nonlocal focusing NLS equation (\ref{eq:nonlocalNLS}). To the best of our knowledge, these exact multiple-pole soliton solutions haven't been reported before.
\end{remark}

\begin{remark}
    The relationship between the highest order terms of $x$ and $t$ in the denominators of $u_2$ and $u_3$ is $t \sim x^2$. If there is a curve $f(x,t)=0$ that $\lim\limits_{t\rightarrow\infty}u(x,t)\neq 0$ on this curve, it requires that the $x$, $t$ terms of the denominators with highest order vanished (since the order of denominator is strictly higher than that of numerator). In this sense, the trajectory of the solution $u_2$ is approximate by
    \begin{equation}\label{eq:contourmapn2nonlocalNLS}
        192 t^2 - 16 x^4 = 0,
    \end{equation}
while the trajectory of the solution $u_3$ is approximate by
    \begin{equation}\label{eq:contourmapn3nonlocalNLS}
        512 x^9 - 1105920 t^4 x  - 36864 t^2 x^5 = 0.
    \end{equation}

 \begin{small}
\begin{figure}[H] 
	\centering  
	\vspace{-0.35cm} 
	\subfigtopskip=2pt 
	\subfigbottomskip=2pt 
	\subfigcapskip=-5pt 
	\subfigure[]{
		\includegraphics[width=0.32\linewidth]{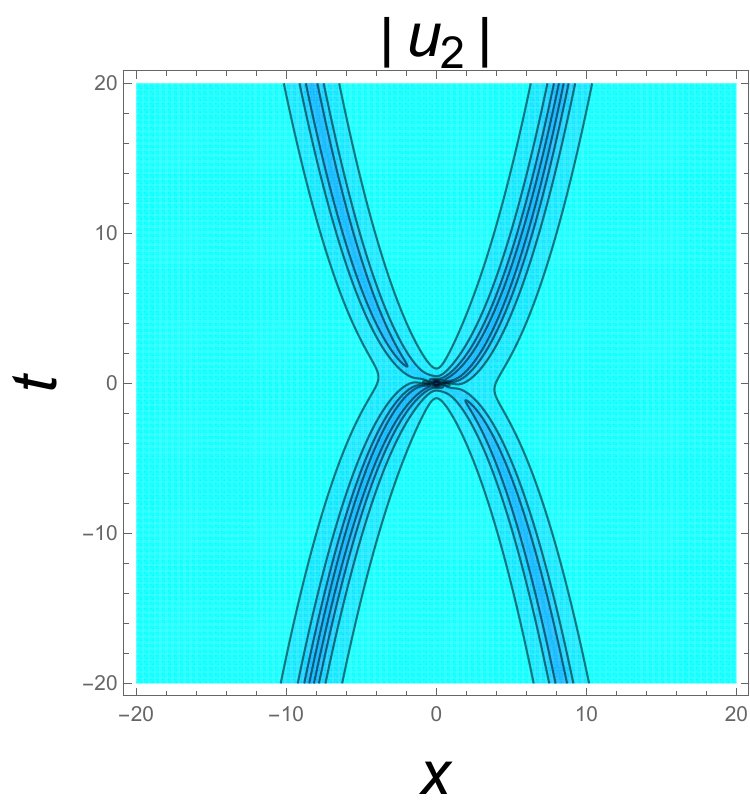}}
	\quad 
	\subfigure[]{
		\includegraphics[width=0.32\linewidth]{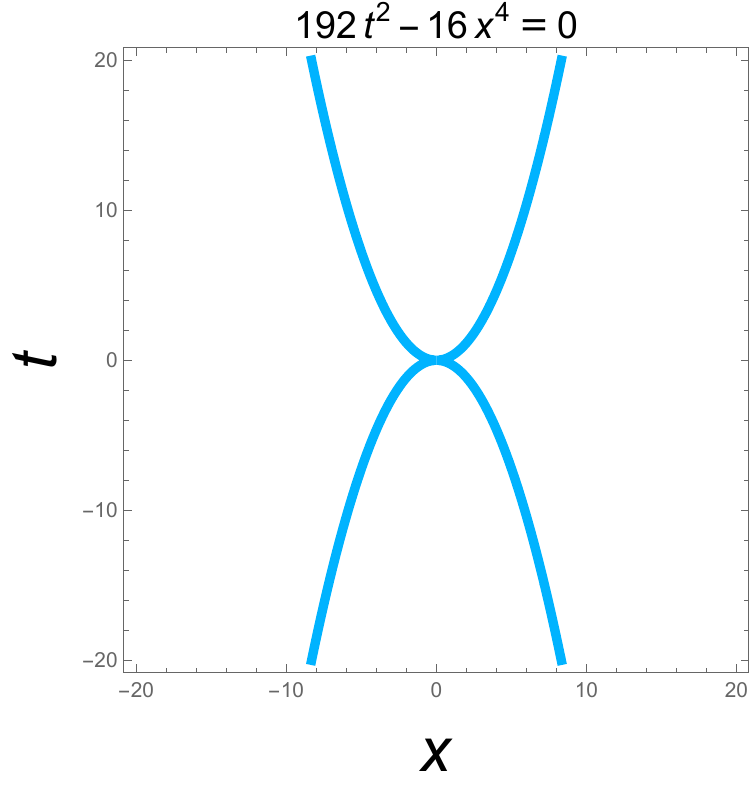}}
\caption{(a) Contour map of the density distribution $|u_2|$. (b) Asymptotic trajectory of the double-pole soliton in (\ref{eq:contourmapn2nonlocalNLS}). }
	\label{2ndnonlocalNLScontour}
\end{figure}
\end{small}

\begin{small}
\begin{figure}[H] 
	\centering  
	\vspace{-0.35cm} 
	\subfigtopskip=2pt 
	\subfigbottomskip=2pt 
	\subfigcapskip=-5pt 
	\subfigure[]{
		\includegraphics[width=0.32\linewidth]{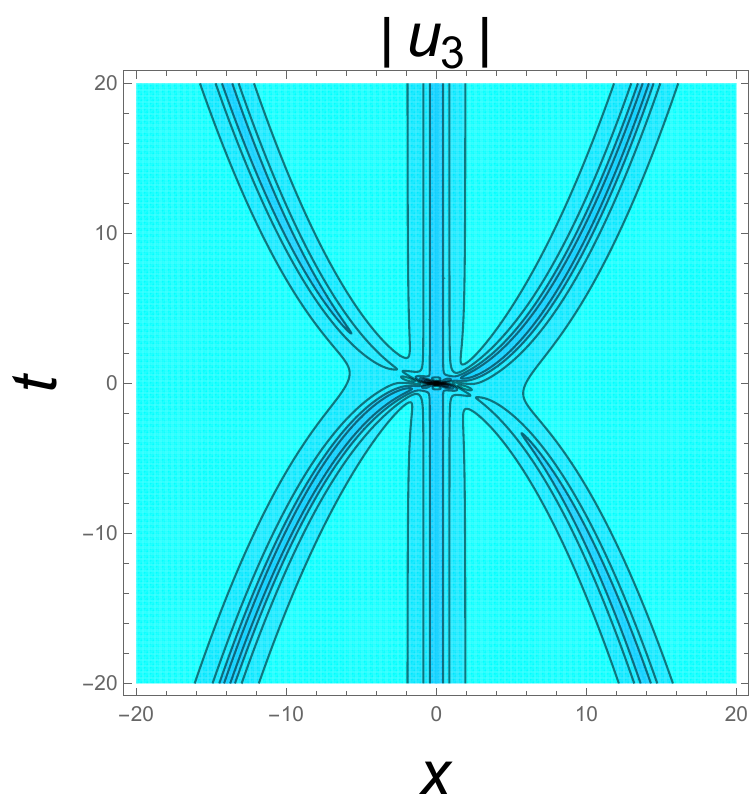}}
	\quad 
	\subfigure[]{
		\includegraphics[width=0.34\linewidth]{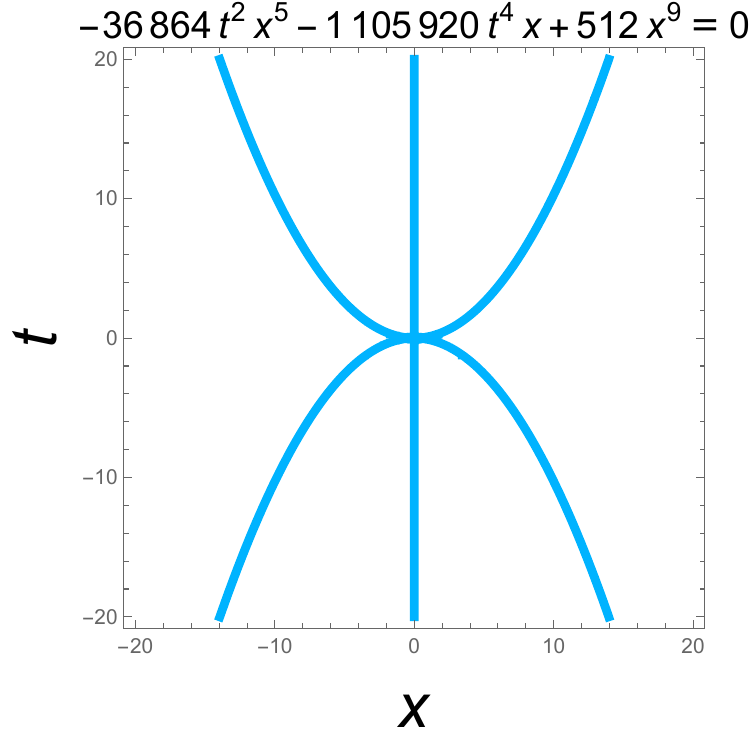}}
\caption{(a) Contour map of the density distribution $|u_3|$. (b) Asymptotic trajectory of the triple-pole soliton in (\ref{eq:contourmapn3nonlocalNLS}).}
	\label{3rdnonlocalNLScontour}
\end{figure}
\end{small}

The contour maps of the density distributions $|u_2|$ and $|u_3|$ are demonstrated in Fig. \ref{2ndnonlocalNLScontour}(a) and Fig. \ref{3rdnonlocalNLScontour}(a), and the asymptotic trajectories of the double-pole and triple-pole solitons are displayed in Fig. \ref{2ndnonlocalNLScontour}(b) and Fig. \ref{3rdnonlocalNLScontour}(b).

\end{remark}

\section{Invariance of DTs regarding poles distribution in $D(\lambda)$ and $D^{-1}(\lambda)$}\label{Section4}

In the previous sections, it has been discussed that how DTs with the same pole can be folded under the same spectral parameter $\lambda_0$. This section considers DTs with more than one pole, i.e., the zeros of ${\rm det}\big(D(\lambda)\big)$ is more than one.
\par
To solve this problem, a possible way is to take the similar approaches in the previous sections. However,
it is found that two DTs may be equivalent although they have different pole forms. Thus this section will introduce a theorem to describe the invariance of Darboux matrices regarding poles distribution in $D(\lambda)$ and $D^{-1}(\lambda)$.
\par
Suppose that $\lambda_1,\lambda_2,\cdots,\lambda_r$ are possible poles of $D(\lambda)$ and $D^{-1}(\lambda)$ with orders $m_j,n_j\in\mathbb{N}$ in Table \ref{Table1}, where $\sum\limits_{j=1}^{r} m_j = \sum\limits_{j=1}^{r} n_j = s$ is the degree of polynomial $\prod\limits _{j=1}^{r}(\lambda-\lambda_j)^{m_j}D(\lambda)$. A distribution of poles of $D(\lambda)$ and $D^{-1}(\lambda)$ means that there exist $\widetilde{m}_j,\widetilde{n}_j\in\mathbb{N}$ such that the orders of poles $\lambda_1,\lambda_2,\cdots,\lambda_r$ of Darboux matrices $\widetilde{D}(\lambda)$ and $\widetilde{D}^{-1}(\lambda)$ are $\widetilde{m}_j$ and $\widetilde{n}_j$ with

\begin{equation}
    \begin{array}{l}
        \text{ (i) Maintain the order of DT invariant, i.e.,} \\
        \qquad\qquad \sum\limits_{i=1}^r \widetilde{n}_i = \sum\limits_{i=1}^r \widetilde{m}_i = s,\\
        \text{(ii) Maintain the sum of the orders of each $\lambda_i$ invariant, i.e.,} \qquad\qquad\\
        \qquad\qquad \widetilde{n}_i+\widetilde{m}_i = n_i + m_i,\quad i=1,\cdots,r.\qquad\qquad
    \end{array}\label{eq:poledistribution}
\end{equation}

The orders of poles of Darboux matrices $\widetilde{D}(\lambda)$ and $\widetilde{D}^{-1}(\lambda)$ are shown in Table \ref{Table2}.\\
\par
\begin{minipage}{\textwidth}
\begin{minipage}[t]{0.48\textwidth}
\renewcommand\arraystretch{1.5}
\makeatletter\def\@captype{table}
\begin{center}
\begin{tabular}{c|c|c|c|c}
\hline
  &$\lambda_1$ & $\lambda_2$ & $\cdots$ & $\lambda_r$\\
\hline
$D(\lambda)$ & $m_1$ & $m_2$ & $\cdots$ & $m_r$   \\
$D^{-1}(\lambda)$ & $n_1$ & $n_2$ & $\cdots$ & $n_r$\\
\hline
\end{tabular}\\
\quad
\par
\end{center}
\caption{Orders of poles of $D(\lambda)$ and $D^{-1}(\lambda)$.}
\label{Table1}
\end{minipage}
\begin{minipage}[t]{0.48\textwidth}
\renewcommand\arraystretch{1.5}
\makeatletter\def\@captype{table}
\begin{center}
\begin{tabular}{c|c|c|c|c}
\hline
  &$\lambda_1$ & $\lambda_2$ & $\cdots$ & $\lambda_r$\\
\hline
$\widetilde{D}(\lambda)$ & $\widetilde{m}_1$ & $\widetilde{m}_2$ & $\cdots$ & $\widetilde{m}_r$   \\
$\widetilde{D}^{-1}(\lambda)$ & $\widetilde{n}_1$ & $\widetilde{n}_2$ & $\cdots$ & $\widetilde{n}_r$\\
\hline
\end{tabular}\\
\end{center}
\caption{Orders of poles of $\widetilde{D}(\lambda)$ and $\widetilde{D}^{-1}(\lambda)$.}
\label{Table2}
\end{minipage}
\end{minipage}

Two Darboux matrices $D(\lambda)$ and $\widetilde{D}(\lambda)$ may be equivalent if the distributions of their poles satisfy (\ref{eq:poledistribution}). The next theorem provides a programme to construct Darboux matrix $\widetilde{D}(\lambda)$ from Darboux matrix $D(\lambda)$.

\begin{theorem}[The existence of DT with arbitrary distribution of poles]\label{thm:DDIeq}
Let $\lambda_{i}\in\mathbb{C}$ $($$i=1,\cdots,r$$)$ and suppose that $D(\lambda)$ and $D^{-1}(\lambda)$ are
\begin{equation}
    \left\{ \begin{array}{l}
    D(\lambda) = I + \sum\limits_{i=1}^{r}\sum\limits_{p_i=0}^{m_i} \frac{A_{i,p_i}}{(\lambda-\lambda_i)^{p_i}},\\
    D^{-1}(\lambda) = I + \sum\limits_{i=1}^{r}\sum\limits_{q_i=0}^{n_i} \frac{B_{i,q_i}}{(\lambda-\lambda_i)^{q_i}},
      \end{array} \right.
\end{equation}
where $n_i,m_i\in\mathbb{N}$, $\sum_{i=1}^r n_i = \sum_{i=1}^r m_i = s\ge 1$, and $A_{i,p_i}$ and $B_{i,q_i}$ are $2\times 2$ matrices independent of $\lambda$. Moreover, set $A_{i,0}=0$, $B_{i,0}=0$.
Then for any distribution of orders $\widetilde{m}_i,\widetilde{n}_i\in\mathbb{N}$ of $\lambda_{i}$ satisfying (\ref{eq:poledistribution}), there exist $\widetilde{A}_{i,\widetilde{p}_i}$ and $\widetilde{B}_{i,\widetilde{q}_i}$ $($$\widetilde{A}_{i,0}=0$, $\widetilde{B}_{i,0}=0$$)$ independent of $\lambda$, such that
\begin{equation}
    \left\{ \begin{array}{l}
    \widetilde{D}(\lambda) = I + \sum\limits_{i=1}^{r}\sum\limits_{\widetilde{p}_i=0}^{\widetilde{m}_i} \frac{\widetilde{A}_{i,\widetilde{p}_i}}{(\lambda-\lambda_i)^{\widetilde{p}_i}},\\
    \widetilde{D}^{-1}(\lambda) = I + \sum\limits_{i=1}^{r}\sum\limits_{\widetilde{q}_i=0}^{\widetilde{n}_i} \frac{\widetilde{B}_{i,\widetilde{q}_i}}{(\lambda-\lambda_i)^{\widetilde{q}_i}},
      \end{array} \right.
\end{equation}
where $\widetilde{D}(\lambda)$ is equivalent to $D(\lambda)$ in sense of
\begin{equation}
    \left\{ \begin{array}{l}
    \prod\limits_{i=1}^r (\lambda-\lambda_i)^{m_i} D(\lambda) = \prod\limits_{i=1}^r  (\lambda-\lambda_i)^{\widetilde{m}_i} \widetilde{D}(\lambda),\\
    \prod\limits_{i=1}^r (\lambda-\lambda_i)^{-m_i} D^{-1}(\lambda) = \prod\limits_{i=1}^{r}  (\lambda-\lambda_i)^{-\widetilde{m}_i} \widetilde{D}^{-1}(\lambda).
      \end{array}\label{eq:DDeqDIDIeq} \right.
\end{equation}

\end{theorem}

\begin{demo}
According to the conditions $\sum_{i=1}^r m_i =\sum_{i=1}^r n_i = \sum_{i=1}^r \widetilde{m}_i = \sum_{i=1}^r \widetilde{n}_i = s$, all of  $\prod_{i=1}^r (\lambda-\lambda_i)^{m_i} D(\lambda)$, $\prod_{i=1}^r  (\lambda-\lambda_i)^{\widetilde{m}_i} \widetilde{D}(\lambda)$, $\prod_{i=1}^{r}  (\lambda-\lambda_i)^{n_i} D^{-1}(\lambda)$ and $\prod_{i=1}^r (\lambda-\lambda_i)^{\widetilde{n}_i} \widetilde{D}^{-1}(\lambda)$ are polynomials of $\lambda$ with degree $s$.

Define
\begin{equation}
    \widetilde{A}_{j,\widetilde{m}_j-k} \triangleq \frac{1}{k!} \left.\frac{\partial^k}{\partial \lambda^k} \left( \prod_{i=1}^{r} (\lambda-\lambda_i)^{m_i}          \left(   I + \sum_{i=1}^{r}\sum_{p_i=0}^{m_i} \frac{A_{i,p_i}}{(\lambda-\lambda_i)^{p_i}} \right)\right)\right|_{\lambda = \lambda_{j}},
\end{equation}
for $1\le j\le r$, $0\le  k \le \widetilde{m}_j-1$, and further denote
\begin{small}
\begin{equation*}
    \Delta(\lambda) \triangleq  \prod_{i=1}^r (\lambda-\lambda_i)^{m_i} \left( I + \sum_{i=1}^{r}\sum_{p_i=0}^{m_i} \frac{A_{i,p_i}}{(\lambda-\lambda_i)^{p_i}} \right) - \prod_{i=1}^r  (\lambda-\lambda_i)^{\widetilde{m}_i} \left( I + \sum_{i=1}^{r}\sum_{\widetilde{p}_i=0}^{\widetilde{m}_i} \frac{\widetilde{A}_{i,\widetilde{p}_i}}{(\lambda-\lambda_i)^{\widetilde{p}_i}}  \right),
\end{equation*}
\end{small}
then it is seen that
\begin{equation}
    \left.\frac{\partial^k}{\partial \lambda^k} \Big( \Delta(\lambda) \Big)\right|_{\lambda = \lambda_{j}} = 0,\quad 1\le j\le r, \; 0\le  k \le \widetilde{m}_j-1.
\end{equation}
This means that $\Delta(\lambda)$ has $\sum_{i=1}^r \widetilde{m}_i=s$ zeros. However, $\Delta(\lambda)$ is a polynomial of $\lambda$ with degree $s-1$ since $A_{i,0}=\widetilde{A}_{i,0}=0$. So $\Delta(\lambda) =0$ by the fundamental theorem of algebra, which leads to the first equation of (\ref{eq:DDeqDIDIeq}).
\par
Notice that the second equation in (\ref{eq:DDeqDIDIeq}) is equivalent to
\begin{equation*}
    \prod_{i=1}^{r}  (\lambda-\lambda_i)^{\widetilde{m}_i-n_i}\prod_{i=1}^{r}  (\lambda-\lambda_i)^{n_i} D^{-1}(\lambda) = \prod_{i=1}^r (\lambda-\lambda_i)^{m_i-\widetilde{n}_i}\prod_{i=1}^r (\lambda-\lambda_i)^{\widetilde{n}_i} \widetilde{D}^{-1}(\lambda),
\end{equation*}
that is
\begin{small}
\begin{equation*}
    \prod_{i=1}^{r}  (\lambda-\lambda_i)^{n_i} D^{-1}(\lambda) = \prod_{i=1}^{r}  (\lambda-\lambda_i)^{\widetilde{m}_i-n_i-m_i+\widetilde{n}_i}\prod_{i=1}^{r}  (\lambda-\lambda_i)^{n_i} D^{-1}(\lambda) =\prod_{i=1}^r (\lambda-\lambda_i)^{\widetilde{n}_i} \widetilde{D}^{-1}(\lambda),
\end{equation*}
\end{small}
by the condition $\widetilde{n}_i+\widetilde{m}_i = n_i + m_i$ for $i=1,2,\cdots,r$. The above equation is a polynomial of $\lambda$, thus the proof of the second equation of (\ref{eq:DDeqDIDIeq}) is similar to the first one. \QED

\end{demo}

In what follows, with the help of DT with the same pole, an example is taken to implement Theorem \ref{thm:DDIeq} in a direct manner, where the rogue waves of the focusing NLS equation are obtained by two DTs with different distributions of poles.\\
\par
{\bf Example: Direct construction of DTs with arbitrary distribution of poles}
\par
The Lax pair of focusing NLS equation
\begin{equation}\label{eq:fNLS}
    {\rm i} u_t = u_{xx} + 2u|u|^2
\end{equation}
is given by
\begin{equation}
\begin{array}{l}
\Phi_x =U(\lambda)\Phi= \left( \begin{matrix} \lambda &  u \\ -\bar u& -\lambda   \end{matrix}\right) \Phi,\\
\Phi_t = V(\lambda)\Phi = \left( \begin{matrix} -2{\rm i} \lambda^2-{\rm i} |u|^2 & -2{\rm i}\lambda u-{\rm i} u_x \\ 2{\rm i}\lambda \bar u-{\rm i} \bar u_x& 2{\rm i} \lambda^2+{\rm i} |u|^2   \end{matrix}\right) \Phi,
\end{array}\label{eq:laxpairNLS}
\end{equation}
where the symmetries of $U(\lambda)$ and $V(\lambda)$ are
\begin{equation}\label{eq:fNLSsymmetry}
    U(-\bar\lambda) = - U^*(\lambda),\qquad V(-\bar\lambda) = - V^*(\lambda),
\end{equation}
which and the fact that $\tr(U)=\tr(V)=0$ indicate that the following lemma.
\begin{lemma}
If $\Phi(\lambda)$ is a fundamental solution matrix of Lax pair (\ref{eq:laxpairNLS}) at $\lambda$,
$\left( \begin{matrix} 0 &  -1 \\ 1& 0  \end{matrix}\right)\bar\Phi(x,t;-\bar\lambda)$ is the other fundamental solution matrix of Lax pair (\ref{eq:laxpairNLS}) at $\lambda$.
\end{lemma}
\par
{\bf Seed solution A: plane wave solution}
\par
Now it is ready to solve the inverse problem of Lax pair (\ref{eq:laxpairNLS}). In doing so,
taking the seed solution of the focusing NLS equation (\ref{eq:fNLS}) as $u_0(x,t)=\frac{1}{2}e^{-\frac{1}{2}i t }$, the corresponding fundamental solution matrix of Lax pair (\ref{eq:laxpairNLS}) is
\begin{small}
\begin{equation*}
    \Phi(x,t;\lambda)=\frac{1}{\alpha}\left( \begin{matrix} e^{-\frac{1}{4}i t} &  0 \\ 0 & e^{\frac{1}{4}i t}  \end{matrix}\right)\left( \begin{matrix} \frac{1}{2}e^{\varphi}(\alpha+\lambda)+\frac{1}{2}e^{-\varphi}(\alpha-\lambda) &  \frac{1}{4}e^{\varphi}-\frac{1}{4}e^{-\varphi} \\ -\frac{1}{4}e^{\varphi}+\frac{1}{4}e^{-\varphi}  & \frac{1}{2}e^{\varphi}(\alpha-\lambda)+\frac{1}{2}e^{-\varphi}(\alpha+\lambda)  \end{matrix}\right),
\end{equation*}
\end{small}
where $\alpha = \sqrt{\lambda^2-\frac{1}{4}}$ and $\varphi = \alpha(x-2{\rm i}\lambda t)$. Notice that $\alpha=0$ if $\lambda=\pm \frac{1}{2}$. Let $\lambda_0 =\frac{1}{2} $, then the solution of the Lax pair (\ref{eq:laxpairNLS}) at $\lambda=\lambda_0$ is rational. It seems that $\Phi(\lambda)$ is not analytic at $\lambda_0$ due to the square root in $\alpha$. However, the integral of elements of $\Phi(\lambda)$ along small circle including $\lambda_0$ is $0$ and $\Phi(\lambda)$ is continuous on the interval $(-\frac{1}{2}, \frac{1}{2})$, which indicates that $\Phi(\lambda)$ is analytic at $\lambda_0$. Some limit processes are needed to get this rational solution and its derivative, i.e.,
\begin{equation}
\begin{array}{l}
    \Phi_0 = \Phi(\lambda=\lambda_0) = \lim\limits_{\lambda\rightarrow\lambda_0} \Phi(\lambda), \\
    \Phi_0^{(1)} = \left.\frac{\partial}{\partial\lambda}\Phi(\lambda)\right|_{\lambda=\lambda_0} = \lim\limits_{\lambda\rightarrow\lambda_0} \left(\frac{\partial}{\partial\lambda}\Phi(\lambda)\right).
\end{array}
\end{equation}
\par
Define two Darboux matrices by the following ways
\begin{equation}
    \begin{array}{l}
    \xi (\lambda) \triangleq \xi_0 + \xi_0^{(1)} (\lambda-\lambda_0),\\
    b(\lambda) \triangleq \Phi(x,t;\lambda) \xi(\lambda),\\
    h(\lambda) \triangleq \left( \begin{matrix} 0 &  -1 \\ 1& 0  \end{matrix}\right)\bar\Phi(x,t;-\bar\lambda) \bar\xi(-\bar\lambda),\\
    H_1 \triangleq (b_0,b_0^{(1)}),\\
    S_1 \triangleq H_1 \left( \begin{matrix}\lambda_0 &  1 \\ 0& \lambda_0  \end{matrix}\right) H_1^{-1},\\
    D_1^{[1]}(\lambda) \triangleq \lambda I - S_1,\\
    h^{[1]}(\lambda)\triangleq D_1^{[1]}(\lambda) h(\lambda),\\
    H_2\triangleq ((h^{[1]})_{0^*},(h^{[1]})_{0^*}^{(1)}),\\
    S_2\triangleq H_2 \left( \begin{matrix}-\bar\lambda_0 &  1 \\ 0& -\bar\lambda_0  \end{matrix}\right) H_2^{-1},\\
    D_1^{[2]}(\lambda) \triangleq \lambda I - S_2,
    \end{array}\label{eq:DTNLSsame}
\end{equation}
where $\xi_0\neq0$ and $\xi_0^{(1)}$ are arbitrary parameters vectors, and the subscript $0^*$ in $H_2$ means $-\bar\lambda_0$. By Theorem~\ref{thm:order1threeequi}, both $D_1^{[1]}(\lambda)$ and $D_1^{[2]}(\lambda)$ are Darboux matrices, so is $D_1^{[2]}(\lambda)D_1^{[1]}(\lambda)$. Thus the exact solution of the focusing NLS equation (\ref{eq:fNLS}) is
\begin{equation}
    u(x,t)= u_0(x,t) + 2S_1(1,2) + 2 S_2 (1,2),
\end{equation}
by comparing the coefficients of $\lambda$ in (\ref{eq:UDDxDU}), where $S_1(1,2)$ and $S_2(1,2)$ are the $(1,2)$ elements of $S_1$ and $S_2$, respectively.
\par
The first-order and second-order rogue waves of the focusing NLS equation (\ref{eq:fNLS}) can be obtained by choosing proper parameters vectors $\xi_0$ and $\xi_0^{(1)}$.
For example, taking $\xi_0 = (1,-1)^T$ and $\xi_0^{(1)} = (1,1)^T$, the first-order rogue wave solution is derived as
\begin{equation}\label{eq:rworder1}
    u(x,t) =\frac{t (4 i + t) + x (2 + x)-2}{2 (2 + t^2 + x (2 + x))}e^{-\frac{i t}{
  2}}.
\end{equation}
\par
\begin{figure}[H] 
	\centering  
	\vspace{-0.35cm} 
	\subfigtopskip=2pt 
	\subfigbottomskip=2pt 
	\subfigcapskip=-5pt 
	\subfigure[]{
		\includegraphics[width=0.35\linewidth]{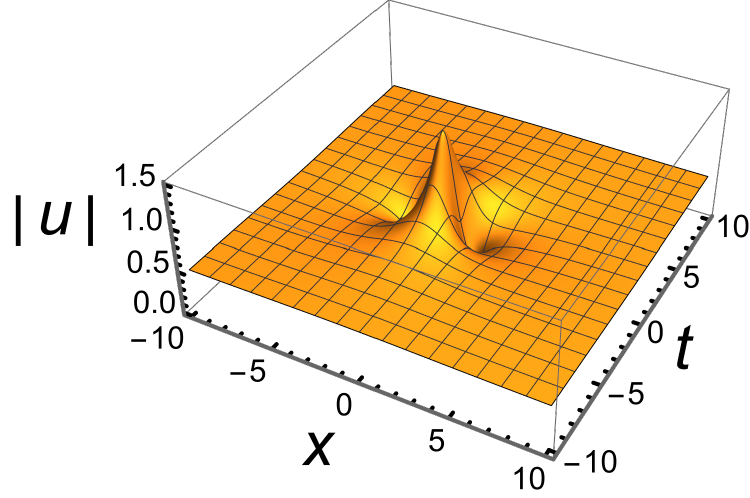}}
\quad
	\subfigure[]{
		\includegraphics[width=0.35\linewidth]{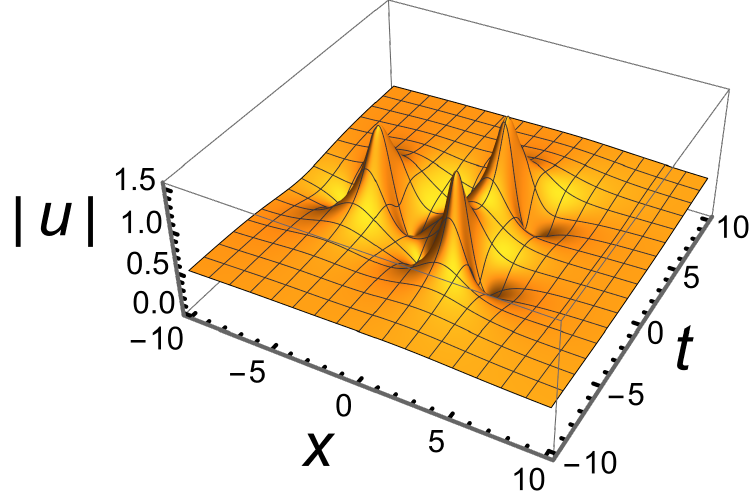}}
\caption{(a) The first-order rogue wave (\ref{eq:rworder1}). (b) The second-order rogue wave (\ref{eq:rwuNLS2nd}). The construction of DT used here, i.e., the two fold DT with the same pole (cf. (\ref{eq:DTNLSsame})), is different from that in Ref. \cite{guo2012nonlinear}.}
	\label{fNLSrw12}
\end{figure}

Taking $\xi_0 = (1,0)^T$ and $\xi_0^{(1)} = (5,9)^T$, the second-order rogue wave solution is derived as
\begin{equation}\label{eq:rwuNLS2nd}
        u(x,t)=\frac{ P_1+P_2 }{P_3+P_4}e^{-\frac{i t}{
  2}},\\
\end{equation}
where
\begin{footnotesize}
\begin{equation*}
\begin{array}{l}
    P_1 =  4248- 1584 i t - 540 t^2 + 48 i t^3 - 30 t^4 +
 12 i t^5 + t^6 + (1224 - 1440 i t - 504 t^2 + 48 i t^3 +
    6 t^4) x,\\
    P_2 = (252 - 72 t^2 + 24 i t^3 + 3 t^4) x^2 + (96 + 48 i t +
    12 t^2) x^3 + (6 + 12 i t + 3 t^2) x^4 + 6 x^5 + x^6,\\
    P_3 = 5904 - 504 t^2 + 60 t^4 +
 2 t^6 + (144 - 720 t^2 + 12 t^4) x,\\
 P_4 = (792 + 6 t^4) x^2 + (288 +
    24 t^2) x^3 + (36 + 6 t^2) x^4 + 12 x^5 + 2 x^6.
 \end{array}
\end{equation*}
\end{footnotesize}
Figure \ref{fNLSrw12} displays the structures of the first-order and second-order rogue waves in (\ref{eq:rworder1}) and (\ref{eq:rwuNLS2nd}), respectively.

The rogue wave solutions (\ref{eq:rworder1}) and (\ref{eq:rwuNLS2nd}) are well-known results by generalized DT \cite{guo2012nonlinear}, however, these solutions can also be derived by a new kind of DT proposed in Theorem~\ref{thm:order1threeequi}.
\par
When deriving the rogue wave solutions (\ref{eq:rworder1}) and (\ref{eq:rwuNLS2nd}), the distributions of poles of the generalized DT in \cite{guo2012nonlinear} and our new kind of DT are different, which provides an example for Theorem~\ref{thm:DDIeq}. Indeed,
our DT $D_1(\lambda)=(\lambda-\lambda_0)^{-1}(\lambda+\bar\lambda_0)^{-1}D_1^{[2]}(\lambda)D_1^{[1]}(\lambda)$, where $D_1^{[1]}(\lambda)$ and $D_1^{[2]}(\lambda)$ are given in (\ref{eq:DTNLSsame}), has the following pole form
\begin{equation}
\begin{array}{l}
    D_1(\lambda) = I - \left(\begin{matrix}b_0 & h_{0^*}\end{matrix}\right)
    \left(\begin{matrix} a_0b_0^{(1)}&   \frac{a_0 h_{0^*}}{-\bar\lambda_0-\lambda_0} \\ \frac{g_{0^*}b_0}{\lambda_0+\bar\lambda_0}  & g_{0^*}h_{0^*}^{(1)}\end{matrix} \right)^{-1} \left(\begin{matrix}\frac{a_0}{\lambda-\lambda_0}\\  \frac{g_{0^*}}{\lambda+\bar\lambda_0}  \end{matrix}\right), \\
    D_1^{-1}(\lambda) = I + \left(\begin{matrix}\frac{b_0}{\lambda-\lambda_0} &  \frac{h_{0^*}}{\lambda+\bar\lambda_0} \end{matrix}\right) \left(\begin{matrix} a_0b_0^{(1)}&   \frac{a_0 h_{0^*}}{-\bar\lambda_0-\lambda_0} \\ \frac{g_{0^*}b_0}{\lambda_0+\bar\lambda_0}  & g_{0^*}h_{0^*}^{(1)}\end{matrix} \right)^{-1} \left(\begin{matrix}a_0\\ g_{0^*} \end{matrix}\right),
\end{array}\label{eq:distributionD1}
\end{equation}
where $b(\lambda)$ and $h(\lambda)$ are the solutions of Lax pair (\ref{eq:laxpairNLS}), and $a(\lambda)$ and $g(\lambda)$ are the solutions of the adjoint Lax pair corresponding to $b(\lambda)$ and $h(\lambda)$, respectively. Moreover, it requires that $a(\lambda)$ is analytic at $\lambda_0$, $g(\lambda)$ is analytic at $-\bar{\lambda}_0$, and $(ab)_0 = (gh)_{0^*} = 0$.
\par
The orders of poles of $D_1(\lambda)$ and $D_1^{-1}(\lambda)$ are shown in Table 2(a).\\
\par

\begin{minipage}{\textwidth}
\begin{minipage}[t]{0.48\textwidth}
\renewcommand\arraystretch{1.2}
\makeatletter\def\@captype{table}
\quad
\begin{center}
\begin{tabular}{c|c|c}
\hline
  &$\lambda_0$ & $-\bar\lambda_0$ \\
\hline
$D_1(\lambda)$ & 1 & 1   \\
$D_1^{-1}(\lambda)$ & 1 & 1\\
\hline
\end{tabular}
\end{center}
\caption{Orders of poles of $D_1(\lambda)$ in (\ref{eq:distributionD1})}
\label{table:orderofpoleD1}
\end{minipage}
\begin{minipage}[t]{0.48\textwidth}
\renewcommand\arraystretch{1.2}
\makeatletter\def\@captype{table}
\quad
\begin{center}
\begin{tabular}{c|c|c}
\hline
  &$\lambda_0$ & $-\bar\lambda_0$ \\
\hline
$D_2(\lambda)$ & 0 & 2   \\
$D_2^{-1}(\lambda)$ & 2 & 0\\
\hline
\end{tabular}
\end{center}
\caption{Orders of poles of $D_2(\lambda)$ in (\ref{eq:distributionD2})}
\label{table:orderofpoleD2}
\end{minipage}
\end{minipage}
\par
Distributing the orders of poles in Table 2(a) according to the rule in (\ref{eq:poledistribution}), another orders of poles are listed in Table 2(b), where the corresponding DT is denoted by $D_2(\lambda)$. Notice that $D_2(\lambda)$ is just the generalized DT in \cite{guo2012nonlinear}, whose pole form is
\begin{equation}
\begin{array}{l}
    D_2(\lambda) = I - \left(\begin{matrix}b_0 & b_{0}^{(1)}\end{matrix}\right)
    \Gamma_2^{-1} \left(\begin{matrix}\frac{g_{0^*}}{\lambda+\bar\lambda_0}\\  \frac{g_{0^*}}{(\lambda+\bar\lambda_0)^2} +  \frac{g_{0^*}^{(1)}}{\lambda+\bar\lambda_0}  \end{matrix}\right), \\
    D_2^{-1}(\lambda) =I + \left(\begin{matrix}\frac{b_0}{\lambda-\lambda_0} & \frac{b_0^{(1)}}{\lambda-\lambda_0}+\frac{b_0}{(\lambda-\lambda_0)^2} \end{matrix}\right)
    \Gamma_2^{-1} \left(\begin{matrix}  g_{0^*} \\  g_{0^*}^{(1)}  \end{matrix}\right), \\
\end{array}\label{eq:distributionD2}
\end{equation}
where $(gh)_{0^*} = (gh)_{0^*}^{(1)} = 0$, and
\begin{equation}
    \Gamma_2 = \left(\begin{matrix} \frac{g_{0^*}b_0}{\lambda_0+\bar\lambda_0}  &   \frac{g_{0^*}b_0^{(1)}}{\lambda_0+\bar\lambda_0} - \frac{g_{0^*}b_0}{(\lambda_0+\bar\lambda_0)^2} \\ \frac{g_{0^*}^{(1)}b_0}{\lambda_0+\bar\lambda_0}+\frac{g_{0^*}b_0}{(\lambda_0+\bar\lambda_0)^2} & \frac{g_{0^*}^{(1)}b_0^{(1)}}{\lambda_0+\bar\lambda_0} - \frac{g_{0^*}^{(1)}b_0}{(\lambda_0+\bar\lambda_0)^2} + \frac{g_{0^*}b_0^{(1)}}{(\lambda_0+\bar\lambda_0)^2} - \frac{2g_{0^*}b_0}{(\lambda_0+\bar\lambda_0)^3} \end{matrix} \right).
\end{equation}
\par
In fact, our new kind of DT $D_1(\lambda)$ is equivalent to the generalized DT $D_2(\lambda)$ in sense of
\begin{equation}\label{eq:D1=D2polynomialsecdistri}
    (\lambda-\lambda_0)(\lambda+\bar\lambda_0) D_1(\lambda) = (\lambda+\bar\lambda_0)^{2} D_2(\lambda),
\end{equation}
since both sides of the above equations are the unique solution of the equations
\begin{equation}
        \left\{
        \begin{array}{l}
            (Db)_0 = 0,\\
            (Db)_0^{(1)} = 0,\\
            (Dh)_{0^*}= 0,\\
            (Dh)_{0^*}^{(1)} = 0,
        \end{array}\label{eq:NLSrw2LEQ}
        \right.
\end{equation}
where for functions $A(x,t)$ and $B(x,t)$ to be determined, $D(\lambda)$ takes the form
\begin{equation}
    D(x,t;\lambda) = \lambda^2 I + \lambda A(x,t) + B(x,t).
\end{equation}
Then the equations in (\ref{eq:NLSrw2LEQ}) are linear algebraic equations of $\Big(A(x,t),B(x,t)\Big)$, namely
\begin{equation}\label{eq:NLSrw2LEQlinear}
    \Big(A(x,t), B(x,t)\Big) M(x,t)  = \Big(   -\lambda_0^2b_0 ,\; -2 \lambda_0b_0 -\lambda_0^2b_0^{(1)} ,\; -\bar\lambda_0^2 h_{0^*} ,\; 2\bar\lambda_0h_{0^*} -\bar\lambda_0^2 h_{0^*}^{(1)}     \Big),
\end{equation}
where the coefficient $M(x,t)$ is a $4\times4$ matrix of form
\begin{equation}\label{eq:coeffmatNLSM}
    M(x,t) = \left( \begin{matrix} \lambda_0b_0 & b_0 + \lambda_0b_0^{(1)} & -\bar\lambda_0 h_{0^*} & h_{0^*} -\bar\lambda_0 h_{0^*}^{(1)} \\ b_0 & b_0^{(1)} &  h_{0^*} & h_{0^*}^{(1)}         \end{matrix}  \right).
\end{equation}
Equation (\ref{eq:NLSrw2LEQlinear}) has unique solution if $M(x,t)$ is invertible. Fortunately, in the case of focusing NLS equation (\ref{eq:fNLS}), it can be checked that ${\rm det}(M(x,t))\neq 0$ if $b_0\neq 0$.
\par
\begin{remark}
The solution $g(x,t;\lambda)$ of adjoint Lax pair in $D_1(\lambda)$ (\ref{eq:distributionD1}) is a little different from that in $D_2(\lambda)$ (\ref{eq:distributionD2}) since the $j$-th derivative in condition $(gh)_{0^*}^{(j)}=0$ is different. However, $g(x,t;\lambda)$ is just an auxiliary function according to equations (\ref{eq:NLSrw2LEQ}), and there is such function meeting the conditions.

\end{remark}

\begin{remark}\label{rmk:singularfirstDTNLS}
The Darboux matrices $D_1(\lambda)$ in (\ref{eq:distributionD1}) and $D_2(\lambda)$ in (\ref{eq:distributionD2}) can be constructed by folding two DTs of order one, i.e.,
\begin{equation}\label{rmk:D1D2}
\begin{array}{l}
    D_1(\lambda) =(\lambda-\lambda_0)^{-1}(\lambda+\bar\lambda_0)^{-1} D_1^{[2]}(\lambda) D_1^{[1]}(\lambda), \\
    D_2(\lambda) =(\lambda+\bar\lambda_0)^{-2} D_2^{[2]}(\lambda) D_2^{[1]}(\lambda),
\end{array}
\end{equation}
where $D_1^{[2]}(\lambda)$ and $D_1^{[1]}(\lambda)$ are given by (\ref{eq:DTNLSsame}), and
\begin{equation}\label{rmk:D1D2-2}
\begin{array}{l}
    D_2^{[1]} = B[\lambda_0,-\bar\lambda_0,1,b(\lambda),h(\lambda)] , \\
    D_2^{[2]} = B[\lambda_0,-\bar\lambda_0,1,\frac{1}{\lambda-\lambda_0}D_2^{[1]}(\lambda)b(\lambda),\frac{1}{\lambda+\bar\lambda_0}D_2^{[1]}(\lambda)h(\lambda)] .
\end{array}
\end{equation}
The Darboux matrices $D_1(\lambda)$ and $D_2(\lambda)$ in (\ref{rmk:D1D2}) with (\ref{rmk:D1D2-2}) can be explained in detail. Taking the free parameter $\xi_0 = (1,-1)^T$, the Darboux matrix $D_2^{[1]}(\lambda)$ transforms the seed solution $u_0(x,t)$ into $-u_0(x,t)$, then the Darboux matrix $D_2^{[2]}(\lambda)$ produces a first-order rogue wave. In addition, $D_1^{[1]}(\lambda)$ doesn't generate a solution to the focusing NLS equation (\ref{eq:fNLS}) because the DT with $D_1^{[1]}(\lambda)$ doesn't keep the symmetries in (\ref{eq:fNLSsymmetry}). However, the DT with $D_1^{[2]}(\lambda) D_1^{[1]}(\lambda)$ keeps the symmetries in (\ref{eq:fNLSsymmetry}), which leads to a first-order rogue wave of the focusing NLS equation (\ref{eq:fNLS}).
\par
Taking the free parameters $\xi_0 = (1,0)^T$ and $\xi_0^{(1)} = (5,9)^T$, it is seen that the determinant of matrix $H_1$ in  (\ref{eq:DTNLSsame})
\begin{equation}\label{eq:detH1NLSblowup}
    {\rm det}(H_1) = \frac{1}{6} (54 + i t^3 + 3 x^2 + x^3 - 3 t^2 (1 + x) -
   3 i t (x^2 + 2 x -2))
\end{equation}
has zeros in $(x,t)\in \mathbb{R}^2$, which means that $D_1^{[1]}(\lambda)$ has singularities. However, iterating $D_1^{[1]}(\lambda)$ once, one can arrive at the matrix $H_2$ in  (\ref{eq:DTNLSsame}), whose determinant is
\begin{equation}
    {\rm det}(H_2) = \frac{G}{-36i \;{\rm det}(H_1)},
\end{equation}
where
\begin{equation}
\begin{array}{l}
    G = i (2952 + t^6 + 72 x + 396 x^2 + 144 x^3 + 18 x^4 + 6 x^5 + x^6 \\
    +
   3 t^4 (10 + 2 x + x^2) + 3 t^2 (-84 - 120 x + 4 x^3 + x^4)),
\end{array}
\end{equation}
doesn't have zeros for  $(x,t)\in\mathbb{R}^2$. In this sense, the singularities of $H_1^{-1}$ and $H_2$ are eliminated in a miracle way, and a global second-order rogue wave solution (\ref{eq:rwuNLS2nd}) is obtained.
\end{remark}

\par
{\bf Seed solution B: zero solution}
\par
In what follows, the zero seed solution and a higher-order DT are taken to illustrate the pole distribution rule in Theorem \ref{thm:DDIeq}. However, the description will be much more concise compared to the case of plane wave seed solution above. We only provide an example to show how the multiple-pole soliton solutions of the focusing NLS equation (\ref{eq:fNLS}) are obtained by a new kind of DT that has a different pole distribution from the generalized DT. To be specific, taking seed solution $u_0(x,t) =0$, construct two Darboux matrices of the forms
\begin{equation}
\begin{array}{l}
    D^{[1]}[\lambda_0,2,h_1(x,t;\lambda)],\\
    D^{[2]}[-\bar\lambda_0,2,h_2(x,t;\lambda)],
\end{array}\label{eq:DTmultipolesolitonNLS}
\end{equation}
where $D^{[1]}[\cdots]$ and $D^{[2]}[\cdots]$ are two DTs defined by (\ref{eq:DTsamepolenn}), and
\begin{equation}
\begin{array}{l}
    h_1(x,t;\lambda) = \Phi(x,t;\lambda)\xi(\lambda),\\
    \xi(\lambda) = \xi_0 + \xi_0^{(1)}(\lambda-\lambda_0) + \frac{1}{2} \xi_0^{(2)}(\lambda-\lambda_0)^2 + \frac{1}{6} \xi_0^{(3)}(\lambda-\lambda_0)^3,\\
    h_2(x,t;\lambda) = D^{[1]}[\lambda_0,2,h_1(x,t;\lambda)](x,t;\lambda)  \left( \begin{matrix} 0 &  -1 \\ 1& 0  \end{matrix}\right)\bar\Phi(x,t;-\bar\lambda) \bar\xi(-\bar\lambda).
\end{array}
\end{equation}
\par
Taking the spectral parameter $\lambda_0 = 1$ and free parameters $\xi_0^{(1)}=(3,6)^T$, $\xi_0^{(2)}=(7,11)^T$ and $\xi_0^{(3)}=(27,365)^T$, the triple-pole soliton solution of the focusing NLS equation (\ref{eq:fNLS}) is obtained, which is listed in (\ref{triple-pole-soliton-fNLS}) of the Appendix.
\par
\begin{figure}[H] 
	\centering  
	\vspace{-0.35cm} 
	\subfigtopskip=2pt 
	\subfigbottomskip=2pt 
	\subfigcapskip=-5pt 
	\subfigure[$\xi_0=(1,0)^T$]{
		\includegraphics[width=0.35\linewidth]{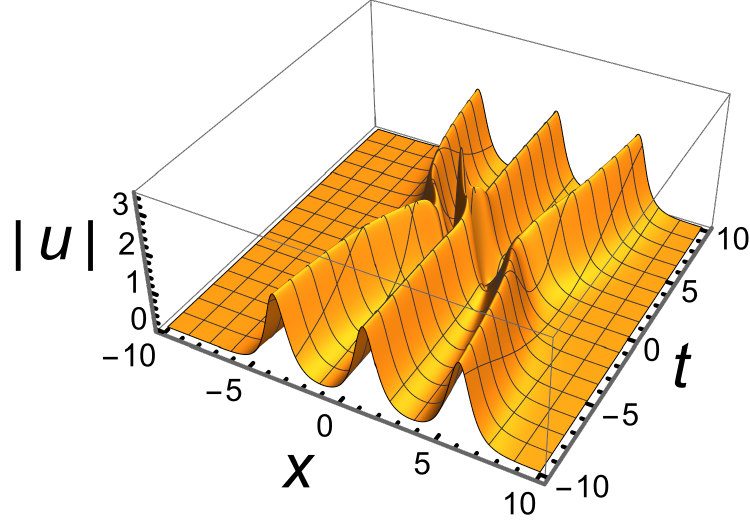}}
\quad
	\subfigure[$\xi_0=(1,1)^T$]{
		\includegraphics[width=0.35\linewidth]{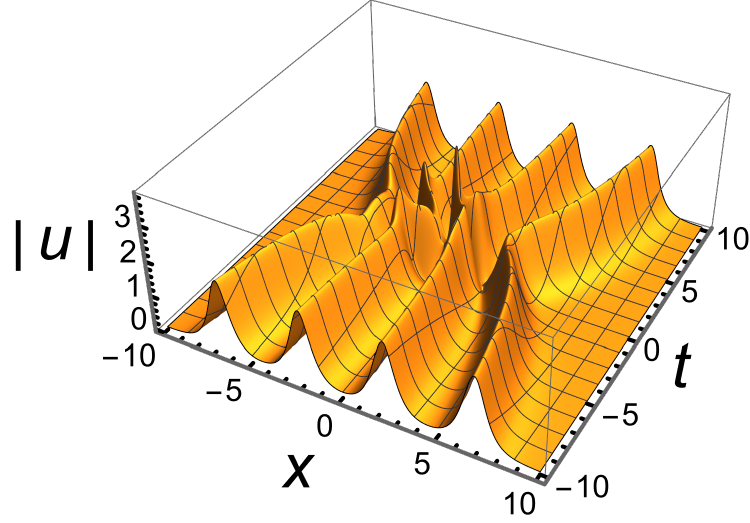}}
\caption{(a) The triple-pole soliton. (b) The quadruple-pole soliton. The DT used here is a fold of two DTs with the same pole of order two, i.e., the terms in (\ref{eq:DTmultipolesolitonNLS}).}
	\label{fNLSsoliton34}
\end{figure}

Take $\xi_0=(1,1)^T$ and the parameters $\xi_0^{(1)}, \xi_0^{(2)}$ and $\xi_0^{(3)}$ remain the same as above, then the quadruple-pole soliton solution of the focusing NLS equation (\ref{eq:fNLS}) is obtained, of the focusing NLS equation (\ref{eq:fNLS}) is obtained, which is listed in (\ref{quadruple-pole-soliton-fNLS}) of the Appendix. The structures of the triple-pole and quadruple-pole soliton solutions are displayed in Fig. \ref{fNLSsoliton34}.

\section{Formal DT in general form}\label{Section5}

The word ``formal" in the title of this section refers to the assumption that the coefficient matrix of the linear equations (\ref{eq:DTgeneral}) below is always invertible. This is a generic condition that is reasonable. For example, the coefficient matrix $M(x,t)$ in (\ref{eq:coeffmatNLSM}) for the focusing NLS equation (\ref{eq:fNLS}) is invertible. However, this is not always true in the reduction of other kinds of integrable systems. In some cases, the obtained solutions will blow up when the coefficient matrix of the linear equations is not invertible.
\par
In the proof of Theorem~\ref{thm:DDIeq}, the construction of $\widetilde{D}(\lambda)$ and $\widetilde{D}^{-1}(\lambda)$ requires a pair of known Darboux matrices $D(\lambda)$ and $D^{-1}(\lambda)$. Moreover, the invariants between $D(\lambda)$ and $\widetilde{D}(\lambda)$ are two quantities: one is the order of the DT, and the other is the sum of the orders of each pole. Thus our task is to construct a Darboux matrix $D(\lambda)$ directly, where the order of $D(\lambda)$ and the zeros (consider the multiplicities) of ${\rm det}\Big( D(\lambda) \Big)$ are arbitrary. This task can be accomplished if the Darboux matrix $D(\lambda)$ can be written as product of some monic Darboux matrices of order one. It can be achieved by combining the classic Darboux matrix and the DT with same pole given in (\ref{eq:DTsamepolenn}).
\par
First of all, the next theorem extends the idea of the equations in (\ref{eq:NLSrw2LEQ}) by considering arbitrary poles.

\begin{theorem}[Construct DT by solving the linear equations in (\ref{eq:DTgeneral})]\label{thm:DTingeneral}
For $j=1,2,\cdots,p,$ let $\lambda_j\in\mathbb{C}$, $m_j\in\mathbb{N}_{+}$ and nonzero $2\times 1$ vector $b_j(x,t;\lambda)$ be the solution of Lax pair (\ref{eq:laxpair1}). Further, assume that $b_j(x,t;\lambda)$ is analytic at $\lambda_j$ and  $s = \frac{1}{2}\sum_{j=1}^{p} m_j$ is an integer. If $D(\lambda)$ is a monic polynomial of degree $s$ in the form
\begin{equation}\label{eq:DT2s}
D(x,t;\lambda) = \lambda^{s} I + \sum_{i=1}^{s} \lambda^{s-i} D_i(x,t)
\end{equation}
and satisfies the linear algebraic equations of $D_i(x,t)~(i=1,2,\cdots,s)$, i.e.,
\begin{equation}\label{eq:DTgeneral}
    \left.\frac{\partial^{k_j}}{\partial \lambda^{k_j}}\Big(   D(x,t;\lambda)    b_j(x,t;\lambda)    \Big) \right|_{\lambda=\lambda_j} =0,
\end{equation}
for $k_j=0,1,\cdots,m_j-1$, then $D(\lambda)$ is a Darboux matrix.
\end{theorem}

\begin{demo}
From a formal point of view, the equations in (\ref{eq:DTgeneral}) have unique solution if the term $2s$ is an even number. So if one can construct a monic Darboux matrix of order $s$ satisfying the equations in (\ref{eq:DTgeneral}), the proof is complete.
\par
Without loss of generality, assume that $m_1,m_2,\cdots,m_{2q}$ are odd, especially, let $m_1=\cdots=m_{r}=1$, $m_{r+1},\cdots,m_{2q}>1$ and $m_{2q+1},\cdots,m_{p}$ be even numbers, where $0\le r \le 2q$ and $0\le q\le \frac{p}{2}$ are two integers.
\par
Using the notations in (\ref{eq:binaryDT1}) and (\ref{eq:DTsamepolenn}), the monic $D(\lambda)$ of order $s$ can be constructed as follows:
\begin{equation}\label{eq:DsfoldofBD}
    D(\lambda) \triangleq B^{[q]}(\lambda) \cdots B^{[1]}(\lambda) D^{[p-r]} (\lambda)\cdots D^{[1]}(\lambda),
\end{equation}
where
\begin{equation*}
    \begin{array}{l}
        D^{[1]}[\lambda_{p},\frac{m_{p}}{2},b_p(\lambda)],\\
        D^{[2]}[\lambda_{p-1},\frac{m_{p-1}}{2},D^{[1]}(\lambda)b_{p-1}(\lambda)],\\
        \cdots\\
        D^{[p-2q]}[\lambda_{2q+1},\frac{m_{2q+1}}{2},D^{[p-2q-1]}(\lambda)\cdots D^{[1]}(\lambda)b_{2q+1}(\lambda)],\\
        D^{[p-2q+1]}[\lambda_{2q},\frac{m_{2q}-1}{2},D^{[p-2q]}(\lambda)\cdots D^{[1]}(\lambda)b_{2q}(\lambda)],\\
        \cdots\\
        D^{[p-r]}[\lambda_{r+1},\frac{m_{r+1}-1}{2},D^{[p-r-1]}(\lambda)\cdots D^{[1]}(\lambda)b_{r}(\lambda)],\\
        h_j(x,t;\lambda) \triangleq \frac{1}{(\lambda-\lambda_0)^{m_j-1}} D^{[p-r]}(\lambda)\cdots D^{[1]}(\lambda) b_j(\lambda) ,\quad j=1,\cdots,2q,\\
        B^{[1]}[\lambda_1,\lambda_2,1,h_1(\lambda),h_2(\lambda)],\\
        B^{[2]}[\lambda_3,\lambda_4,1,B^{[1]}(\lambda)h_3(\lambda),B^{[1]}(\lambda)h_4(\lambda)],\\
        \cdots \\
        B^{[q]}[\lambda_{2q-1},\lambda_{2q},1,B^{[q-1]}(\lambda)\cdots B^{[1]}(\lambda)h_{2q-1}(\lambda),B^{[q-1]}(\lambda)\cdots B^{[1]}(\lambda)h_{2q}(\lambda)].
    \end{array}
\end{equation*}
 \QED
\end{demo}

\begin{corollary}[The DT constructed by solving (\ref{eq:DTgeneral}) is decomposable]\label{corollary:DTtonDet}
The Darboux matrix $D(x,t;\lambda)$ in Theorem \ref{thm:DTingeneral} is the $s$ iterations of monic Darboux matrices of order one, and
    \begin{equation}\label{eq:detDnfrac}
        {\rm det}\Big( D(x,t;\lambda) \Big) = \prod_{j=1}^{p}(\lambda-\lambda_j)^{m_j}.
    \end{equation}
\end{corollary}

\begin{demo}
By the equations in (\ref{eq:Dlambdaitera}), $D[\lambda_{0},m,b(\lambda)]$ is the $m$ iterations of monic Darboux matrices of order one. Then $D(x,t;\lambda)$ is the $s$ iterations of monic Darboux matrices of order one from the equation (\ref{eq:DsfoldofBD}). Notice that
\begin{equation*}
    \begin{array}{l}
{\rm det}\Big( D[\lambda_0,1,h(x,t;\lambda)] \Big) = (\lambda-\lambda_0)^2,\\
{\rm det}\Big( B[\lambda_1,\lambda_2,1,h_1(x,t;\lambda),h_2(x,t;\lambda)] \Big) = (\lambda-\lambda_1)(\lambda-\lambda_2),
\end{array}
\end{equation*}
then the equation (\ref{eq:detDnfrac}) holds immediately. \QED
\end{demo}

\begin{corollary}\label{corollary:polyDTtopoleDT}
For any given $s_i\in\mathbb{N}$ for $i=1,2,\cdots,p$, if they satisfy $0\le s_i \le m_i$ and $\sum\limits_{i=1}^{p}s_i=s$, then the Darboux matrix $D(x,t;\lambda)$ in Theorem \ref{thm:DTingeneral} admits the following pole form
    \begin{equation}\label{eq:Dformalpoleform}
        \left\{\begin{array}{l}
            D(x,t;\lambda) = \prod\limits_{i=1}^{p}(\lambda-\lambda_i)^{s_i} \left( I + \sum\limits_{i=1}^{p}\sum\limits_{k_i=0}^{s_i} \frac{A_{i,k_i}(x,t)}{(\lambda-\lambda_i)^{k_i}}\right) ,\\
            D^{-1}(x,t;\lambda) = \prod\limits_{i=1}^{p}(\lambda-\lambda_i)^{-s_i} \left( I + \sum\limits_{i=1}^{p}\sum\limits_{k_i=0}^{m_i-s_i} \frac{B_{i,k_i}(x,t)}{(\lambda-\lambda_i)^{k_i}}\right),
                                    \end{array} \right.
    \end{equation}
    where $A_{i,0}=B_{i,0}=0$ and
    \begin{equation}
        A_{i,s_i-k} = \frac{1}{k!} \left.\frac{\partial^k}{\partial \lambda^k} \Big( D(\lambda)\Big)\right|_{\lambda = \lambda_{i}},\quad 0\le k\le s_i-1,
    \end{equation}
    \begin{equation}
        B_{i,m_i-s_i-k} = \frac{1}{k!} \left.\frac{\partial^k}{\partial \lambda^k} \Big(\prod_{j=1}^{p}(\lambda-\lambda_j)^{m_j} D^{-1}(\lambda)\Big)\right|_{\lambda = \lambda_{i}},\quad 0\le k\le m_i-s_i-1.
    \end{equation}
\end{corollary}

\begin{demo}
Notice that both sides of the first equation in (\ref{eq:Dformalpoleform}) are monic polynomials of $\lambda$ with degree $s$. Define
    \begin{equation}
        A_{i,s_i-k} \triangleq \frac{1}{k!} \left.\frac{\partial^k}{\partial \lambda^k} \Big( D(\lambda)\Big)\right|_{\lambda = \lambda_{i}},\quad 1\le i \le p,\;0\le k\le s_i-1.
    \end{equation}
Then following the same proof as Theorem~\ref{thm:DDIeq}, the first equation in (\ref{eq:Dformalpoleform}) holds.
\par
On the other hand, the second equation in (\ref{eq:Dformalpoleform}) is equivalent to
\begin{equation}\label{eq:whyD-1polynomial}
        {\rm det}\Big( D(x,t;\lambda) \Big)D^{-1}(x,t;\lambda) = \prod\limits_{i=1}^{p}(\lambda-\lambda_i)^{m_i-s_i} \left( I + \sum\limits_{i=1}^{p}\sum\limits_{k_i=0}^{m_i-s_i} \frac{B_{i,k_i}(x,t)}{(\lambda-\lambda_i)^{k_i}}\right),
\end{equation}
according to the equation (\ref{eq:detDnfrac}).
\par
The left hand side of (\ref{eq:whyD-1polynomial}) is the adjoint matrix of $D(x,t;\lambda)$, which is a monic polynomial of $\lambda$ with degree $s$ since $D(x,t;\lambda)$ is a $2\times 2$ matrix. The right hand side of (\ref{eq:whyD-1polynomial}) is also a monic polynomial of $\lambda$ with degree $s$. Then by the same proof as Theorem~\ref{thm:DDIeq}, it can be verified that the second equation in (\ref{eq:Dformalpoleform}) holds. \QED

\end{demo}

\begin{remark}[Special cases of DT in Theorem~\ref{thm:DTingeneral}]\label{rmk:gDTreduce} Some remarks for Theorem~\ref{thm:DTingeneral} are given below:
\par
(i) When $s=1$ and $p=2$, the $D(\lambda)$ in (\ref{eq:DT2s}) provides a classic Darboux transformation.
\par
(ii) If there exists a subset $\alpha \subset \{1,2,\cdots,p\}$ such that
    \begin{equation}
        \sum_{j\in \alpha} m_j = s,
    \end{equation}
    the $D(\lambda)$ in (\ref{eq:DT2s}) provides a generalized Darboux transformation. For example, for the focusing NLS equation (\ref{eq:fNLS}) in Section \ref{Section4}, the spectral parameters $\lambda_j~(j=1,2,\cdots,n)$ and $-\bar\lambda_j$ always appear in pairs, that is, $\lambda_j$ are the poles of $D(\lambda)$ while $-\bar\lambda_j$ are the poles of $D^{-1}(\lambda)$. A common condition for this type of generalized Darboux transformation is
    \begin{equation}
        \{\lambda_1,\cdots,\lambda_n \}\cap  \{-\bar\lambda_1,\cdots,-\bar\lambda_n \} = \emptyset.
    \end{equation}
\par
(iii) When $s=1$ and $p=1$, the $D(\lambda)$ in (\ref{eq:DT2s}) just provides the Darboux transformation in Theorem~\ref{thm:DT1} or Theorem~\ref{thm:order1threeequi}.
\par
    (iv) When $s$ is any positive integer and $p=1$, the $D(\lambda)$ in (\ref{eq:DT2s}) provides the Darboux transformation in Theorem~\ref{thm:Generaln}.
\end{remark}

Theorem~\ref{thm:DTingeneral} gives a comprehensive framework for DTs and their explicit construction. Additionally, Remark~\ref{rmk:gDTreduce} informs us of which DTs are included in Theorem~\ref{thm:DTingeneral}. However, there is still a lack of an appropriate criterion to judge the generality of the DTs given by Theorem~\ref{thm:DTingeneral}. Can the $D(\lambda)$ in (\ref{eq:DT2s}) provide all monic polynomial DTs of $\lambda$, if not, what kind of DTs can it construct? The next theorem answers this question. In brief, a Darboux matrix can be decomposed into the product of $n$ first-order monic Darboux matrices if and only if it is the product of Darboux matrix given by Theorem~\ref{thm:DTingeneral} and some trivial scalar Darboux matrices. According to Corollary \ref{corollary:DTtonDet}, the sufficiency is obvious. The necessity is proved in Theorem \ref{thm:DTall}.

\begin{theorem}[The DT can be constructed by solving (\ref{eq:DTgeneral}) if it is decomposable]\label{thm:DTall}
    For a monic Darboux matrix of order $n$ for the Lax pair $($\ref{eq:laxpair1}$)$, i.e.,
    \begin{equation}\label{thm:Darboux512}
        D(x,t;\lambda) = \lambda^n I + \lambda^{n-1}D_1 + \cdots + D_n,
    \end{equation}
    if at some point $(x_0,t_0)\in\mathbb{R}^2$ it can be decomposed into
    \begin{equation}\label{eq:DTinitialfrac}
        D(x_0,t_0;\lambda) = (\lambda I - K_n) (\lambda I - K_{n-1}) \cdots (\lambda I - K_1),
    \end{equation}
    where $K_1,K_2,\cdots,K_n\in {\rm Mat}(2,\mathbb{C})$, then there exist distinct $\lambda_j\in\mathbb{C}$, arbitrary $\mu_k\in\mathbb{C}$, $m_j\in\mathbb{N}_+$ and nonzero $2\times 1$ solutions $b_j(x,t;\lambda)$ of the Lax pair $($\ref{eq:laxpair1}$)$ which are analytic at $\lambda_j$ for $j=1,2,\cdots,p$ and $k=1,2,\cdots,q$, such that $\frac{1}{2}\sum_{j=1}^{p} m_j + q =n$ and the Darboux matrix can be constructed by
    \begin{equation}\label{eq:Dclassn}
        D(x,t;\lambda) = \prod_{k=1}^{q}(\lambda-\mu_k) G(x,t;\lambda),
    \end{equation}
    where $G(x,t;\lambda)=\lambda^{n-q} I+\sum_{i=1}^{n-q} \lambda^{n-q-i} G_{i}(x,t)  $ is uniquely determined by Theorem \ref{thm:DTingeneral} with coefficients $G_1,G_2,\cdots,G_{n-q}$ solving the following linear equations
    \begin{equation}\label{eq:lineareqG}
        \left.\frac{\partial^{k_j}}{\partial \lambda^{k_j}}\Big(   G(x,t;\lambda)    b_j(x,t;\lambda)    \Big) \right|_{\lambda=\lambda_j} =0, \quad 0\le k_j \le m_j-1,\;j=1,2,\cdots,p.
    \end{equation}
\end{theorem}

\begin{demo} This theorem is proved by induction. For $n=1$, according to Theorem~\ref{thm:classificationDT1}, there are three types of DT, namely,
\par
    (i) $(\lambda -c) I$;
\par
    (ii) $B[\lambda_1,\lambda_2,1,h_1(x,t;\lambda),h_2(x,t;\lambda)]$;
\par
    (iii) $D[\lambda_0,1,h(x,t;\lambda)]$.
\par
The first one is the form $(\lambda - \mu_1)I$ with $p=0$, the second one is the solution of
    \begin{equation}\label{eq:516}
        \left\{ \begin{array}{l} B(x,t;\lambda_1) h_1(x,t;\lambda_1) = 0,\\ B(x,t;\lambda_2) h_2(x,t;\lambda_2) = 0,
    \end{array}\right.
    \end{equation}
    and the third one is the solution of
    \begin{equation}\label{eq:517}
        \left\{ \begin{array}{l} D(\lambda_0) h(\lambda_0) = 0,\\ \left.\frac{\partial}{\partial \lambda}\Big(   D(x,t;\lambda)    h(x,t;\lambda)    \Big) \right|_{\lambda=\lambda_0} =0.  \end{array}\right.
    \end{equation}
Both (\ref{eq:516}) and (\ref{eq:517}) take the form of (\ref{eq:lineareqG}), thus they satisfy the Theorem~\ref{thm:DTall} with $q=0$. Then the theorem holds for $n=1$.
\par
By the equation (\ref{thm:Darboux512}), comparing the coefficients of $\lambda$ in
    \begin{equation}
    \begin{array}{l}
        \widetilde{U}(\lambda) D(\lambda) = D_x(\lambda)  +   D(\lambda) U(\lambda), \\
        \widetilde{V}(\lambda) D(\lambda) = D_t(\lambda)  +   D(\lambda) V(\lambda),
    \end{array}
    \end{equation}
yields the definition of $\widetilde{U}(\lambda)$, $\widetilde{V}(\lambda)$ and the compatible ODEs for the elements of $D_j(x,t)$. Then (\ref{thm:Darboux512}) provides a Darboux matrix if and only if the compatible ODEs hold.
\par
Notice that each of the compatible ODEs is a first order nonlinear ODEs, then for any initial data at $(x_0,t_0)$, there exists unique solution to the compatible ODEs.
\par
Now suppose that this theorem holds for $1,2,\cdots,n-1$, then consider a monic Darboux matrix $D^{\{n-1\}}(x,t;\lambda)$ of order $n-1$ for the Lax pair (\ref{eq:laxpair1}) with initial value
    \begin{equation}
        D^{\{n-1\}}(x_0,t_0;\lambda) = (\lambda I - K_{n-1}) (\lambda I - K_{n-2}) \cdots (\lambda I - K_{1}).
    \end{equation}
Induction hypothesis shows that there exist $\lambda_j$, $\mu_k$, $m_j$ and $b_j(x,t;\lambda)$ for $j=1,2,\cdots,p$ and $k=1,2,\cdots,q$), such that
    \begin{equation}\label{eq:DTn-1fac}
        D^{\{n-1\}}(x,t;\lambda) = \prod_{k=1}^{q}(\lambda-\mu_k) G(x,t;\lambda)
    \end{equation}
where $\frac{1}{2}\sum_{j=1}^{p} m_j + q =n-1$ and $G(x,t;\lambda)$ is a monic polynomial of $\lambda$ with degree $n-1-q$ which satisfies
    \begin{equation}\label{eq:Glinearpartproof}
        \left.\frac{\partial^{k_j}}{\partial \lambda^{k_j}}\Big(   G(x,t;\lambda)    b_j(x,t;\lambda)    \Big) \right|_{\lambda=\lambda_j} =0, \quad 0\le k_j \le m_j-1,\;j=1,2,\cdots,p.
    \end{equation}
\par
If $q\ge 1$ in (\ref{eq:DTn-1fac}), consider a Darboux matrix of order $n-q$ for the Lax pair (\ref{eq:laxpair1}) with initial value
    \begin{equation}
        D_a^{\{n-q\}}(x_0,t_0;\lambda) = (\lambda I - K_{n}) G(x_0,t_0;\lambda).
    \end{equation}
Notice that the scalar multiplier $\prod_{k=1}^{q}(\lambda-\mu_k)I$ is always a Darboux matrix which does not change the Lax pair (\ref{eq:laxpair1}) since it transforms $U(\lambda)$, $V(\lambda)$ to $U^{\{q\}}(\lambda)=U(\lambda)$, $V^{\{q\}}(\lambda)=V(\lambda)$. By the equation (\ref{eq:DTn-1fac}), the commutative diagram below can be checked
    \begin{displaymath}
    \xymatrix{
        U(\lambda) \ar[r]^{D^{\{n-1\}}(\lambda)} \ar[d]^{{\rm scalar}} & U^{\{n-1\}}(\lambda)   \\
           U^{\{q\}}(\lambda)=U(\lambda)    \ar[ur]^{G(\lambda)}     &  }
    \end{displaymath}
where $G(\lambda)=G(x,t;\lambda)$ is a Darboux matrix of the Lax pair (\ref{eq:laxpair1}) since $D^{\{n-1\}}(x,t;\lambda)$ is a Darboux matrix and $U^{\{n-1\}}(\lambda)$ is a polynomial of $\lambda$.
\par
According to Corollary \ref{corollary:DTtonDet}, the Darboux matrix $G(x,t;\lambda)$ in (\ref{eq:DTn-1fac}) is the $n-q-1$ folds of some Darboux matrixes of order one since $G(x,t;\lambda)$ is the unique solution of (\ref{eq:Glinearpartproof}). Considering the initial values at $(x_0,t_0)$, $G(x_0,t_0;\lambda)$ is the product of $n-q-1$ matrices of the form $\lambda I - C$, where $C\in {\rm Mat}(2,\mathbb{C})$. Thus we have
    \begin{equation}
        D_a^{\{n-q\}}(x,t;\lambda) = \prod_{k=1}^{\tilde{q}}(\lambda-\tilde{\mu}_k) \tilde{G}(x,t;\lambda),
    \end{equation}
where $\tilde{G}(x,t;\lambda)$ is a solution of a linear system (\ref{eq:lineareqG}) by induction. Let
    \begin{equation}\label{eq:Dan}
        D_a^{\{n\}}(x,t;\lambda) = \prod_{k=1}^{q}(\lambda-\mu_k) D_a^{\{n-q\}}(x,t;\lambda) = \prod_{k=1}^{q}(\lambda-\mu_k)\prod_{k=1}^{\tilde{q}}(\lambda-\tilde{\mu}_k) \tilde{G}(x,t;\lambda),
    \end{equation}
    then it is a Darboux matrix of order $n$ with initial value
    \begin{equation*}
    \begin{array}{l}
        D_a^{\{n\}}(x_0,t_0;\lambda) =  \prod\limits_{k=1}^{q}(\lambda-\mu_k) D_a^{\{n-q\}}(x_0,t_0;\lambda) =\prod\limits_{k=1}^{q}(\lambda-\mu_k) (\lambda I - K_{n}) G(x_0,t_0;\lambda) \\
        = (\lambda I - K_{n})D^{\{n-1\}}(x_0,t_0;\lambda) = (\lambda I - K_n) (\lambda I - K_{n-1}) \cdots (\lambda I - K_1).
    \end{array}
    \end{equation*}
Thus $D_a^{\{n\}}(x,t;\lambda)$ is the unique monic polynomial DT of degree $n$ with initial value (\ref{eq:DTinitialfrac}), and it can be constructed by (\ref{eq:Dan}).
\par
If $q=0$ in (\ref{eq:DTn-1fac}), consider the monic Darboux matrix $D^{\{1\}}(x,t;\lambda)$ of order one for the Lax pair of triplet $\Big(U^{\{n-1\}}(\lambda),V^{\{n-1\}}(\lambda),\Phi^{\{n-1\}}(\lambda)\Big)$ with initial value $\lambda I- K_n$.
\par
{\bf Case 1}: If $D^{\{1\}}(x,t;\lambda)=(\lambda - c) I$, then $(\lambda - c)D^{\{n-1\}}(x,t;\lambda)$ is the DT with initial value (\ref{eq:DTinitialfrac}), where $D^{\{n-1\}}(x,t;\lambda)$ is expressed by (\ref{eq:DTn-1fac}).
\par
For the other two cases, it is necessary to obtain the fundamental solution matrix of Lax pair with triplet $\Big(U^{\{n-1\}}(\lambda),$ $V^{\{n-1\}}(\lambda),\Phi^{\{n-1\}}(\lambda)\Big)$.
\par
Suppose that the fundamental solution matrix of Lax pair (\ref{eq:laxpair1}) is $\Phi(x,t;\lambda)$ that is analytic at $\lambda=\lambda_0$, then it follows
\par
    (i) If $\lambda_0\notin \{  \lambda_1,\lambda_2,\cdots,\lambda_p \}$, then $\Phi^{\{n-1\}}(\lambda) = D^{\{n-1\}}(\lambda)\Phi(\lambda)$ is the fundamental solution matrix of Lax pair with triplet $\Big(U^{\{n-1\}}(\lambda),$ $V^{\{n-1\}}(\lambda),\Phi^{\{n-1\}}(\lambda)\Big)$, which is analytic at $\lambda=\lambda_0$.
\par
(ii) If $\lambda_0 \in \{  \lambda_1,\lambda_2,\cdots,\lambda_p \}$, without loss of generality, assume that $\lambda_0 = \lambda_1$. Since $b_1(x,t;\lambda)$ is a solution of Lax pair (\ref{eq:laxpair1}) and is analytic at $\lambda = \lambda_1$, there exists $(x,t)$-independent $2\times 1$ vector $\xi_1(\lambda)$ satisfying
    \begin{equation*}
        h_1(x,t;\lambda) = \Phi(x,t;\lambda) \xi_1(\lambda).
    \end{equation*}
    Let $\xi_1^{\bot}(\lambda)$ be a $2\times 1$ vector with two properties:
    (a) $\xi_1^{\bot}(\lambda)$ is analytic at $\lambda = \lambda_0$,
    (b) ${\rm det}\left.\Big( \begin{matrix}   \xi_1(\lambda) &   \xi_1^{\bot}(\lambda)     \end{matrix}  \Big)\right|_{\lambda = \lambda_0}\neq0$.
\par
Then it is claimed that
    \begin{equation}\label{eq:Phivarbn-1}
    \Phi^{\{n-1\}}(\lambda)=\left(\begin{matrix} \frac{1}{(\lambda-\lambda_1)^{m_1}}D^{\{n-1\}}(\lambda)\Phi(\lambda)\xi_1(\lambda) & D^{\{n-1\}}(\lambda)\Phi(\lambda)\xi_1^{\bot}(\lambda)  \end{matrix}\right)
    \end{equation}
    is a fundamental solution matrix for triplet $\Big(U^{\{n-1\}}(\lambda),V^{\{n-1\}}(\lambda),\Phi^{\{n-1\}}(\lambda)\Big)$.
\par
Since
    \begin{equation*}
        \left.\frac{\partial^{k_1}}{\partial \lambda^{k_1}}\Big(   D^{\{n-1\}}(x,t;\lambda) \Phi(x,t;\lambda) \xi_1(\lambda)    \Big) \right|_{\lambda=\lambda_j} =0, \quad 0\le k_1 \le m_1-1,
    \end{equation*}
    it is clear that $\lambda=\lambda_1$ is a removable singularity for $\frac{1}{(\lambda-\lambda_1)^{m_1}}D^{\{n-1\}}(\lambda)\Phi(\lambda)\xi_1(\lambda)$. Then the equation (\ref{eq:Phivarbn-1}) and Corollary~\ref{corollary:DTtonDet} imply
    \begin{equation*}
    \begin{array}{l}
        {\rm det}\Big( \Phi^{\{n-1\}}(\lambda) \Big) = {\rm det}\left( D^{\{n-1\}}(\lambda)\Phi(\lambda)\left(\begin{matrix} \xi_1(\lambda) & \xi_1^{\bot}(\lambda)  \end{matrix}\right) \left(\begin{matrix} \frac{1}{(\lambda-\lambda_1)^{m_1}} & 0\\ 0 &1  \end{matrix}\right) \right)\\
        =\prod_{j=2}^{p}(\lambda-\lambda_j)^{m_j} {\rm det}\Big( \Phi(\lambda) \Big) {\rm det}\Big( \begin{matrix}   \xi_1(\lambda) &   \xi_1^{\bot}(\lambda)     \end{matrix}  \Big) \neq 0,
    \end{array}
    \end{equation*}
    at certain neighbourhood of $\lambda_1$. By the definition of Darboux matrix, $\Phi^{\{n-1\}}(\lambda)$ in (\ref{eq:Phivarbn-1}) is a fundamental solution matrix of triplet $\Big(U^{\{n-1\}}(\lambda),V^{\{n-1\}}(\lambda),\Phi^{\{n-1\}}(\lambda)\Big)$ in the punctured neighbourhood of $\lambda_1$. At point $\lambda_1$, the proof is similar to Lemma~\ref{thm:induction} by using the fact that the $\lim\limits_{\lambda\rightarrow\lambda_0}$ and $\frac{\partial}{\partial x}$ are exchangeable. So the claim is proved.
\par
Now it is ready to consider the cases that $D^{\{1\}}(x,t;\lambda)$ is classic Darboux matrix (Case 2) and Darboux matrix with the same poles (Case 3).
\par
    {\bf Case 2}: Assume that $D^{\{1\}}(x,t;\lambda)$ is a classic Darboux matrix, i.e., $K_n\sim \left( \begin{matrix}  \mu_1 & 0 \\ 0 & \mu_2  \end{matrix} \right)$ with $\mu_1\neq\mu_2$. Let $\Phi_1(x,t;\lambda)$ and $\Phi_2(x,t;\lambda)$ be two fundamental solution matrices for the Lax pair (\ref{eq:laxpair1}), which are analytic at $\lambda=\lambda_1$ and $\lambda=\lambda_2$, respectively.
\par
    {\bf Subcase 2.1}: If $\mu_1,\mu_2\notin\{ \lambda_1,\cdots,\lambda_p \}$, then $D^{\{n-1\}}\Phi_1$ and $D^{\{n-1\}}\Phi_2$ are fundamental solution matrices for triplet $\Big(U^{\{n-1\}}(\lambda),V^{\{n-1\}}(\lambda),\Phi^{\{n-1\}}(\lambda)\Big)$ and analytic at $\lambda_1$ and $\lambda_2$, respectively. So there are two $2\times1$ vectors $\eta_1(\lambda)$ and $\eta_2(\lambda)$ such that
    \begin{equation}
        D^{\{ 1 \}} = B[\mu_1,\mu_2,1,D^{\{ n-1 \}}\Phi_1\eta_1,D^{\{ n-1 \}}\Phi_2\eta_2],
    \end{equation}
which results in
    \begin{equation}\label{eq:527}
        \left\{ \begin{array}{l}\left. D^{\{1\}}D^{\{n-1\}}\Phi_1\eta_1 \right|_{\lambda = \mu_1} = 0,\\  \left. D^{\{1\}}D^{\{n-1\}}\Phi_2\eta_2 \right|_{\lambda = \mu_2} = 0.             \end{array}\right.
    \end{equation}
Furthermore, notice that $q=0$, then (\ref{eq:Glinearpartproof}) indicates that
    \begin{equation}\label{eq:528}
        \left.\frac{\partial^{k_j}}{\partial \lambda^{k_j}}\Big(   D^{\{1\}}(\lambda)D^{\{n-1\}}(\lambda)    b_j(\lambda)    \Big) \right|_{\lambda=\lambda_j} =0, \quad 0\le k_j \le m_j-1,\;j=1,\cdots,p.
    \end{equation}
Letting $\lambda_{p+1}=\mu_1$, $\lambda_{p+2}=\mu_2$, $b_{p+1}=\Phi_1\eta_1$, $b_{p+2}=\Phi_2\eta_2$, it is obvious that the monic Darboux matrix $D^{\{1\}}(\lambda)D^{\{n-1\}}(\lambda)$ of order $n$ is uniquely determined by the equations (\ref{eq:527}) and (\ref{eq:528}), where the equations have the same form as equations in (\ref{eq:lineareqG}). So $D^{\{1\}}(\lambda)D^{\{n-1\}}(\lambda)$ can be constructed by the way provided in Theorem~\ref{thm:DTall} with $q=0$.
\par
    {\bf Subcase 2.2}: If $\mu_1\in\{\lambda_1,\cdots,\lambda_p\}$ and $\mu_2\notin\{\lambda_1,\cdots,\lambda_p\}$, without loss of generality, let $\mu_1 = \lambda_1$. Then the fundamental solution matrix analytical at $\mu_1$ is
    \begin{equation}
        \left(\begin{matrix} \frac{1}{(\lambda-\lambda_1)^{m_1}}D^{\{n-1\}}(\lambda)\Phi_1(\lambda)\xi_1(\lambda) & D^{\{n-1\}}(\lambda)\Phi_1(\lambda)\xi_1^{\bot}(\lambda)  \end{matrix}\right),
    \end{equation}
    while the fundamental solution matrix analytical at $\mu_2$ is $D^{\{n-1\}}(\lambda)\Phi_2(\lambda)$.
\par
    Then there exist two $2\times1$ vectors $\eta_1(\lambda)$ and $\eta_2(\lambda)$ which are analytic at $\mu_1$ and $\mu_2$, respectively, such that
    \begin{equation}\label{eq:D1case2.2}
        D^{\{1\}} = B[\mu_1,\mu_2,1,\left(\begin{matrix} \frac{1}{(\lambda-\lambda_1)^{m_1}}D^{\{n-1\}}\Phi\xi_1 & D^{\{n-1\}}\Phi\xi_1^{\bot} \end{matrix}\right)\eta_1, D^{\{n-1\}}\Phi_2\eta_2].
    \end{equation}
\par
    Notice that
    \begin{equation*}
        D^{\{n-1\}}(\lambda)\Phi_1(\lambda)\xi_1(\lambda) = \sum_{j=m_1}^{\infty} \frac{1}{j!} \left.\frac{\partial^{j}}{\partial\lambda^{j}}\Big(     D^{\{n-1\}}(\lambda)\Phi_1(\lambda)\xi_1(\lambda)\Big)\right|_{\lambda=\lambda_1}(\lambda-\lambda_1)^{j},
    \end{equation*}
    which leads to
    \begin{scriptsize}
    \begin{equation}\label{eq:lambdatolambda0isder1}
        \lim_{\lambda\rightarrow\lambda_1} \frac{1}{(\lambda-\lambda_1)^{m_1}}D^{\{n-1\}}(\lambda)\Phi_1(\lambda)\xi_1(\lambda) = \frac{1}{m_1!}    \left.\frac{\partial^{m_1}}{\partial\lambda^{m_1}}\Big(     D^{\{n-1\}}(\lambda)\Phi_1(\lambda)\xi_1(\lambda)\Big)\right|_{\lambda=\lambda_1},
    \end{equation}
    \begin{equation}\label{eq:lambdatolambda0isder2}
        \lim_{\lambda\rightarrow\lambda_1}\frac{\partial}{\partial \lambda}\left( \frac{1}{(\lambda-\lambda_1)^{m_1}}D^{\{n-1\}}(\lambda)\Phi_1(\lambda)\xi_1(\lambda)\right) = \frac{1}{(m_1+1)!}    \left.\frac{\partial^{m_1+1}}{\partial\lambda^{m_1+1}}\Big(     D^{\{n-1\}}(\lambda)\Phi_1(\lambda)\xi_1(\lambda)\Big)\right|_{\lambda=\lambda_1}.
    \end{equation}
    \end{scriptsize}
    \par
The $D^{\{1\}}(\lambda)$ in (\ref{eq:D1case2.2}) satisfies
    \begin{small}
    \begin{equation}
    \left\{
    \begin{array}{l}
        \left.D^{\{1\}}(\lambda) \left(\begin{matrix} \frac{1}{(\lambda-\lambda_1)^{m_1}}D^{\{n-1\}}(\lambda)\Phi_1(\lambda)\xi_1(\lambda) & D^{\{n-1\}}(\lambda)\Phi_1(\lambda)\xi_1^{\bot}(\lambda)  \end{matrix}\right) \eta_1(\lambda) \right|_{\lambda=\mu_1} = 0,\\
        \left.D^{\{1\}}(\lambda)D^{\{n-1\}}(\lambda) \Phi_2(\lambda)\eta_2(\lambda)\right|_{\lambda=\mu_2} = 0.
    \end{array}\label{eq:D1Dn-1Case22}
    \right.
    \end{equation}
    \end{small}
    \par
    Denote $\eta_1(\lambda) = (\eta_{1a}(\lambda),\eta_{1b}(\lambda))^T$. If $\eta_{1a}(\lambda_1)\neq 0$, the first equation in (\ref{eq:D1Dn-1Case22}) and the equation (\ref{eq:Glinearpartproof}) lead to
    \begin{equation}
        \left. \frac{\partial^{m_1}}{\partial\lambda^{m_1}}\Big( D^{\{1\}} D^{\{n-1\}} \Phi_1 \xi_1 \Big) \right|_{\lambda=\lambda_1} + m_1!\frac{\eta_{1b}}{\eta_{1a}}D^{\{1\}}D^{\{n-1\}}\Phi_1\xi_1^{\bot} =0,
    \end{equation}
   that is,
    \begin{equation}
        \left.\frac{\partial^{m_1}}{\partial\lambda^{m_1}}\Big( D^{\{1\}} D^{\{n-1\}} \Phi_1 \tilde{\xi}_1  \Big)\right|_{\lambda = \lambda_1} = 0,
    \end{equation}
    where
    \begin{equation}
        \tilde{\xi}_1(\lambda) = \sum_{j=0}^{m_1-1}\frac{1}{j!}(\lambda-\lambda_1)^j(\xi_1)_0^{(j)} + \frac{1}{m_1!}(\lambda-\lambda_1)^{m_1}\Big((\xi_1)_0^{(m_1)}+m_1!\frac{\eta_{1b,0}}{\eta_{1a,0}}\xi_{1,0}^{\bot}\Big).
    \end{equation}
    \par
    Notice that $(\xi_1)_0^{(j)} =(\tilde{\xi}_1)_0^{(j)}$  for $j=0,1,\cdots,m_1-1$, then we have
    \begin{equation}
        \left.\frac{\partial^{j}}{\partial\lambda^{j}}\Big( D^{\{1\}} D^{\{n-1\}} \Phi_1 \tilde{\xi}_1  \Big)\right|_{\lambda = \lambda_1} = 0, \quad j=0,1,\cdots,m_1.
    \end{equation}
    Redefine $\tilde{\lambda}_1=\lambda_1$, $\tilde{m}_1=m_1+1$, $\tilde{b}_1 = \Phi_1\tilde{\xi}_1$, and let $\lambda_{p+1}=\mu_2$, $m_{p+1}=1$, $b_{p+1}=\Phi_2\eta_2$, then $D^{\{1\}} D^{\{n-1\}}$ is a DT of the form (\ref{eq:Dclassn}) with initial value (\ref{eq:DTinitialfrac}).
\par
    If $\eta_{1a}(\lambda_1)= 0$, the product $D^{\{1\}} D^{\{n-1\}}$ satisfies
    \begin{equation}\label{eq:D1Dn-1mixeq}
        \left\{
        \begin{array}{l}
            \left.D^{\{1\}} D^{\{n-1\}} \Phi_1 \xi_1^{\bot}\right|_{\lambda=\lambda_1} = 0,\\
            \left.D^{\{1\}} D^{\{n-1\}} \Phi_2 \eta_2\right|_{\lambda=\mu_2} = 0, \\
            \left. \frac{\partial^{k_j}}{\partial\lambda^{k_j}}\Big(D^{\{1\}} D^{\{n-1\}}b_j\Big)\right|_{\lambda=\lambda_j} = 0,\quad 0\le k_j \le m_j-1,\; 1\le j\le p.
        \end{array} \right.
    \end{equation}
    However, the above equations are not the standard form (\ref{eq:lineareqG}) since it contains two solutions $\Phi_1 \xi_1^{\bot}$ and $b_1$ of Lax pair (\ref{eq:laxpair1}), which are analytic at $\lambda=\lambda_1$.
\par
    Now construct a Darboux matrix by $T^{\{1\}}=(\lambda-\lambda_1)I$, which is the unique solution of
    \begin{equation}
        \left\{
        \begin{array}{l}
            \left.T^{\{1\}} \Phi_1 \xi_1^{\bot}\right|_{\lambda=\lambda_1} = 0,\\
            \left.T^{\{1\}} \Phi_1 \xi_1\right|_{\lambda=\lambda_1} = \left.T^{\{1\}} b_1\right|_{\lambda=\lambda_1} =0. \\
        \end{array} \right.
    \end{equation}
Similarly, construct another Darboux matrix by $T^{\{n-1\}}$ for the Lax pair (\ref{eq:laxpair1}), which is uniquely determined by
    \begin{equation}\label{eq:Tn-1standard}
        \left\{
        \begin{array}{l}
            \left.T^{\{n-1\}} \Phi_2 \eta_2\right|_{\lambda=\mu_2} = 0,\\
            \left.\frac{\partial^{k}}{\partial\lambda^{k}}\Big(T^{\{n-1\}}b_1\Big)\right|_{\lambda=\lambda_1} = 0, \quad 0\le k_j \le m_j-2, \\
            \left. \frac{\partial^{k_j}}{\partial\lambda^{k_j}}\Big(T^{\{n-1\}}b_j\Big)\right|_{\lambda=\lambda_j} = 0,\quad 0\le k_j \le m_j-1,\; 2\le j\le p.
        \end{array} \right.
    \end{equation}
    Since $T^{\{1\}}$ is a scalar DT, which turns Lax pair (\ref{eq:laxpair1}) to Lax pair (\ref{eq:laxpair1}) itself, then $T^{\{n-1\}}T^{\{1\}}$ is a DT for Lax pair (\ref{eq:laxpair1}). It can be checked directly that $T^{\{n-1\}}T^{\{1\}}$ solves the equations in (\ref{eq:D1Dn-1mixeq}) as $D^{\{1\}} D^{\{n-1\}}$ does. Since the solution of the equations in (\ref{eq:D1Dn-1mixeq}) is unique, it follows
    \begin{equation}
        D^{\{1\}} D^{\{n-1\}} = T^{\{n-1\}}T^{\{1\}} = (\lambda-\lambda_1) T^{\{n-1\}},
    \end{equation}
in which the equations in (\ref{eq:Tn-1standard}) for DT $T^{\{n-1\}}$ is a standard form as (\ref{eq:lineareqG}).
\par
   {\bf Subcase 2.3}: If $\mu_1,\mu_2\in\{ \lambda_1,\cdots,\lambda_p\}$, the proof is similar to Subcase 2.2, so we omit it for simplicity.
\par
    {\bf Case 3}: Assume that $D^{\{1\}}(x,t;\lambda)$ is a DT with same pole, i.e., $K_n\sim \left( \begin{matrix}  \lambda_0 & 1 \\ 0 & \lambda_0  \end{matrix} \right)$.
\par
    {\bf Subcase 3.1}: If $\lambda_0\notin\{ \lambda_1,\cdots,\lambda_p\}$, the proof is similar to Subcase 2.1.
\par
    {\bf Subcase 3.2}: If $\lambda_0\in\{ \lambda_1,\cdots,\lambda_p\}$, without loss of generality, let $\lambda_0 = \lambda_1$, then the fundamental solution matrix that is analytic at $\lambda_0$ is
    \begin{equation}
        \left( \begin{matrix}\frac{1}{(\lambda-\lambda_0)^{m_1}}D^{\{n-1\}}(\lambda)\Phi(\lambda)\xi_1(\lambda) & D^{\{n-1\}}(\lambda)\Phi(\lambda)\xi_1^{\bot}(\lambda)                \end{matrix} \right).
    \end{equation}
So there exists a $2\times 1$ vector $\eta(\lambda)$ which is analytic at $\lambda_0$ and is independent of $x,t$, such that
    \begin{equation*}
        D^{\{1\}} = D[ \lambda_0 , 1 , \left( \begin{matrix}\frac{1}{(\lambda-\lambda_0)^{m_1}}D^{\{n-1\}}(\lambda)\Phi(\lambda)\xi_1(\lambda) & D^{\{n-1\}}(\lambda)\Phi(\lambda)\xi_1^{\bot}(\lambda)                \end{matrix} \right)\eta(\lambda) ],
    \end{equation*}
which satisfies the following equations by Theorem~\ref{thm:order1threeequi}
    \begin{small}
    \begin{equation*}
    \left\{ \begin{array}{l}  \lim\limits_{\lambda\rightarrow\lambda_0} D^{\{1\}}(\lambda)   \left( \begin{matrix}\frac{1}{(\lambda-\lambda_0)^{m_1}}D^{\{n-1\}}(\lambda)\Phi(\lambda)\xi_1(\lambda) & D^{\{n-1\}}(\lambda)\Phi(\lambda)\xi_1^{\bot}(\lambda)                \end{matrix} \right)\eta(\lambda) =0,\\
    \lim\limits_{\lambda\rightarrow\lambda_0}\frac{\partial}{\partial\lambda}\left( D^{\{1\}}(\lambda)   \left( \begin{matrix}\frac{1}{(\lambda-\lambda_0)^{m_1}}D^{\{n-1\}}(\lambda)\Phi(\lambda)\xi_1(\lambda) & D^{\{n-1\}}(\lambda)\Phi(\lambda)\xi_1^{\bot}(\lambda)                \end{matrix} \right)\eta(\lambda)\right)=0.
    \end{array}  \right.
    \end{equation*}
    \end{small}
\par
Denote $\eta(\lambda) = \Big(\eta_a(\lambda),\eta_b(\lambda)\Big)^T$. Reminding $ D^{\{n-1\}}$ in (\ref{eq:DTn-1fac}) with (\ref{eq:Glinearpartproof}) and the limits in (\ref{eq:lambdatolambda0isder1}) and (\ref{eq:lambdatolambda0isder2}), the above equations are equivalent to

    \begin{equation}
    \left\{ \begin{array}{l}\frac{1}{m_1!}  (\eta_a)_0 \Big( D^{\{1\}}D^{\{n-1\}}\Phi_1\xi_1  \Big)_0^{(m_1)} + (\eta_b)_0\Big( D^{\{1\}}D^{\{n-1\}}\Phi_1\xi_1^{\bot}  \Big)_0 = 0,\\
    \frac{1}{(m_1+1)!}  (\eta_a)_0 (D^{\{1\}})_0 \Big( D^{\{n-1\}}\Phi_1\xi_1  \Big)_0^{(m_1+1)} + \frac{1}{m_1!}  (\eta_a D^{\{1\}})_0^{(1)} \Big( D^{\{n-1\}}\Phi_1\xi_1  \Big)_0^{(m_1)}\\
    \quad +\Big(\eta_b D^{\{1\}}D^{\{n-1\}}\Phi_1\xi_1^{\bot}  \Big)_0^{(1)} = 0.
                           \end{array}  \right.
    \end{equation}
\par
    If $(\eta_a)_0\neq0$, the above equations are equivalent to
    \begin{equation}
    \left\{\begin{array}{l}
        \left.\frac{\partial^{m_1}}{\partial\lambda^{m_1}}\Big( D^{\{1\}} D^{\{n-1\}} \Phi_1 \tilde{\xi}_1  \Big)\right|_{\lambda = \lambda_0} = 0,\\
        \left.\frac{\partial^{m_1+1}}{\partial\lambda^{m_1+1}}\Big( D^{\{1\}} D^{\{n-1\}} \Phi_1 \tilde{\xi}_1  \Big)\right|_{\lambda = \lambda_0} = 0,
    \end{array}\right.
    \end{equation}
    where
    \begin{equation*}
    \begin{array}{l}
        \tilde{\xi}_1(\lambda) = \sum_{j=0}^{m_1-1}\frac{1}{j!}(\lambda-\lambda_0)^j(\xi_1)_0^{(j)} + \frac{1}{m_1!}(\lambda-\lambda_0)^{m_1}\Big((\xi_1)_0^{(m_1)}+m_1!\left(\frac{\eta_{b}}{\eta_{a}}\xi_{1}^{\bot}\right)_0\Big)\\
        \qquad+\frac{1}{(m_1+1)!}(\lambda-\lambda_0)^{m_1+1}\Big((\xi_1)_0^{(m_1+1)}+(m_1+1)!\left(\frac{\eta_{b}}{\eta_{a}}\xi_{1}^{\bot}\right)_0^{(1)}\Big).
    \end{array}
    \end{equation*}
    Notice that $(\xi_1)_0^{(j)} =(\tilde{\xi}_1)_0^{(j)}$  for $j=0,1,\cdots,m_1-1$, then one has
    \begin{equation}
        \left.\frac{\partial^{j}}{\partial\lambda^{j}}\Big( D^{\{1\}} D^{\{n-1\}} \Phi_1 \tilde{\xi}_1  \Big)\right|_{\lambda = \lambda_1} = 0, \quad j=0,1,\cdots,m_1+1.
    \end{equation}
    Redefine $\tilde{\lambda}_1=\lambda_1$, $\tilde{m}_1=m_1+2$, $\tilde{b}_1 = \Phi_1\tilde{\xi}_1$, then $D^{\{1\}} D^{\{n-1\}}$ is a DT of the form (\ref{eq:Dclassn}) with initial value (\ref{eq:DTinitialfrac}).
\par
If $(\eta_a)_0=0, (\eta_a)_0^{(1)}\neq 0$, then the product $D^{\{1\}}D^{\{n-1\}}$ satisfies
    \begin{equation}
        \left\{
            \begin{array}{l}
                \Big(D^{\{1\}}D^{\{n-1\}}\Phi_1\xi_1^{\bot}\Big)_0 = 0,\\
                \Big(D^{\{1\}}D^{\{n-1\}}\Phi_1\xi_1\Big)_0^{(m_1)} + m_1!\frac{(\eta_b)_0}{(\eta_a)_0^{(1)}}\Big(D^{\{1\}}D^{\{n-1\}}\Phi_1\xi_1^{\bot}\Big)_0^{(1)} = 0,\\
                \left.\frac{\partial^{k_j}}{\partial \lambda^{k_j}}\Big(   D^{\{1\}}D^{\{n-1\}}   b_j    \Big) \right|_{\lambda=\lambda_j} =0, \quad 0\le k_j \le m_j-1,\;j=1,\cdots,p.
            \end{array}\label{eq:D1Dn-1case3.20n0}
        \right.
    \end{equation}
\par
    Now construct a DT by $(\lambda-\lambda_0)T^{\{n-1\}}$, where $T^{\{n-1\}}$ is determined by
    \begin{equation}
    \left\{
    \begin{array}{l}
        \Big(  T^{\{n-1\}}\Phi_1\tilde{\xi}_1  \Big)_0^{(k)} = 0 ,\quad 0\le k\le m_1-1,\\
        \left.\frac{\partial^{k_j}}{\partial \lambda^{k_j}}\Big(   T^{\{n-1\}}   b_j    \Big) \right|_{\lambda=\lambda_j} =0, \quad 0\le k_j \le m_j-1,\;j=2,\cdots,p,
    \end{array}
    \right.
    \end{equation}
where
    \begin{small}
    \begin{equation*}
        \tilde{\xi}_1(\lambda) = \sum_{j=0}^{m_1-2}\frac{1}{j!}(\lambda-\lambda_0)^j(\xi_1)_0^{(j)} + \frac{1}{(m_1-1)!}(\lambda-\lambda_0)^{m_1-1}\Big((\xi_1)_0^{(m_1-1)}+(m_1-1)!\frac{(\eta_{b})_0}{(\eta_{a})_0^{(1)}}\left(\xi_{1}^{\bot}\right)_0\Big).
    \end{equation*}
    \end{small}
    It can be checked directly that $(\lambda-\lambda_0)T^{\{n-1\}} =D^{\{1\}}D^{\{n-1\}} $, since $(\lambda-\lambda_0)T^{\{n-1\}} $ satisfies the equations in (\ref{eq:D1Dn-1case3.20n0}) where $D^{\{1\}}D^{\{n-1\}}$ is replaced with $(\lambda-\lambda_0)T^{\{n-1\}}$.
\par
If $(\eta_a)_0=(\eta_a)_0^{(1)}= 0$ and $m_1=1$, one can change the definition of $\xi_1$ and $\xi_1^{\bot}$. Then the product $D^{\{1\}}D^{\{n-1\}}$ satisfies the equations in (\ref{eq:D1Dn-1mixeq}) and the case is solved.
\par
If $(\eta_a)_0=(\eta_a)_0^{(1)}= 0$ and $m_1\ge2$, then the product $D^{\{1\}}D^{\{n-1\}}$ satisfies
    \begin{equation}
        \left\{
            \begin{array}{l}
                \Big(D^{\{1\}}D^{\{n-1\}}\Phi_1\xi_1^{\bot}\Big)_0 = 0,\\
                \Big(D^{\{1\}}D^{\{n-1\}}\Phi_1\xi_1^{\bot}\Big)_0^{(1)} = 0,\\
                \left.\frac{\partial^{k_j}}{\partial \lambda^{k_j}}\Big(   D^{\{1\}}D^{\{n-1\}}   b_j    \Big) \right|_{\lambda=\lambda_j} =0, \quad 0\le k_j \le m_j-1,\;j=1,\cdots,p.
            \end{array}\label{eq:D1Dn-1case3.200}
        \right.
    \end{equation}
Now consider a DT defined by $(\lambda-\lambda_0)^2 T^{\{n-2\}}$, where $T^{\{n-2\}}$ is determined by the following equations
    \begin{equation}
    \left\{
    \begin{array}{l}
        \Big(  T^{\{n-2\}}\Phi_1\xi_1  \Big)_0^{(k)} = 0 ,\quad 0\le k\le m_1-3,\\
        \left.\frac{\partial^{k_j}}{\partial \lambda^{k_j}}\Big(   T^{\{n-2\}}   b_j    \Big) \right|_{\lambda=\lambda_j} =0, \quad 0\le k_j \le m_j-1,\;j=2,\cdots,p.
    \end{array}
    \right.
    \end{equation}
It can be checked directly that $(\lambda-\lambda_0)^2 T^{\{n-2\}} =D^{\{1\}}D^{\{n-1\}}$, since $(\lambda-\lambda_0)^2 T^{\{n-2\}} $ satisfies the equations in (\ref{eq:D1Dn-1case3.200}) once the term $D^{\{1\}}D^{\{n-1\}}$ is replaced with $(\lambda-\lambda_0)^2 T^{\{n-2\}}$. This ends the proof of the theorem. \QED
\end{demo}

In what follows, the Kaup-Boussinesq equation is taken as an example to illustrate the application of Theorem~\ref{thm:DTingeneral}.

\subsection{An example to apply the Theorem~\ref{thm:DTingeneral}}\
\par
The Kaup-Boussinesq (KB) equation is an approximate of Euler equation arising from incompressible fluid \cite{ivanov2009two}, which takes the form
\begin{equation}
\begin{array}{l}
    u_t + w_x + \frac{3}{2} uu_x = 0,\\
    w_t  - \frac{1}{4} u_{xxx} + \frac{1}{2}uw_x+ wu_x = 0.
\end{array}\label{eq:KBw}
\end{equation}
In certain setting, the equation (\ref{eq:KBw}) is also referred as the good KB equation because it has global bounded travelling wave solutions and its solution is stable under some perturbations \cite{gong2022linear}. The good KB equation (\ref{eq:KBw}) has Lax pair in matrix form
\begin{equation}
    \begin{array}{l}
        \Phi_x = U\Phi = \left(\begin{matrix}  0 & 1 \\ w-\lambda^2+\lambda u &  0    \end{matrix} \right)\Phi,\\
        \Phi_t = V\Phi = \left(\begin{matrix}  \frac{1}{4} u_x & -\lambda-\frac{1}{2}u \\ (\lambda+\frac{1}{2}u)(\lambda^2-\lambda u - w)+\frac{1}{4} u_{xx} & -\frac{1}{4} u_x    \end{matrix} \right)\Phi,
    \end{array}\label{eq:LaxpairKBw}
\end{equation}
which is a energy-dependent system as that in \cite{lin2007darboux,schulze2008darboux}.
\par
Direct calculations show that it is impossible to construct a Darboux matrix of the form (\ref{eq:DT2s}) with $s=1,2$ for the good KB equation (\ref{eq:KBw}) with constant seeds $u=u_0$ and $w=w_0$. Thus a Darboux matrix with other leading term instead of unit matrix $I$ is considered. To do so, take
\begin{equation}
    G(x,t;\lambda) =L(x,t;\lambda)D(x,t;\lambda)=\left(\begin{matrix}  b(x,t) & 0 \\ \; \lambda\; a(x,t) + d(x,t) &  c(x,t)    \end{matrix} \right)  D(x,t;\lambda),
\end{equation}
where $D(x,t;\lambda)$ is a Darboux matrix of the form (\ref{eq:DT2s}). Theorem~\ref{thm:DTall} has told us what such DT likes, where the scalar functions $a(x,t)$, $b(x,t)$, $c(x,t)$ and $d(x,t)$ are determined by keeping the symmetries. Since the spectral parameter $\lambda$ is absent in the determinant ${\rm det}(L(x,t;\lambda)) = b(x,t)  c(x,t)$, then $L(x,t;\lambda)$ is not a typical Darboux transformation. In fact, the matrix $L(\lambda)$ can be regard as certain elementary operation on the polynomial matrix $D(x,t;\lambda)$, whose inverse is
\begin{equation}
    L^{-1}(x,t;\lambda)= \frac{1}{b(x,t)c(x,t)}\left(\begin{matrix}  c(x,t) & 0 \\ \; -\lambda\; a(x,t) - d(x,t) &  b(x,t)    \end{matrix} \right),
\end{equation}
which is still a polynomial of $\lambda$. Thus it is obvious that
\begin{equation}
    \widetilde{U}(x,t;\lambda)\triangleq G_xG^{-1}+GUG^{-1} = L_xL^{-1} +  L\left( D_xD^{-1} + DUD^{-1} \right) L^{-1},
\end{equation}
is also a polynomial of $\lambda$ if $D(x,t;\lambda)$ is a Darboux matrix of order $n$.
\par
If $G(x,t;\lambda)$ provides a Darboux transformation for the good KB equation (\ref{eq:KBw}), there exist two new potential functions $\tilde{u}(x,t)$ and $\tilde{w}(x,t)$ such that
\begin{equation}
    \widetilde{U}(x,t;\lambda) =  \left(\begin{matrix}  0 & 1 \\ \tilde{w}-\lambda^2+\lambda \tilde{u} &  0    \end{matrix} \right).
\end{equation}
\par
Comparing the coefficients of $\lambda$ in $\widetilde{U}G=G_x+GU$, yields
\begin{equation}
    \begin{array}{l}
        \lambda^{n+2}: \quad  c-b + a D_1(1,2)=0,\\
        \lambda^{n+1}: \quad a+b D_1(1,2) =0,\\
                             d D_1(1,2)-b D_1(1,1) + c D_1(2,2) +  aD_2(1,2) -cu - aD_1(1,2)u + b \tilde{u} - a_x =0, \\
        \lambda^{n}:\quad d + a D_1(1,1) + b D_2(1,2) - b D_1(1,2)u-b_x = 0,\\
                d D_2(1,2)-b D_2(1,1)+ c D_2(2,2) + aD_3(1,2) - d D_1(1,2)u - cD_1(2,2)u - aD_2(1,2)u \\
                + bD_1(1,1)\tilde{u}  - cw - aD_1(1,2)w +b\tilde{w} - D_1(1,1) a_x - d_x - a D_{1,x}(1,1)=0,\\
                bD_1(1,2) \tilde{u}-d - aD_1(1,1) -b D_2(1,2) - D_1(1,2) a_x - c_x - a D_{1,x}(1,2)=0,
    \end{array}\label{eq:abcdKB}
\end{equation}
where $D_i(k,j)$ are the $(k,j)$ elements of matrices $D_i$ for $i=1,\cdots,n$ and $k,j=1,2$ and $D_i$ are the coefficients of the Darboux matrix
\begin{equation*}
    D(x,t;\lambda) = \lambda^{n} I + \sum_{i=1}^{n} \lambda^{n-i} D_i(x,t).
\end{equation*}
\par
The scalar functions $a$, $c$ and $d$ can be solved by the first, second and forth equations in (\ref{eq:abcdKB}) as
\begin{equation}
    \begin{array}{l}
        a = -b D_1(1,2),\\
        c  = b+  b D_1(1,2)^2,\\
        d = b D_1(1,1) D_1(1,2) -  bD_2(1,2) + b D_1(1,2) u + b_x,
    \end{array}
\end{equation}
then eliminating $\tilde{u}$ by the third and sixth equations in (\ref{eq:abcdKB}), a linear first order ODE about scalar function $b$ is derived
\begin{equation}\label{eq:equation-b-function}
\begin{array}{l}
    2 (1 + D_1(1,2)^2) b_x +
   D_1(1,2) \Big( D_1(1,1) (D_1(1,2)^2-1) + D_1(2,2)\\
   + D_1(1,2)^2D_1(2,2)
     -2 D_1(1,2) D_2(1,2) + D_1(1,2)^2 u +
     D_{1,x}(1,2)  \Big)b = 0.
\end{array}
\end{equation}
\par
Solving $b$ by the equation (\ref{eq:equation-b-function}) and substituting it into the third and fifth equations in (\ref{eq:abcdKB}), the new potential functions $\tilde{u}$ and $\tilde{w}$ are obtained below
\begin{small}
\begin{equation}\label{eq:uwilKB}
    \tilde{u}  = -\frac{ D_1(1,1)(D_1(1,2)^2-1) + ( 1+D_1(1,2)^2 )D_1(2,2) - 2 D_1(1,2) D_2(1,2) - u + D_{1,x}(1,2)     }{ 1+ D_1(1,2)^2 },
\end{equation}
\end{small}
\begin{small}
\begin{equation}
\begin{array}{l}
   \tilde{w} = \frac{1}{4 (1 +
    D_1(1,2)^2)^2}\Big(D_1(1,1)^2 (-4 + D_1(1,2)^2 + 2 D_1(1,2)^4 + D_1(1,2)^6)\\ +
   4 D_2(1,1) + 4 D_2(1,2)^2 - 4 D_2(2,2) + 4 D_1(2,2) u\\
   +D_1(1,2)^6 (D_1(2,2)^2 - 4 D_2(2,2) + 2 D_1(2,2) u + u^2) \\
   +4 w +
   2 D_1(1,2)^4 (D_1(2,2)^2 + 3  D_1(2,2)u +
      2 (D_2(1,1) - 3 D_2(2,2) + u^2 + w))\\
      -2 D_1(2,2) D_{1x}(1,2) +
   4 u D_{1x}(1,2) - 2 D_{1x}(1,2)^2\\
   +2 D_{1}(1,1) \big((2 + 3 D_1(1,2)^2 + 2 D_1(1,2)^4 + D_1(1,2)^6) D_1(2,2) \\- 8 D_1(1,2) D_2(1,2) - 6 D_1(1,2)^3 D_2(1,2) -
      2 D_1(1,2)^5 D_2(1,2) - 2 u + 3 D_1(1,2)^4 u\\
       +
      D_1(1,2)^6 u +
      D_1(1,2)^2 (2 u - 3 D_{1x}(1,2)) +
      3 D_{1x}(1,2)\big) \\   +
   D_1(1,2)^2 (D_1(2,2)^2 + 8 D_2(1,1) - 12 D_2(2,2) + 4 u^2 +
      8 w + D_1(2,2) (8 u - 2 D_{1x}(1,2))\\ -
      2 u D_{1x}(1,2) + 3 D_{1x}(1,2)^2 -
      4 D_{2x}(1,2)) - 4 D_{2x}(1,2) \\+
   2 D_1(1,2)^5 (2 D_3(1,2) - 2 D_2(1,2) u -
      D_{1x}(1,1) - D_{1x}(2,2) +
      u_x)\\ +
   2 D_1(1,2) (2 D_1(2,2) D_2(1,2) + 2 D_3(1,2) - 6D_2(1,2) u\\ +
      D_{1x}(1,1) + 6 D_2(1,2) D_{1,x}(1,2) -
      D_{1x}(2,2) +
      2 u_x - D_{1xx}(1,2))\\ +
   2 D_1(1,2)^3 (2 D_1(2,2) D_2(1,2) + 4 D_3(1,2) - 6D_2(1,2) u -
      2 D_{1x}(2,2) +
      3 u_x - D_{1xx}(1,2))\Big).
\end{array}\label{eq:wwilKB}
\end{equation}
\end{small}

\begin{remark}
Notice that the elements of $D_2$ and $D_3$ appear in the expressions $\tilde{u}$ and $\tilde{w}$, thus it seems that the equations in (\ref{eq:uwilKB}) and (\ref{eq:wwilKB}) hold only for $n\ge 3$. However, it is seen that if one sets $D_3=0$ for $n=2$ and $D_2=D_3=0$ for $n=1$, the expressions $\tilde{u}$ and $\tilde{w}$ in (\ref{eq:uwilKB}) and (\ref{eq:wwilKB}) still work.
\end{remark}
\par
Now it is ready to solve the inverse problem of Lax pair (\ref{eq:LaxpairKBw}) and derive the exact solutions of the good KB equation (\ref{eq:KBw}). Take the constant seed solution of (\ref{eq:KBw}) as
\begin{equation}
    u_0 = c,\qquad w_0 = d,
\end{equation}
then the fundamental solution matrix of the Lax pair (\ref{eq:LaxpairKBw}) is
\begin{equation}
    \Phi(x,t;\lambda) = \frac{1}{2}e^{-\frac{1}{2}\varphi(ct+2\lambda t + 2x)}\left(\begin{matrix}  e^{(c+2\lambda)\varphi t} + e^{2\varphi x} & \frac{1}{\varphi} \left( e^{2\varphi x}-e^{(c+2\lambda)\varphi t}  \right) \\ \varphi \left(e^{2\varphi x} -e^{(c+2\lambda)\varphi t} \right)   &       e^{(c+2\lambda)\varphi t} + e^{2\varphi x}     \end{matrix}   \right),
\end{equation}
where $\varphi = \sqrt{d+(c-\lambda)\lambda}$.
\par
Since the main focus of the current work is on DT itself, only three exact solutions to the good KB equation (\ref{eq:KBw}) are listed below.
\par
Firstly, taking $c=0$, $d=1$, $s=1$, $m_1=2$, $\lambda_1 = 0 $ and $h_1(\lambda) = \Phi(\lambda)\left( \begin{matrix}  1\\ 0  \end{matrix} \right)$  in Theorem~\ref{thm:DTingeneral}, the exact soliton solution to the good KB equation (\ref{eq:KBw}) is obtained as
\begin{equation}\label{eq:first-soliton-KB}
\begin{array}{l}
    u = -\frac{16 t {\rm sinh}(2 x)}{3 + 8 t^2 + 4 {\rm cosh}(2 x) + {\rm cosh}(4 x)}, \\
w = \frac{32 t^2 (5 + 4 t^2) -45 - 64 {\rm cosh}(2 x) - 20 (1 + 8 t^2) {\rm cosh}(4 x) +
 {\rm cosh}(8 x)}{2 (3 + 8 t^2 + 4 {\rm cosh}(2 x) + {\rm cosh}(4 x))^2},
\end{array}
\end{equation}
whose structures are displayed in Fig. \ref{DarkwexpmulsolKBSP}.

\setcounter{subfigure}{0}
\begin{small}
\begin{figure}[H] 
	\centering  
	\vspace{-0.35cm} 
	\subfigtopskip=2pt 
	\subfigbottomskip=2pt 
	\subfigcapskip=-5pt 
	\subfigure[$u$]{
		\includegraphics[width=0.32\linewidth]{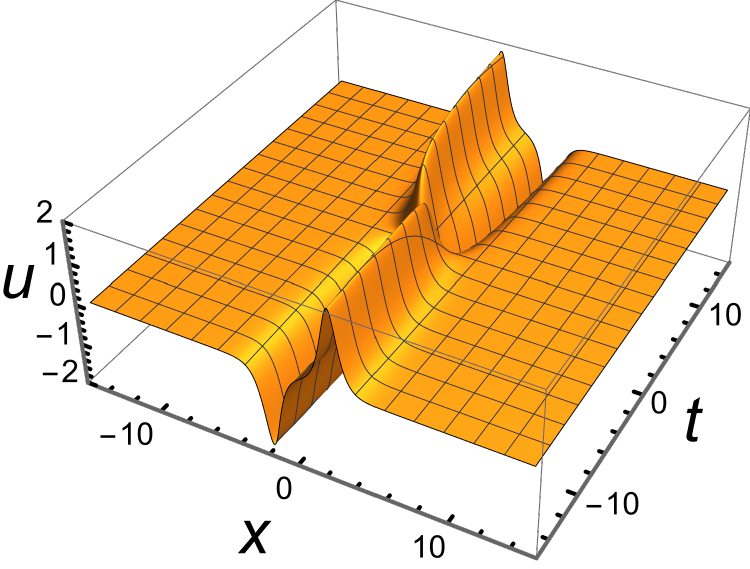}}
	\quad 
	\subfigure[$w$]{
		\includegraphics[width=0.32\linewidth]{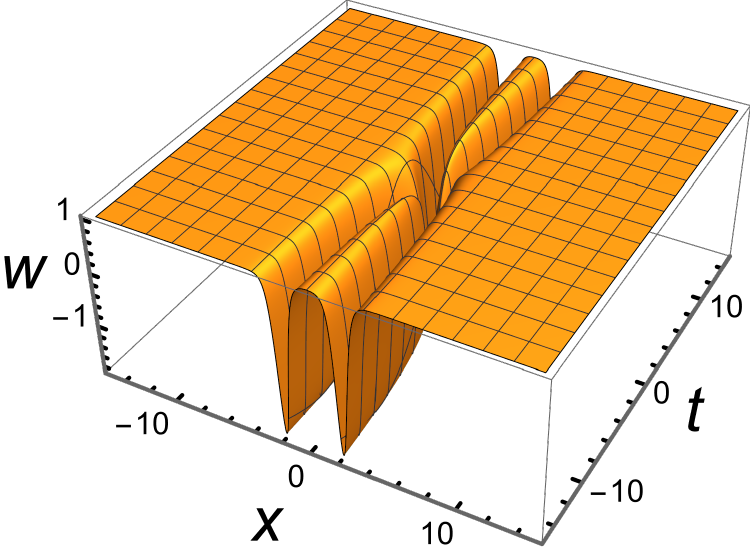}}
\caption{The multi-pole soliton (\ref{eq:first-soliton-KB}) of the KB equation (\ref{eq:KBw}) obtained by the DT with the same pole.}
	\label{DarkwexpmulsolKBSP}
\end{figure}
\end{small}
\par
\par
Secondly, taking $c=0$, $d=1$, $s=1$, $m_1=1$, $m_2=1$, $\lambda_1 = \frac{1}{2} $, $\lambda_2=-\frac{1}{2}$ and $h_1(\lambda) = h_2(\lambda)= \Phi(\lambda)\left( \begin{matrix}  1\\ 0  \end{matrix} \right)$ in Theorem~\ref{thm:DTingeneral}, i.e., the classic DT of the form $B[\lambda_1,\lambda_2,1,h_1(\lambda),h_2(\lambda)]$ in (\ref{eq:binaryDT1}) and (\ref{eq:binaryDT2}), the exact soliton solution to the good KB equation (\ref{eq:KBw}) is obtained as
\begin{small}
\begin{equation}\label{eq:first-soliton-classic-KB}
\begin{array}{l}
    u = -\frac{6 e^{\frac{1}{2} \sqrt{3} (t+2 x)} \left(e^{\sqrt{3} t}-1\right) \left(e^{2 \sqrt{3} x}-1\right)}{e^{\sqrt{3}
   t}+4 e^{2 \sqrt{3} x}+4 e^{2 \sqrt{3} (t+x)}+2 e^{\frac{1}{2} \sqrt{3} (t+2 x)}-2 e^{\sqrt{3} (t+2 x)}+2
   e^{\frac{3}{2} \sqrt{3} (t+2 x)}+2 e^{\frac{1}{2} \sqrt{3} (3 t+2 x)}+e^{\sqrt{3} (t+4 x)}+2 e^{\frac{1}{2}
   \sqrt{3} (t+6 x)}}, \\
w = \frac{w_1}{\left((e^{\sqrt{3}
   t}+4 e^{2 \sqrt{3} x}+4 e^{2 \sqrt{3} (t+x)}+2 e^{\frac{1}{2} \sqrt{3} (t+2 x)}-2 e^{\sqrt{3} (t+2 x)}+2
   e^{\frac{3}{2} \sqrt{3} (t+2 x)}+2 e^{\frac{1}{2} \sqrt{3} (3 t+2 x)}+e^{\sqrt{3} (t+4 x)}+2 e^{\frac{1}{2}
   \sqrt{3} (t+6 x)}\right)^2},
\end{array}
\end{equation}
\end{small}
where
\begin{equation*}
    \begin{array}{l}
        w_1 = e^{2 \sqrt{3} t}+16 e^{4 \sqrt{3} x}+34 e^{2 \sqrt{3} (t+x)}+16 e^{4 \sqrt{3} (t+x)}-21 e^{\sqrt{3} (t+2 x)}\\
        -21
   e^{\frac{3}{2} \sqrt{3} (t+2 x)}-54 e^{2 \sqrt{3} (t+2 x)}-21 e^{\frac{5}{2} \sqrt{3} (t+2 x)}-21 e^{3 \sqrt{3}
   (t+2 x)}\\
   +e^{\frac{1}{2} \sqrt{3} (3 t+2 x)}-21 e^{\sqrt{3} (3 t+2 x)}+e^{\frac{1}{2} \sqrt{3} (5 t+2 x)}+34 e^{2
   \sqrt{3} (t+3 x)}\\
   -14 e^{\sqrt{3} (t+4 x)}+e^{2 \sqrt{3} (t+4 x)}-14 e^{\sqrt{3} (3 t+4 x)}+4 e^{\frac{1}{2}
   \sqrt{3} (t+6 x)}\\
   -21 e^{\sqrt{3} (t+6 x)}-21 e^{\frac{1}{2} \sqrt{3} (5 t+6 x)}+4 e^{\frac{1}{2} \sqrt{3} (7 t+6
   x)}\\
   +4 e^{\frac{1}{2} \sqrt{3} (t+10 x)}-21 e^{\frac{1}{2} \sqrt{3} (3 t+10 x)}+4 e^{\frac{1}{2} \sqrt{3} (7 t+10
   x)}+e^{\frac{1}{2} \sqrt{3} (3 t+14 x)}+e^{\frac{1}{2} \sqrt{3} (5 t+14 x)},
    \end{array}
\end{equation*}
whose structures are displayed in Fig. \ref{DarkwexpmulsolKBclassic}.

\setcounter{subfigure}{0}
\begin{small}
\begin{figure}[H] 
	\centering  
	\vspace{-0.35cm} 
	\subfigtopskip=2pt 
	\subfigbottomskip=2pt 
	\subfigcapskip=-5pt 
	\subfigure[$u$]{
		\includegraphics[width=0.32\linewidth]{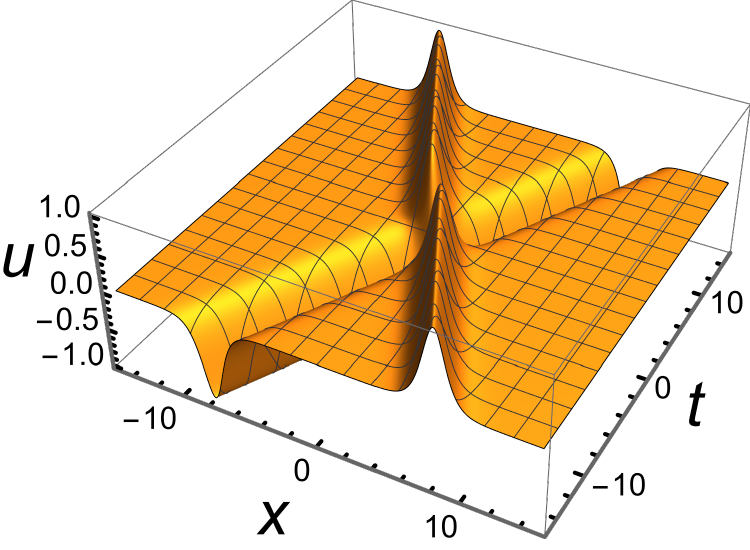}}
	\quad 
	\subfigure[$w$]{
		\includegraphics[width=0.32\linewidth]{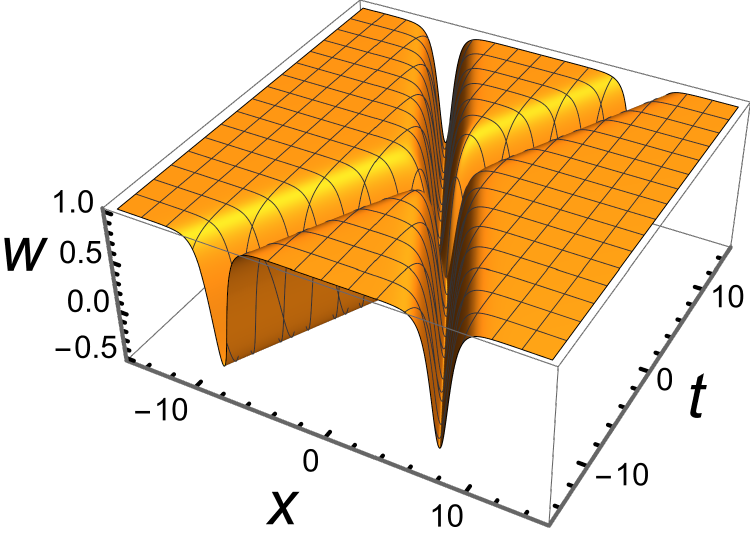}}
\caption{The multi-soliton solution (\ref{eq:first-soliton-classic-KB}) of the KB equation (\ref{eq:KBw}) obtained by a classic DT.}
	\label{DarkwexpmulsolKBclassic}
\end{figure}
\end{small}

Finally, taking $c=0$, $d=0$, $s=2$, $m_1=4$, $\lambda_1 = 0 $ and $h_1(\lambda) = \Phi(\lambda)\left( \begin{matrix}  1\\ 0  \end{matrix} \right)$ in Theorem~\ref{thm:DTingeneral}, another exact solution to the good KB equation (\ref{eq:KBw}) is obtained as
\begin{equation}\label{eq:KBu2samepoleration}
\begin{array}{l}
    u = \frac{4 t x}{t^2 + x^4},\\
w = \frac{2 x^2 (-5 t^2 + x^4)}{(t^2 + x^4)^2},
\end{array}
\end{equation}
which is a rational solution with singularity at point $(0,0)$. However, this singularity point is not just a blow up point because $\lim\limits_{t\rightarrow 0}u(0,t)=0$ and $\lim\limits_{x\rightarrow 0}u(x,x^2)=\infty$, that is, the limit $\lim\limits_{(x,t)\rightarrow 0}u(x,t)$ does not exist. The plots of $u$ for different time $t$ is demonstrated in Fig. \ref{illrationmulpolesolKBSP}.
\begin{small}
\begin{figure}[H] 
	\centering  
	\vspace{-0.35cm} 
	\subfigtopskip=2pt 
	\subfigbottomskip=2pt 
	\subfigcapskip=-5pt 
	\subfigure[$u(x,-10)$]{
		\includegraphics[width=0.25\linewidth]{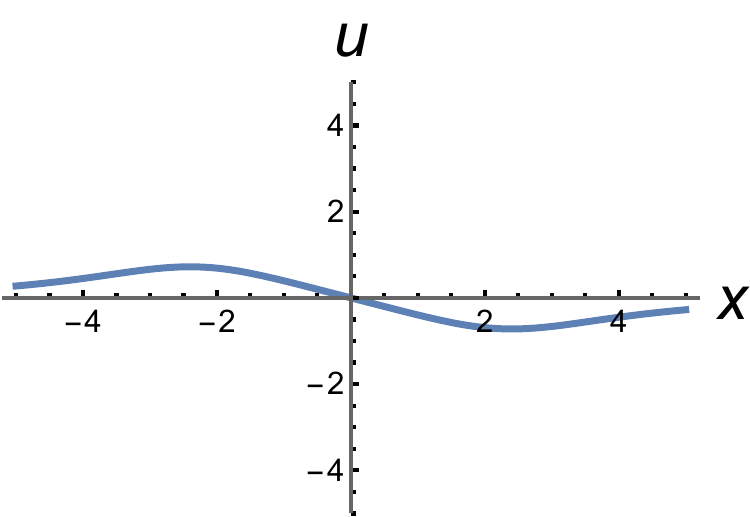}}
	\subfigure[$u(x,-5)$]{
		\includegraphics[width=0.25\linewidth]{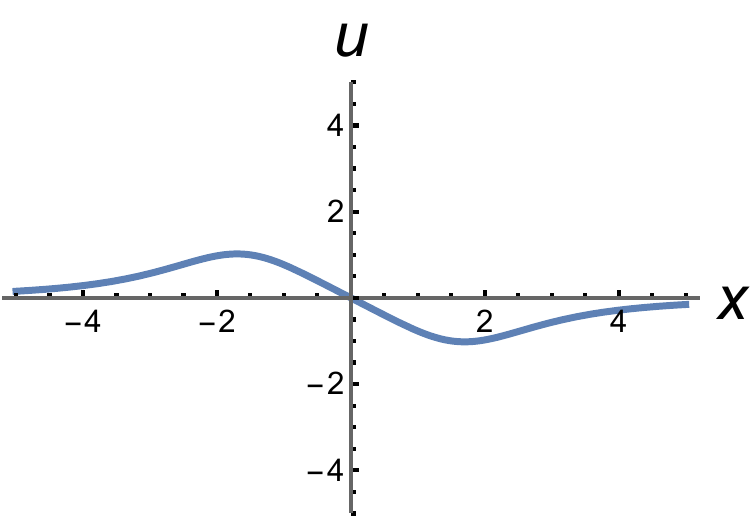}}
	\subfigure[$u(x,-0.25)$]{
		\includegraphics[width=0.25\linewidth]{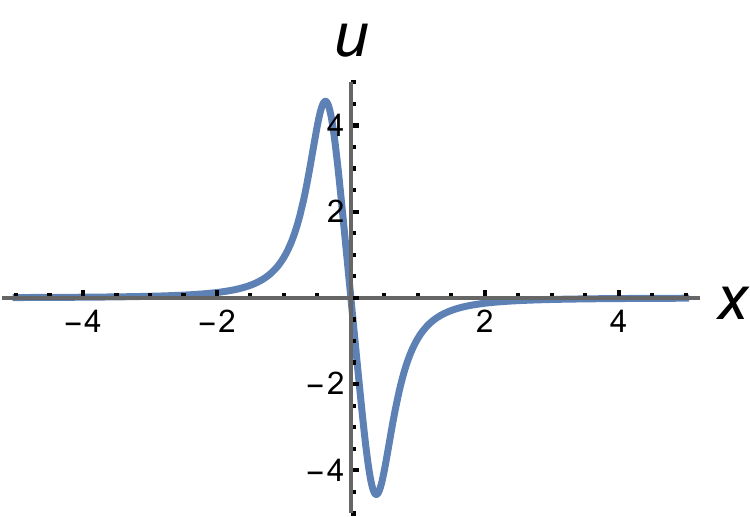}}\\
\subfigure[$u(x,0)$]{
		\includegraphics[width=0.25\linewidth]{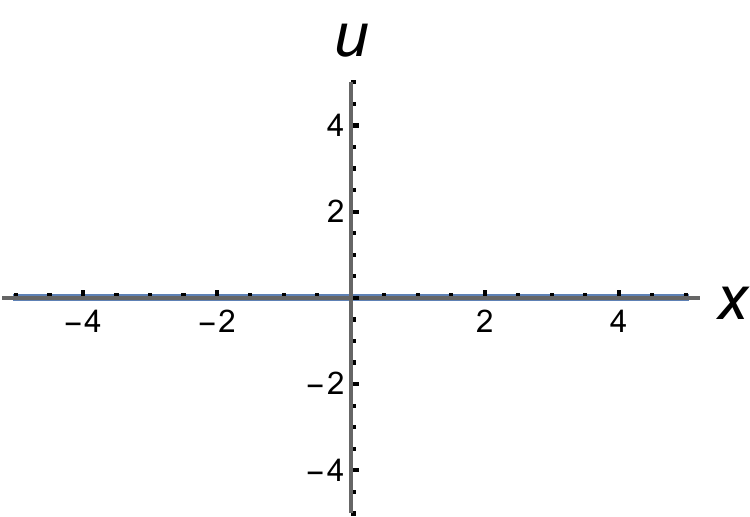}}
	\subfigure[$u(x,0.25)$]{
		\includegraphics[width=0.25\linewidth]{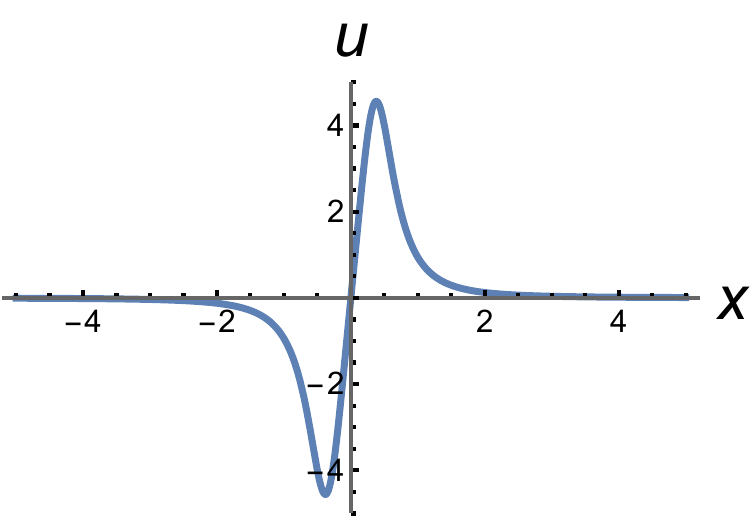}}
	\subfigure[$u(x,10)$]{
		\includegraphics[width=0.25\linewidth]{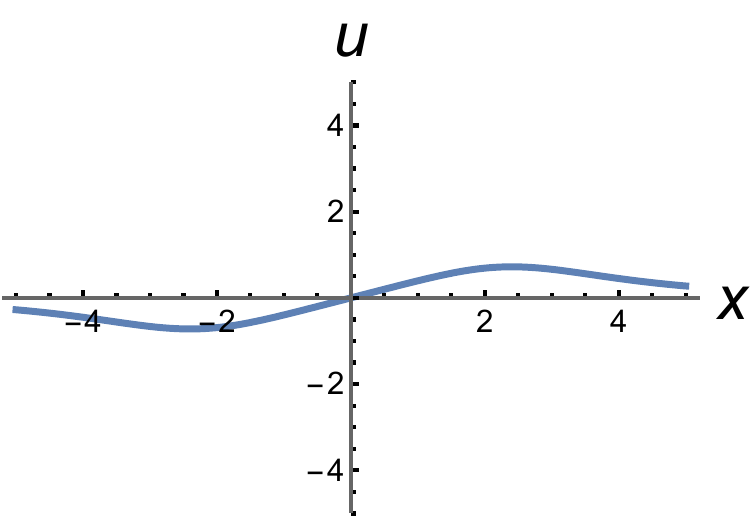}}
\caption{The rational solution of the good KB equation (\ref{eq:KBw}) for different time. }
	\label{illrationmulpolesolKBSP}
\end{figure}
\end{small}

\subsection{Examples of ``irreducible" Darboux matrices}
\par
Recalling Theorem~\ref{thm:DTall}, the Darboux matrix which can be decomposed into the product of $n$ monic Darboux matrices are constructed explicitly. However, is there a Darboux matrix that is monic polynomial of $\lambda$ but can not be decomposed into the product of $n$ monic Darboux matrices?
More generally, in this subsection, we define the irreducible polynomial Darboux matrix (or irreducible polynomial matrix) as the monic polynomial Darboux matrix (or matrix) of $\lambda$ that can not be decomposed into product of lower order polynomial monic Darboux matrices (or matrices).
In what follows, some examples of irreducible Darboux matrices are proposed.
\par
A monic square matrix polynomial $A(\lambda)=\lambda^n I + A_{1} \lambda^{n-1} + \cdots + A_{n}$ is called post-divisible without remainder by $\lambda I-B$ if there exists a polynomial matrix $Q(\lambda)$ such that
\begin{equation}
    A(\lambda) = Q(\lambda) (\lambda I - B).
\end{equation}
A corollary of the generalised Bezoute theorem \cite{kaczorek2007polynomial} shows that monic square matrix polynomial $A(\lambda)$ is post-divisible without remainder by $\lambda I-B$ if and only if the unilateral matrix equation $B^n  + A_{1} B^{n-1} + \cdots + A_{n-1}B + A_{n}=0$ has a solution $B$. There are many studies on the solution of this unilateral equation \cite{macduffee1933theory,kaczorek2007polynomial}, and this corollary leads to the following example that a monic polynomial matrix of degree two can not be decomposed into the product of two monic polynomial matrices of degree one.

\begin{example}\label{eq:example-irreducible}
The polynomial matrix $\lambda^2I+ P \left( \begin{matrix} 0& 1\\0 & 0  \end{matrix} \right)P^{-1}$ is a monic irreducible polynomial matrix.
\end{example}

\begin{demo}
It can be verified that both matrix equation $X^2+\left( \begin{matrix} 0& 1\\0 & 0  \end{matrix} \right)=0$ and $X^2+ P \left( \begin{matrix} 0& 1\\0 & 0  \end{matrix} \right)P^{-1}=0$ with $P\in {\rm GL}(2,\mathbb{C})$ have no solution for $X\in {\rm Mat}(2,\mathbb{C})$.
\par
If $\lambda^2I+ P \left( \begin{matrix} 0& 1\\0 & 0  \end{matrix} \right)P^{-1}$ is not a monic irreducible polynomial matrix,  there exists matrices $B$ and $C$ independent of $\lambda$, such that
    \begin{equation}
        \lambda^2I+ P \left( \begin{matrix} 0& 1\\0 & 0  \end{matrix} \right)P^{-1} = \left(\lambda I - C \right) \left(\lambda I -B \right),
    \end{equation}
     which means that $\lambda^2I+ P \left( \begin{matrix} 0& 1\\0 & 0  \end{matrix} \right)P^{-1}$ is  post-divisible without remainder by $\lambda I -B$. According to the equivalent condition mentioned above, it is obvious that
     \begin{equation*}
        B^2+  P \left( \begin{matrix} 0& 1\\0 & 0  \end{matrix} \right)P^{-1} =0,
     \end{equation*}
     which is in contradiction with the fact that the equation $X^2+ P \left( \begin{matrix} 0& 1\\0 & 0  \end{matrix} \right)P^{-1}=0$ has no solution for $X\in {\rm Mat}(2,\mathbb{C})$.\quad \QED

\end{demo}
\par
Next we aim to find a Darboux matrix $D(x,t;\lambda)=\lambda^2 I + \lambda D_1(x,t) +D_2(x,t)$, which satisfies the initial condition
\begin{equation}\label{eq:initialvalueirrDT}
    D_1(0,0) = 0, \qquad D_2(0,0) = P \left( \begin{matrix} 0& 1\\0 & 0  \end{matrix} \right)P^{-1}.
\end{equation}
Example \ref{eq:example-irreducible} indicates that the initial value $D(0,0;\lambda)=\lambda^2 I +P \left( \begin{matrix} 0& 1\\0 & 0  \end{matrix} \right)P^{-1}$ of $D(x,t;\lambda)$ is irreducible, so
it is difficult to get $D_1(x,t)$ and $D_2(x,t)$ from $\widetilde{U}D=D_x+DU$ directly. However, it is observed that $\lambda^2+ \lambda \epsilon I  + P \left( \begin{matrix} 0& 1\\0 & 0  \end{matrix} \right)P^{-1}$ is reducible for any $\epsilon\neq 0$, that is
\begin{equation*}
    \lambda^2+ \lambda \epsilon I  + P \left( \begin{matrix} 0& 1\\0 & 0  \end{matrix} \right)P^{-1} =  \left( \lambda I - P_2 \left( \begin{matrix} 0& 1\\0 & 0  \end{matrix} \right)P_2^{-1} \right) \left( \lambda I - P_1 \left( \begin{matrix} -\epsilon & 1\\0 & -\epsilon  \end{matrix} \right)P_1^{-1} \right),
\end{equation*}
where $P_2 = P \left( \begin{matrix} 1& 0\\0 & -\epsilon  \end{matrix} \right)$, $P_1 =  P \left( \begin{matrix} -1& 0\\0 & -\epsilon  \end{matrix} \right)$.
\par
A Darboux matrix with initial value $\lambda^2+ \lambda \epsilon I  + P \left( \begin{matrix} 0& 1\\0 & 0  \end{matrix} \right)P^{-1}$ can be constructed below
\begin{equation*}
\begin{array}{l}
    D^{[1]}[-\epsilon,1,h_1(x,t;\lambda)],\\
    D^{[2]}[0,1,h_2(x,t;\lambda)],
\end{array}
\end{equation*}
where
\begin{equation*}
\begin{array}{l}
    h_1(x,t;\lambda) = \Phi(x,t;\lambda)\left( \xi_1 + (\lambda+\epsilon) \xi_1^{(1)}  \right),\\
    \xi_1 = \Phi^{-1}(0,0,-\epsilon)P_1\left( \begin{matrix} 1\\0 \end{matrix} \right),\\
    \xi_1^{(1)} = \Phi^{-1}(0,0,-\epsilon)P_1\left( \begin{matrix} 0\\1 \end{matrix} \right) - \Phi^{-1}(0,0,-\epsilon)\Phi^{(1)}(0,0,-\epsilon)\Phi^{-1}(0,0,-\epsilon)P_1\left( \begin{matrix} 1\\0 \end{matrix}\right) ,\\
    \Psi(x,t;\lambda) = D^{[1]}(x,t;\lambda)\Phi(x,t;\lambda),\\
    h_2(x,t;\lambda) = \Psi(x,t;\lambda)\left( \xi_2 + \lambda \xi_2^{(1)}  \right),\\
    \xi_2 = \Psi^{-1}(0,0,0)P_2\left( \begin{matrix} 1\\0 \end{matrix} \right),\\
    \xi_2^{(1)} = \Psi^{-1}(0,0,0)P_2\left( \begin{matrix} 0\\1 \end{matrix} \right) - \Psi^{-1}(0,0,0)\Psi^{(1)}(0,0,0)\Psi^{-1}(0,0,0)P_2\left( \begin{matrix} 1\\0 \end{matrix}\right) .
    \end{array}
\end{equation*}
Obviously, taking $\epsilon\rightarrow 0$, the limit of $D^{[2]}D^{[1]}$ is the Darboux matrix with the irreducible initial value $\lambda^2+ P \left( \begin{matrix} 0& 1\\0 & 0  \end{matrix} \right)P^{-1}$ at $(x,t)=(0,0)$.
\par
\begin{figure}[H] 
	\centering  
	\vspace{-0.35cm} 
	\subfigtopskip=2pt 
	\subfigbottomskip=2pt 
	\subfigcapskip=-5pt 
	\subfigure[$P=I$]{
		\includegraphics[width=0.35\linewidth]{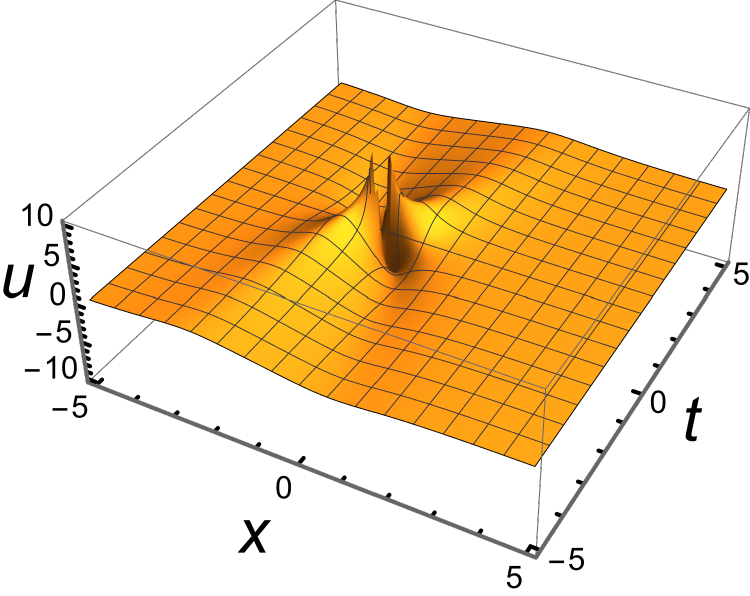}}
\quad
	\subfigure[$P=\sigma_1$]{
		\includegraphics[width=0.35\linewidth]{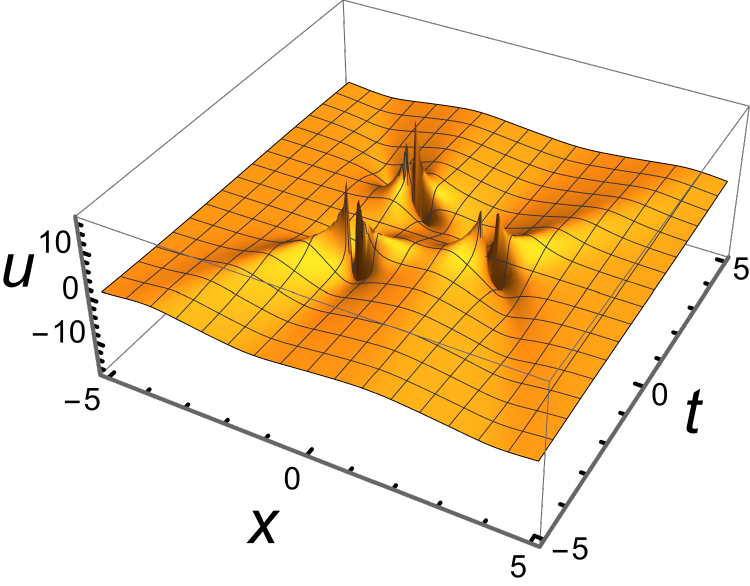}}
\caption{Rational solutions of the good KB equation (\ref{eq:KBw}) by the second-order irreducible DT for different matrix $P$ in initial values (\ref{eq:initialvalueirrDT}). (a) The solution is given by (\ref{eq:uuKBirreduci1}), and the DT of this solution is given by (\ref{eq:DTglobalirr}), which is irreducible for all $(x,t)\in\mathbb{R}^2$. (b) The solution is given by (\ref{eq:uuKBirreduci2}),  and the DT of this solution is given by (\ref{eq:DDKB2Dxt-lambda}), which is irreducible only for $t=0$. }
	\label{irreducibleKBSP}
\end{figure}

Select seed solution of the good KB equation (\ref{eq:KBw}) as $u_0=0$ and $w_0=0$, and take $P=I$, then the limit of $D(x,t;\lambda)=D^{[2]}D^{[1]}$ for $\epsilon\rightarrow 0$ is
\begin{equation}\label{eq:DTglobalirr}
    D(x,t;\lambda) = \lambda^2 I + \lambda  \left( \begin{matrix} 0& \frac{t}{(1+x)^2}\\0 & 0  \end{matrix} \right) +  \left( \begin{matrix} 0& \frac{1}{1+x}\\0 & 0  \end{matrix} \right),
\end{equation}
which is irreducible for all $(x,t)\in\mathbb{R}^2$. Then reminding the reconstruction solution in (\ref{eq:uwilKB}) and (\ref{eq:wwilKB}), the exact rational solution of the good KB equation (\ref{eq:KBw}) is
\begin{equation}
    \begin{array}{l}
        u_{{\rm sin}}(x,t) = \frac{4 t (1 + x)}{t^2 + (1 + x)^4},\\
        w_{{\rm sin}}(x,t) = \frac{2 (1 + x)^2 (-5 t^2 + (1 + x)^4)}{(t^2 + (1 + x)^4)^2},
    \end{array}\label{eq:uuKBirreduci1}
\end{equation}
which can also be derived from (\ref{eq:KBu2samepoleration}) with $x\rightarrow x+1$, and is displayed in Fig. \ref{irreducibleKBSP}(a) for component $u$. This rational solution has singularity at point $(x,t)=(-1,0)$, which has been discussed above. Surprisedly, this kind of singularity possesses higher-order structure by taking different matrix $P$. For example, taking $P=\sigma_1=\left( \begin{matrix} 0& 1\\1 & 0  \end{matrix} \right)$, the limit of $D^{[2]}D^{[1]}$ for $\epsilon\to 0$ is
\begin{equation}\label{eq:DDKB2Dxt-lambda}
     D(x,t;\lambda) = \lambda^2 I + \lambda D_1  + D_2,
\end{equation}
where
\begin{equation*}
\begin{array}{l}
    D_1 =\left( \begin{matrix} \frac{3 t (3 + 2 x^3)}{
  9 t^2 x^2 - (x^3-3)^2}& -\frac{3 t (3 t^2 + 6 x + x^4)}{
   9 t^2 x^2 - (x^3-3)^2)}\\
   \frac{9 t x^2}{
  9 t^2 x^2 - (x^3-3)^2}& \frac{
  3 t (3 + 2 x^3)}{-9 t^2 x^2 + (x^3-3)^2}  \end{matrix} \right),\\
  D_2 =  \left( \begin{matrix} \frac{9 t^2 + 9 x - 3 x^4}{-9 t^2 x^2 + (x^3-3)^2}& \frac{
 3 x (-3 t^2 - 3 x + x^4)}{-9 t^2 x^2 + (x^3-3)^2}\\ \frac{
 9 - 3 x^3}{-9 t^2 x^2 + (x^3-3)^2} & \frac{
 3 x (x^3-3)}{-9 t^2 x^2 + (x^3-3)^2}  \end{matrix} \right).
\end{array}
\end{equation*}
\par
Notice that for $t\neq0$, $D(x,t;\lambda)$ can be decomposed into
\begin{small}
\begin{equation*}
    D(x,t;\lambda) = \left( \lambda I + \left( \begin{matrix}   \frac{(x^3-3 ) (x^4-3 t^2 - 3 x )}{
 t (9 t^2 x^2 - (x^3-3 )^2)} &  -\frac{(x^4-3 t^2 - 3 x )^2}{
  t (9 t^2 x^2 - (x^3-3 )^2)}\\
  \frac{(x^3-3 )^2}{
 t (9 t^2 x^2 - (x^3-3 )^2)}& \frac{(x^3-3 ) (3 t^2 + 3 x - x^4)}{
 t (9 t^2 x^2 - (x^3-3 )^2)}   \end{matrix} \right)   \right) \left( \lambda I - \left( \begin{matrix}   -\frac{x}{t} & \frac{x^2}{t} \\  -\frac{1}{t} & \frac{x}{t}\end{matrix} \right)   \right),
\end{equation*}
\end{small}
which means that $D(x,t;\lambda)$ in (\ref{eq:DDKB2Dxt-lambda}) is irreducible for $t=0$ and is reducible for others.
\par
The exact second-order rational solution of equation (\ref{eq:KBw}) corresponding to matrix $P=\sigma_1$ is
\begin{equation}\label{eq:uuKBirreduci2}
    u_{{\rm tri}}(x,t) = \frac{12 t (18 t^2 x^2 (x^3-3 )-27 t^4 x
   + (x^3-3 ) (9 + 21 x^3 + x^6))}{
81 t^6 + (x^3-3)^4 + 27 t^4 x (12 + 5 x^3) -
 9 t^2 x^2 (x^6-18 - 24 x^3)},
\end{equation}
which is displayed in Fig. \ref{irreducibleKBSP}(b). The component $w$ is very complicated and we omit it for simplicity.

\begin{remark}
If the rational solution in (\ref{eq:uuKBirreduci1}) is regarded as a fundamental singular solution of the good KB equation (\ref{eq:KBw}), then the solution in (\ref{eq:uuKBirreduci2}) is the higher-order version of this fundamental solution. This can be explained as follows. Imagine that there are two coordinate systems on Figure \ref{irreducibleKBSP}(a) and Figure \ref{irreducibleKBSP}(b), respectively, in which the origin of the first coordinate system is located at the singularity $(-1,0)$ of solution (\ref{eq:uuKBirreduci1}), while the origin of the second coordinate system is located at one of the three singularities of the solution (\ref{eq:uuKBirreduci2}), say $S=(\sqrt[3]{3},0)$. For each fixed $x$ and $t$, stretching the entire coordinate systems, the positions of the other two singularities in the second coordinate system will move further and further away, and the function value of (\ref{eq:uuKBirreduci2}) at point $S$ will increasingly influence the stretched system, eventually becoming completely the same as the situation after stretching the first coordinate system, whose formula is
\begin{equation*}
     u_{{\rm tri}}\big(\frac{x}{h}+ \sqrt[3]{3},\; \frac{t}{h}\big) -  u_{{\rm sin}}\big( \frac{x}{h}-1,\; \frac{t}{h} \big) =\mathcal{O}\left(\frac{1}{h^2}\right),\quad h\to \infty,
\end{equation*}
where $h$ is the stretching parameter, and $u_{{\rm sin}}(x,t)$ and $u_{{\rm tri}}(x,t)$ are the rational solutions in (\ref{eq:uuKBirreduci1}) and (\ref{eq:uuKBirreduci2}), respectively.
\end{remark}

\section{Conclusions}

In this paper, a novel generalized Darboux matrix for the $2\times2$ Lax pairs, i.e., the ``Darboux matrix with the same poles", is proposed algebraically. As an application, the stationary single soliton solution and global multi-pole soliton solutions of the PT-symmetric focusing NLS equation are proposed, in which the spectral parameter corresponding to these solitons is exactly $\lambda=0$. In the ``Darboux matrix with the same poles", $D(\lambda)$ and $D^{-1}(\lambda)$ have the identical poles, and the first-order case has a Jordan canonical form consisting of a single Jordan block. Moreover, the first-order monic Darboux matrices have been fully classified, including the trivial ones, classic Darboux matrices, and the ``Darboux matrix with the same poles". A unified theorem is introduced to state that, in generic, constructing a Darboux matrix that can be decomposed into the product of the first-order monic Darboux matrices is equivalent to solving a system of linear algebraic equations. Subsequently, the Kaup-Boussinesq equation serves as an illustration of the theorem, with several new exact solutions presented explicitly. Finally, examples of ``irreducible" Darboux matrices are also studied for the Kaup-Boussinesq equation, where the Darboux matrices can not be expressed by the product of the first-order monic Darboux matrices.

\section*{Acknowledgements}
The authors acknowledge financial support from the National Natural Science Foundation of
China, Grant No. 12371247 and No. 12431008.  The authors would like to thank Prof. Zixiang Zhou for inspiring discussions and helpful suggestions.

\section*{Data availability statement}

All data that support the findings of this study are included within the article.

\section*{Appendix}
The triple-pole soliton of the focusing NLS equation (\ref{eq:fNLS}) derived by (\ref{eq:DTmultipolesolitonNLS}) is
\begin{equation}\label{triple-pole-soliton-fNLS}
u(x,t) = \frac{96 e^{2 x-4 i t} \left(e^{8 x} M_1-4 e^{4 x} M_2 -2592 M_3\right)}{9 e^{12 x}+8 e^{8 x} N_1+144 e^{4 x}
   N_2+26873856},
\end{equation}
where
\begin{equation*}
    \begin{array}{l}
        M_1 =36 x^2 -576 t^2+12 i t (24 x-5)+3 x+184,\\
        M_2 = 2654208 t^4+576 t^2 \left(576 x^2+912 x+433\right)
        +36 i t
   \left(2304 x^2+28056 x+21343\right)\\
   \qquad+10368 x^4+32832 x^3+44172 x^2-26133 x-328021,\\
        M_3 = 4608 t^2+96 i t
   (24 x+43)-288 x^2-888 x+895,\\
   N_1 = 2654208 t^4+1152 t^2 \left(288 x^2+24 x-1471\right)+10368 x^4\\
        \qquad+22464 x^3+131976
   x^2+132384 x+320543,\\
   N_2 = 21233664 t^4+73728 t^2 \left(36 x^2+111 x+283\right)+82944 x^4\\
        \qquad+345600 x^3-307584 x^2-1411536
   x+1545241.
    \end{array}
\end{equation*}
\par
The quadruple-pole soliton of the focusing NLS equation (\ref{eq:fNLS}) derived by (\ref{eq:DTmultipolesolitonNLS}) is
\begin{equation}\label{quadruple-pole-soliton-fNLS}
u(x,t) = \frac{16 e^{2 x-4 i t} \left(-3 e^{12 x} M_1 +e^{8 x} M_2+e^{4 x}
   M_3 +3 M_4\right)}{9 e^{16 x}+ 4 e^{12 x} N_1 +2 e^{8 x}
   N_2+4 e^{4 x} N_3+9},
\end{equation}
where
\begin{small}
\end{small}
\begin{small}
\begin{equation*}
    \begin{array}{l}
    M_1 =8 x^3 -512 i t^3-384 t^2 (x-3)+48 i t \left(2 x^2-11 x+3\right)-60
   x^2+6 x-347,\\
        M_2 = 8388608 i t^7+2097152
   t^6 (x-2)+786432 i t^5 \left(2 x^2-9 x+22\right)\\
   \qquad+49152 t^4 \left(8 x^3-60 x^2+294 x-1067\right)\\
   \qquad+1536 i t^3 \left(64
   x^4-448 x^3+464 x^2+1136 x+559\right)\\
   \qquad+384 t^2 \left(64 x^5-704 x^4+3760 x^3+656 x^2-31009 x-21214\right)\\
   \qquad+16 i t
   \left(128 x^6-960 x^5-4128 x^4-24880 x^3+255138 x^2-593085 x-1016125\right)\\
   \qquad+512 x^7-6400 x^6+26112 x^5+13568
   x^4\\
   \qquad-254872 x^3-817500 x^2-2071678 x+4316249,\\
        M_3 = 8388608 i t^7-2097152 t^6 (x-1)+786432 i t^5 \left(2 x^2-3 x-10\right)\\
        \qquad+47872 x^4-49152 t^4 \left(8 x^3-12 x^2-126
   x-515\right)\\
   \qquad+1536 i t^3 \left(64 x^4-320 x^3+1360 x^2-2416 x+3439\right)\\
   \qquad-384 t^2 \left(64 x^5-256 x^4-400 x^3+13664
   x^2-37601 x+84457\right)\\
   \qquad+16 i t \left(128 x^6-1344 x^5+12000 x^4-114448 x^3+246882 x^2-133743 x-1901221\right)\\
   \qquad-512
   x^7+4352 x^6-25344 x^5+50648 x^3-1529460 x^2+8028466 x-2777365,\\
        M_4 = 512 i t^3-384 t^2 x-48 i t \left(2 x^2-x+11\right)+8 x^3-12 x^2+138 x-647,\\
        N_1 = 1048576 t^6+24576 t^4 \left(8 x^2-48
   x+167\right)\\
   \qquad+768 t^2 \left(16 x^4-160 x^3+600 x^2+1328 x-3467\right)\\
   \qquad+256 x^6-3072 x^5+6624 x^4-8000 x^3+149736
   x^2+136080 x+530959,\\
        N_2 = 134217728 t^8+4194304 t^6 \left(8 x^2-24 x+19\right)\\
        \qquad+98304 t^4 \left(32 x^4-192 x^3+512 x^2-6056
   x+9497\right)\\
   \qquad+1024 t^2 \left(128 x^6-1152 x^5+4464 x^4-23680 x^3+37296 x^2-8244 x+833345\right)\\
   \qquad+2048 x^8-24576
   x^7+133120 x^6+171008 x^5-3191040 x^4\\
   \qquad+10058752 x^3+28408928 x^2-59790144 x+29904955,\\
        N_3 = 1048576 t^6+24576 t^4 \left(8 x^2-89\right)+768 t^2 \left(16 x^4-32
   x^3+24 x^2+2312 x+293\right)\\
   \qquad+256 x^6-1536 x^5+12768 x^4-78272 x^3+246120 x^2-931056 x+2109199.
    \end{array}
\end{equation*}
\end{small}

\newcommand{\etalchar}[1]{$^{#1}$}
\par

\end{document}